\documentclass[final,5p,times,twocolumn,authoryear]{elsarticle}

\usepackage{amssymb}
\usepackage{amsmath}

\usepackage{lineno}

\usepackage{upgreek}

\usepackage{pifont}

\usepackage{lipsum}

\usepackage{xcolor}

\journal{Earth and Planetary Science Letters}

\begin{document}

\begin{frontmatter}



\title{Formation of Ryugu's parent planetesimal beyond the CO$_{2}$ snow line from small pebbles: insights from thermal evolution modeling} 


\author[label1]{Sota Arakawa\corref{cor1}} 
\author[label2]{Hidenori Genda} 
\author[label3]{Noriyuki Kawasaki} 
\author[label4]{Wataru Fujiya} 
\author[label5]{Yosei Iwasaki} 
\author[label6]{Shigeru Wakita} 

\cortext[cor1]{Corresponding author: Sota Arakawa (arakawas@jamstec.go.jp)}

\affiliation[label1]{organization={Japan Agency for Marine-Earth Science and Technology},
            addressline={3173-25 Showa-machi, Kanazawa-ku}, 
            city={Yokohama},
            postcode={236-0001}, 
            country={Japan}}

\affiliation[label2]{organization={Earth-Life Science Institute, Institute of Science Tokyo},
            addressline={2-12-1 Ookayama, Meguro-ku}, 
            city={Tokyo},
            postcode={152-8550}, 
            country={Japan}}

\affiliation[label3]{organization={Department of Earth and Planetary Sciences, Hokkaido University},
            addressline={Kita-10 Nishi-8, Kita-ku}, 
            city={Sapporo},
            postcode={060-0810}, 
            country={Japan}}

\affiliation[label4]{organization={Faculty of Science, Ibaraki University},
            addressline={2-1-1 Bunkyo}, 
            city={Mito},
            postcode={310-8512}, 
            country={Japan}}

\affiliation[label5]{organization={School of Science, Institute of Science Tokyo},
            addressline={2-12-1 Ookayama, Meguro-ku}, 
            city={Tokyo},
            postcode={152-8551}, 
            country={Japan}}

\affiliation[label6]{organization={Department of Earth, Atmospheric, and Planetary Sciences, Purdue University},
            city={West Lafayette},
            postcode={47907-2051}, 
            state={IN},
            country={USA}}

\begin{abstract}

Astronomical observations of planet-forming circumstellar disks indicate that planetesimals form from 0.1-mm- to 1-cm-sized dust aggregates, commonly called ``pebbles.''
When such pebbles accrete beyond the water snow line, the resulting icy planetesimals undergo water--rock differentiation and develop porous pebble-pile cores whose voids are saturated with liquid water.
Circulation of this water enhances heat transport in the core, suppressing the temperature rise caused by the decay of radionuclides.
In this study, we constrain the accretion age and constituent pebble size of Ryugu's parent planetesimal by modeling the thermal evolution of icy planetesimals and comparing the results with the precipitation ages and temperatures of aqueously formed minerals identified in samples returned from asteroid Ryugu.
Our numerical results suggest that Ryugu's parent planetesimal accreted within 2.0 Myr of the formation of calcium--aluminum-rich inclusions.
The inferred early accretion age supports the hypothesis that Ryugu's parent planetesimal formed earlier than most chondrule-bearing carbonaceous chondrite parent planetesimals, potentially explaining the absence of chondrules in Ryugu samples.
We also found that the core of the parent planetesimal was composed of pebbles no larger than a few millimeters.
Such small pebble sizes are consistent with theoretical predictions for planetesimals formed beyond the CO$_{2}$ snow line.

\end{abstract}



\begin{keyword}

Aqueous alteration \sep Carbonates \sep Hayabusa2 \sep Planetesimals \sep Ryugu




\end{keyword}

\end{frontmatter}



\section{Introduction}
\label{sec:introduction}

Hydrous carbonaceous (C-type) asteroids are thought to be the primary source of water and volatiles for the Earth and terrestrial planets \citep[e.g.,][]{2016GeocJ..50...27G, 2023ASPC..534.1031K}.
The JAXA Hayabusa2 spacecraft explored the near-Earth C-type asteroid (162173) Ryugu and delivered the collected samples to Earth.
These samples are dominated by hydrous phyllosilicates and contain coarse-grained minerals formed through aqueous alteration \citep[e.g.,][]{2022PJAB...98..227N, 2023Sci...379.8671N, 2023NatAs...7..398Y}.
The mineralogical, petrological, and chemical characteristics of the Ryugu samples closely resemble those of CI (Ivuna-type) carbonaceous chondrites \citep[e.g.,][]{2022NatAs...6.1163I, 2023Sci...379.8671N, 2023Sci...379.7850Y} as well as samples collected from another C-type asteroid, (101955) Bennu \citep[e.g.,][]{2024M&PS...59.2453L}.
Ryugu is a 1-km-sized rubble-pile asteroid thought to have originated from the collisional disruption of a parent planetesimal with a radius greater than 10 km \citep[e.g.,][]{2019Sci...364..252S, 2019Sci...364..268W, 2020Natur.579..518O}.
The timing and duration of fluid activity within the parent body have been constrained by $^{53}$Mn--$^{53}$Cr dating of aqueously formed dolomite \citep[][]{2022PJAB...98..227N, 2023NatAs...7..309M, 2023Sci...379.7850Y, 2023SciA....9I7048Y, 2024GeCoA.382...40S, 2024ApJ...965...52T, kawasaki2025dolomite}.
Furthermore, the internal temperature of the parent planetesimal has been estimated from equilibrium oxygen isotope fractionation between dolomite and magnetite \citep[][]{2023Sci...379.7850Y, kawasaki2025dolomite}.

The parent planetesimal have formed through the accumulation of dust aggregates in the solar protoplanetary disk \citep[e.g.,][]{2024A&A...691A.147M}.
Astronomical observations of extrasolar protoplanetary disks have revealed that 0.1-mm- to 1-cm-sized dust aggregates are abundant in the midplanes of disks \citep[e.g.,][]{2023ASPC..534..501M}.
These aggregates, commonly referred to as ``pebbles,'' are therefore considered the building blocks of planetesimals.
Various mechanisms have been proposed for the accumulation of pebbles \citep[e.g.,][]{2023ASPC..534..465L}, and the size and composition of the resulting planetesimals depend on the specific processes by which pebbles are accumulated \citep[e.g.,][]{2014prpl.conf..547J}.
The size of pebbles evolves through sequential collisions within the disk \citep[e.g.,][]{2023ApJ...951L..16A}, and thus varies with both time and location in the disk \citep[e.g.,][]{2023ASPC..534..717D}.

For a planetesimal composed of rock and ice, water--rock differentiation occurs when the internal temperature reaches the melting point of ice.
If pebbles survive the differentiation process without being destroyed, the resulting rocky core would not be consolidated but instead form a pebble-pile structure, with water filling the pores between the aggregates (Figure \ref{fig:1}).
Water circulation enhances heat transfer within such a porous core \citep[e.g.,][]{1989Icar...82..244G, 2003E&PSL.213..249Y, 2023PSJ.....4..144T}, and the thermal history of the planetesimal differs significantly depending on whether or not water circulation occurs.
Moreover, the efficiency of heat transfer depends on the permeability of the porous pebble-pile core, which in turn is a function of the pebble radius \citep[e.g.,][]{2009E&PSL.287..559B, dullien2012porous}.
Therefore, it is of great interest whether the thermal history of Ryugu's parent planetesimal can constrain the size of pebbles present during the first few million years of the solar protoplanetary disk.

In this study, we investigate the thermal history of Ryugu's parent planetesimal based on the precipitation ages and temperatures of dolomite in the returned samples.
The precipitation conditions of dolomite found in the Ryugu samples are summarized in Section \ref{sec:dolomite}.
An evolutionary model for icy planetesimals is introduced in Section \ref{sec:models}.
We perform one-dimensional numerical simulations to investigate the temperature and structural evolution of icy planetesimals (Section \ref{sec:results}), and compare the results with sample analysis data reported by \citet{kawasaki2025dolomite}.
From these comparisons, we derive constraints on the pebble radius ($r_{\rm peb}$), the planetesimal radius ($R_{\rm p}$), and the accretion age ($t_{\rm acc}$) of Ryugu's parent body.
We also present an analytical solution for the temperature evolution at the center of planetesimal (Section \ref{sec:analytic}).
This study offers new insights into the formation and thermal evolution of icy planetesimals in the early solar system.

\section{Precipitation Ages and Temperatures of Dolomite in Ryugu Samples}
\label{sec:dolomite}

The Ryugu samples contain abundant and large aqueously formed carbonates, including dolomite, which can be dated to determine the timing of fluid activity.
In this study, we use the data for the A0058 and C0002 samples reported in \citet{kawasaki2025dolomite} and compare them with our numerical results.
The A0058 and C0002 samples were collected from the first and second touchdown sites, respectively.
The $^{53}$Mn--$^{53}$Cr ages of dolomite in the Ryugu samples have been determined either by in situ secondary ion mass spectrometry (SIMS) analyses \citep{2022PJAB...98..227N, 2023NatAs...7..309M, 2023Sci...379.7850Y} or by whole-rock sample analyses \citep{2023SciA....9I7048Y, 2024ApJ...965...52T}.
Since the precipitation age can vary among individual grains, in situ SIMS analyses are suitable for determining the $^{53}$Mn--$^{53}$Cr ages of single dolomite grains.
However, \citet{2024GeCoA.382...40S} claimed that the SIMS-based Mn--Cr ages of Ryugu dolomite reported in earlier studies were affected by artificial errors caused by the use of inappropriate standards for data correction.
To address this issue, \citet{2024GeCoA.382...40S} synthesized homogeneous crystalline dolomite to be used as standard materials, and evaluated the relative sensitivity factor (RSF) of Mn/Cr for SIMS analysis, that is, the ratio of Mn/Cr measured by SIMS to the true Mn/Cr ratio.
Using this improved standard, \citet{kawasaki2025dolomite} determined the Mn--Cr ages of dolomite in the Ryugu samples.
They analyzed dolomite grains from two Ryugu samples, C0002 and A0058, and found that their precipitation ages are $2.0^{+0.6}_{-0.6}~{\rm Myr}$ and $4.1^{+0.9}_{-0.8}~{\rm Myr}$ after the formation of calcium--aluminum-rich inclusions (CAIs), respectively.

The precipitation temperature of dolomite is estimated based on the equilibrium oxygen isotope fractionation between dolomite and magnetite.
Dolomite and magnetite grains are occasionally found in close proximity within the same lithological context in Ryugu samples, and their oxygen isotopic compositions suggest that they precipitated from the same aqueous fluid under oxygen isotopic equilibrium conditions \citep[][]{2023Sci...379.7850Y, kawasaki2025dolomite}.
Therefore, the temperature-dependent nature of equilibrium oxygen isotope fractionation can be used to infer their precipitation temperatures.
Using this approach, the precipitation temperatures of dolomite in the C0002 and A0058 samples were estimated to be ${(365 \pm 21)}~{\rm K}$ and ${(310 \pm 10)}~{\rm K}$, respectively \citep[][]{2023Sci...379.7850Y, kawasaki2025dolomite}.

In addition to constraints on its precipitation age and temperature, dolomite is thought to have formed during the retrograde cooling stage of the parent planetesimal.
\citet{2023NatGe..16..675F} conducted carbon and oxygen isotope analyses of calcite and dolomite in Ryugu samples.
They found that variations in both carbon and oxygen isotope ratios in dolomite are smaller than those in calcite, suggesting that dolomite precipitated when the aqueous fluid and silicates were approaching oxygen isotope equilibrium, likely during retrograde cooling of the parent body.
\citet{2024M&PS...59.2097K} also performed oxygen isotope analyses of carbonates in Ryugu samples and found that dolomite and breunnerite, along with some magnetite, precipitated during the late stage of aqueous alteration, when oxygen isotopic equilibrium had nearly been achieved between the fluid and solid phases.
This finding further supports the conclusion of \citet{2023NatGe..16..675F}.

\section{Models}
\label{sec:models}

\begin{figure*}[]
\centering
\includegraphics[width = 0.9\textwidth]{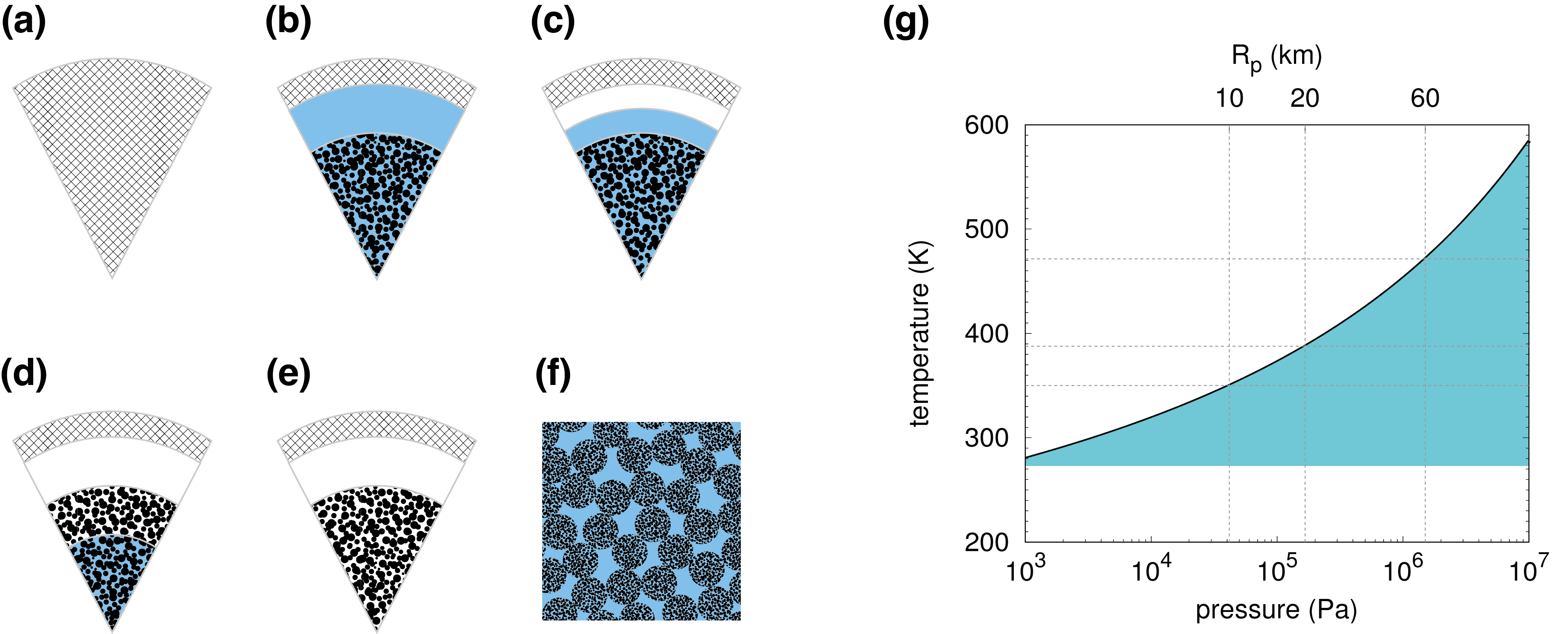}
\caption{
{\it Left}.~(a)--(f) Schematics of the interior structure evolution of parent planetesimals.
(a) Planetesimals are initially undifferentiated, with interiors composed of a mixture of anhydrous silicates and ice.
(b) Water--rock differentiation and aqueous alteration of anhydrous silicates are assumed to occur instantaneously upon ice melting.
(c) The water mantle freezes from the top downward.
(d) Water within the pebble-pile core also freezes from the top downward.
(e) Final structure.
(f) Zoomed-in view of the pebble-pile core.
Pebbles are composed of hydrous phyllosilicates with an intra-pebble packing fraction $\phi_{\rm peb}$.
The structural (inter-pebble) packing fraction within the core is $\phi_{\rm str}$, yielding a volume fraction of phyllosilicates within the core of $\chi_{\rm phy} = \phi_{\rm peb} \phi_{\rm str}$.
{\it Right}.~(g) Phase diagram of ${\rm H}_{2}{\rm O}$.
The shaded region represents the condition under which water is stable as a liquid.
The black curve shows the saturation pressure \citep{2002JPCRD..31..387W}.
The central pressure ($P_{\rm center}$) is related to the planetesimal radius ($R_{\rm p}$) by Equation (\ref{eq:P_center}), and the central boiling temperature is expressed as a function of $R_{\rm p}$.
}
\label{fig:1}
\end{figure*}

To model the thermal evolution of Ryugu's parent planetesimal, we adopt a previously developed model that assumes a spherically symmetric small body \citep[e.g.,][]{2011EPS...63.1193W, 2021ApJ...917L...5K}.
The original numerical code used in this study was developed by \citet{2024PASJ...76..130A}, which we have extended to simulate water--rock differentiation processes.
The spatial and temporal evolution of temperature, $T$, within a planetesimal is described as a function of time, $t$, and radial distance from the center, $R$, using the heat conduction equation:
\begin{equation}
\rho c \frac{{\partial}T}{{\partial}t} = - \frac{1}{4 \uppi R^{2}} \frac{{\partial}F}{{\partial}R} + Q_{\rm decay},
\label{eq:conduction}
\end{equation}
where $\rho$ is the bulk density, $c$ is the specific heat capacity, and $Q_{\rm decay}$ is the heat production due to the decay of radioactive nuclides.
The outward energy flux at radius $R$, $F = F (R)$, is given by
\begin{equation}
F = - 4 \uppi R^{2} k \frac{{\partial}T}{{\partial}R},
\end{equation}
where $k$ is the thermal conductivity.
We solve the heat conduction equation using the finite-difference method with an explicit time integration scheme, as in our previous work \citep{2024PASJ...76..130A}.
For simplicity, we assume that the planetesimal formed instantaneously and had a spatially uniform temperature of $T = 70~{\rm K}$ at the accretion time, $t = t_{\rm acc}$.
The origin of time ($t = 0$) is defined as the timing of CAI formation.
The surface temperature at $r = R_{\rm p}$ is fixed at $T_{\rm surf} = 70~{\rm K}$, corresponding to the temperature at which ${\rm C}{\rm O}_{2}$ remains in the solid phase \citep[e.g.,][]{2023Sci...379.8671N}.
We do not consider ice sublimation from the surface \citep[e.g.,][]{2023ApJ...956L..25Z}.
The material properties ($c$, $k$, and $Q_{\rm decay}$) are listed in Table S1 (see Supplementary Materials).

\subsection{Structure evolution}
\label{sec:structure}

Figure \ref{fig:1} illustrates the schematic evolution of the internal structure of the parent planetesimals.
Initially, planetesimals are undifferentiated, and their interiors consist of a mixture of anhydrous silicates and ${\rm H}_{2}{\rm O}$ ice (Panel (a)).
The internal temperature gradually increases mainly due to the decay heat of short-lived radioactive nuclide $^{26}{\rm Al}$.
Once the temperature reaches the melting point of ice, $T_{\rm melt} = 273~{\rm K}$, structural evolution begins through water--rock differentiation (Panel (b)).
We also assume that hydrothermal reactions forming phyllosilicates occur at $T = T_{\rm melt}$.
Because the sedimentation timescale for 1-mm-sized pebbles is $\lesssim 1~{\rm yr}$ \citep{2011EPS...63.1193W}, differentiation is considered to occur instantaneously.
We further assume that the resulting pebble-pile core is composed of phyllosilicates.
As the radiogenic heating diminishes over time, the internal temperature begins to decline.
This leads to the top-down freezing of the water mantle (Panel (c)), followed by the progressive freezing of water within the pebble-pile core (Panels (d) and (e)).
The time evolution of each layer's depth is calculated by accounting for the heat consumption and release associated with ice melting/freezing and aqueous alteration (i.e., phyllosilicate formation).
Further details of our numerical modeling are provided in Supplementary Materials (Texts S1--S4 and Figure S1).

We assume that the density is homogeneous within each layer.
The bulk density of the undifferentiated layer is given by
\begin{equation}
\bar{\rho} = \chi_{\rm sil} \rho_{\rm sil} + {( 1 - \chi_{\rm sil} )} \rho_{\rm ice},
\end{equation}
where $\rho_{\rm sil} = 3100~{\rm kg}~{\rm m}^{-3}$ and $\rho_{\rm ice} = 1000~{\rm kg}~{\rm m}^{-3}$ are the material densities of anhydrous silicate and ice, respectively.
The volume fraction of anhydrous silicate in the undifferentiated layer, $\chi_{\rm sil}$, is given by $\chi_{\rm sil} = f_{\rm sil} \bar{\rho} / \rho_{\rm sil}$, where $f_{\rm sil}$ is the mass fraction of anhydrous silicate in the undifferentiated layer.
In this study, we adopt $f_{\rm sil} = 0.5$ as the fiducial value, which corresponds to $\chi_{\rm sil} = 0.244$ and $\bar{\rho} = 1512~{\rm kg}~{\rm m}^{-3}$.
The bulk density of the water/ice mantle is equal to its material density.
The material density of water is set to $\rho_{\rm wtr} = 1000~{\rm kg}~{\rm m}^{-3}$.
We assume that all pores within the undifferentiated layer are completely filled with ${\rm H}_{2}{\rm O}$ ice.
Numerical simulations of the porosity evolution of Ryugu's parent planetesimal \citep{2021Icar..35814166N, 2021NatAs...5..766S} indicate that a planetesimal with $R_{\rm p} \gtrsim 10~{\rm km}$ would not retain high porosity in the undifferentiated layer, which support our assumption.

We also assume that the pebble-pile core contains phyllosilicates and water, and water freezes when temperature falls below $T_{\rm melt}$.
The material density and the volume fraction of phyllosilicate in the pebble-pile core are $\rho_{\rm phy} = 2600~{\rm kg}~{\rm m}^{-3}$ \citep{2015JGRE..120..123N} and $\chi_{\rm phy}$, respectively.
We consider two types of pores: intra-pebble micropores and inter-pebble macropores (Figure \ref{fig:1}(f)).
The pebble density is set to $\rho_{\rm peb} = 1800~{\rm kg}~{\rm m}^{-3}$, which corresponds to the average density of Ryugu samples \citep[e.g.,][]{2023Sci...379.8671N}.
The volume filling factor of phyllosilicates within individual pebbles is then given by $\phi_{\rm peb} = \rho_{\rm peb} / \rho_{\rm phy} = 69.2\%$.
The volume packing fraction of pebbles within the pebble-pile structure is denoted by $\phi_{\rm str}$.
We adopt $\phi_{\rm str} = 80\%$ as the fiducial value, which corresponds to the random packing of polydisperse spheres \citep[e.g.,][]{2020JGRE..12506519G}.
Accordingly, the volume fraction of phyllosilicates within the core is $\chi_{\rm phy} = \phi_{\rm peb} \phi_{\rm str} = 0.554$, and the bulk density of the pebble-pile core is given by
\begin{equation}
\rho_{\rm core} = \chi_{\rm phy} \rho_{\rm phy} + {( 1 - \chi_{\rm phy} )} \rho_{\rm wtr},
\end{equation}
which yields $\rho_{\rm core} = 1886~{\rm kg}~{\rm m}^{-3}$ for $\phi_{\rm str} = 80\%$.

The assumption of a pebble-pile structure within the core would be valid if the pebbles can withstand disruption caused by gravitational compression.
Pebbles with a packing fraction of $\phi_{\rm peb} \ge 60\%$, composed of submicron-sized grains, exhibit yield strengths exceeding $10~{\rm MPa}$ \citep[e.g.,][]{2024ApJ...974...76M}, suggesting that the pebbles in the core would remain intact (or at least survive) gravitational compression within planetesimals with radii $R_{\rm p} \lesssim 200~{\rm km}$ (see Section \ref{sec:stability} and Equation (\ref{eq:P_center})).

\subsection{Thermophysical stability of liquid ${\rm H}_{2}{\rm O}$}
\label{sec:stability}

In our thermal evolution model, ${\rm H}_{2}{\rm O}$ exists in liquid state when the temperature exceeds $T_{\rm melt} = 273~{\rm K}$.
However, water cannot remain in the interior of a planetesimal if the pressure is lower than the saturation pressure of ${\rm H}_{2}{\rm O}$, which depends on $T$ \citep{2002JPCRD..31..387W}.
The shaded region in Figure \ref{fig:1}(g) indicates the condition under which liquid ${\rm H}_{2}{\rm O}$ is stable, and the black line represents the temperature dependence of the saturation pressure.
We also evaluate the relationship between $R_{\rm p}$ and the hydrostatic pressure at the center, $P_{\rm center}$.
Assuming that the planetesimal is nearly fully differentiated, $P_{\rm center}$ is given by
\begin{equation}
P_{\rm center} = 1.50~{\left( \frac{R_{\rm p}}{60~{\rm km}} \right)}^{2}~{\rm MPa}
\label{eq:P_center}
\end{equation}
(see Supplementary Text S5).
Given that the precipitation temperature of dolomite in the C0002 grain is $\approx 365~{\rm K}$ \citep{kawasaki2025dolomite}, the parent planetesimal of the Ryugu C0002 grain must have had a radius larger than $10~{\rm km}$.

\subsection{Effective thermal conductivity within the pebble-pile core}
\label{sec:k_cw}

In a water-saturated pebble-pile core, water circulation through the pore network enhances heat transfer.
The effective thermal conductivity of such a core, $k_{\rm c(w)}$, is given by the following equation \citep[e.g.,][]{1989Icar...82..244G, 2023PSJ.....4..144T}:
\begin{equation}
k_{\rm c(w)} = {\rm Nu} k_{\rm wtr},
\label{eq:Nu}
\end{equation}
where $k_{\rm wtr} = 0.56~{\rm W}~{\rm m}^{-1}~{\rm K}^{-1}$ is the thermal conductivity of water, and ${\rm Nu}$ is the Nusselt number.
${\rm Nu}$ depends on the Rayleigh--Darcy number, ${\rm Ra}$ \citep[e.g.,][]{2021JFM...911R...4P}:
\begin{equation}
{\rm Nu} = 
\begin{cases}
1                                        & \text{(${\rm Ra} < 42.3$),} \\
0.0081 {\rm Ra} + 0.067 {\rm Ra}^{0.61}  & \text{(${\rm Ra} \ge 42.3$).}
\end{cases}
\label{eq:NuRa}
\end{equation}
${\rm Ra}$ is given by
\begin{equation}
{\rm Ra} = \frac{ \alpha_{\rm wtr} {\Delta T} g_{\rm c(w)} R_{\rm c(w)} }{ \nu_{\rm wtr} \kappa_{\rm wtr} } K,
\label{eq:Ra}
\end{equation}
where $\alpha_{\rm wtr}$, $\nu_{\rm wtr}$, and $\kappa_{\rm wtr}$ are the thermal expansion coefficient, kinematic viscosity, and thermal diffusivity of water, respectively \citep[e.g.,][]{2002gedy.book.....T}.
Values for $\alpha_{\rm wtr}$ and $\nu_{\rm wtr}$ used in this study are summarized in the Supplementary Table S1, and $\kappa_{\rm wtr}$ is given by $\kappa_{\rm wtr} = k_{\rm wtr} / {( \rho_{\rm wtr} c_{\rm wtr} )}$, where $c_{\rm wtr}$ is the specific heat capacity of water.
The core structure evolves over time.
$R_{\rm c(w)}$ is the outer radius of the water-saturated pebble-pile core, and the gravitational acceleration at $R = R_{\rm c(w)}$ is given by $g_{\rm c(w)}  = {( 4 \uppi / 3 )} {\mathcal G} R_{\rm c(w)} \rho_{\rm core}$, where ${\mathcal G}$ is the gravitational constant.
The temperature difference between $R = 0$ and $R_{\rm c(w)}$ is defined as ${\Delta T} = T_{\rm center} - T_{\rm melt}$.
The permeability of the water-saturated pebble-pile core, $K$, is proportional to the square of $r_{\rm peb}$ \citep[e.g.,][]{2009E&PSL.287..559B, dullien2012porous}:
\begin{equation}
K = \frac{ {\left( 1 - \phi_{\rm str} \right)}^{3} }{ 9 h_{\rm CK} {\phi_{\rm str}}^{2} } {r_{\rm peb}}^{2} = 2.8 \times 10^{-10}~{\left( \frac{r_{\rm peb}}{1~{\rm mm}} \right)}^{2}~{\rm m}^{2},
\label{eq:K_pmb}
\end{equation}
where $h_{\rm CK} = 5$ is the Carman--Kozeny constant \citep[e.g.,][]{dullien2012porous}.

\section{Results}
\label{sec:results}

\begin{figure*}[]
\centering
\includegraphics[width = 0.9\textwidth]{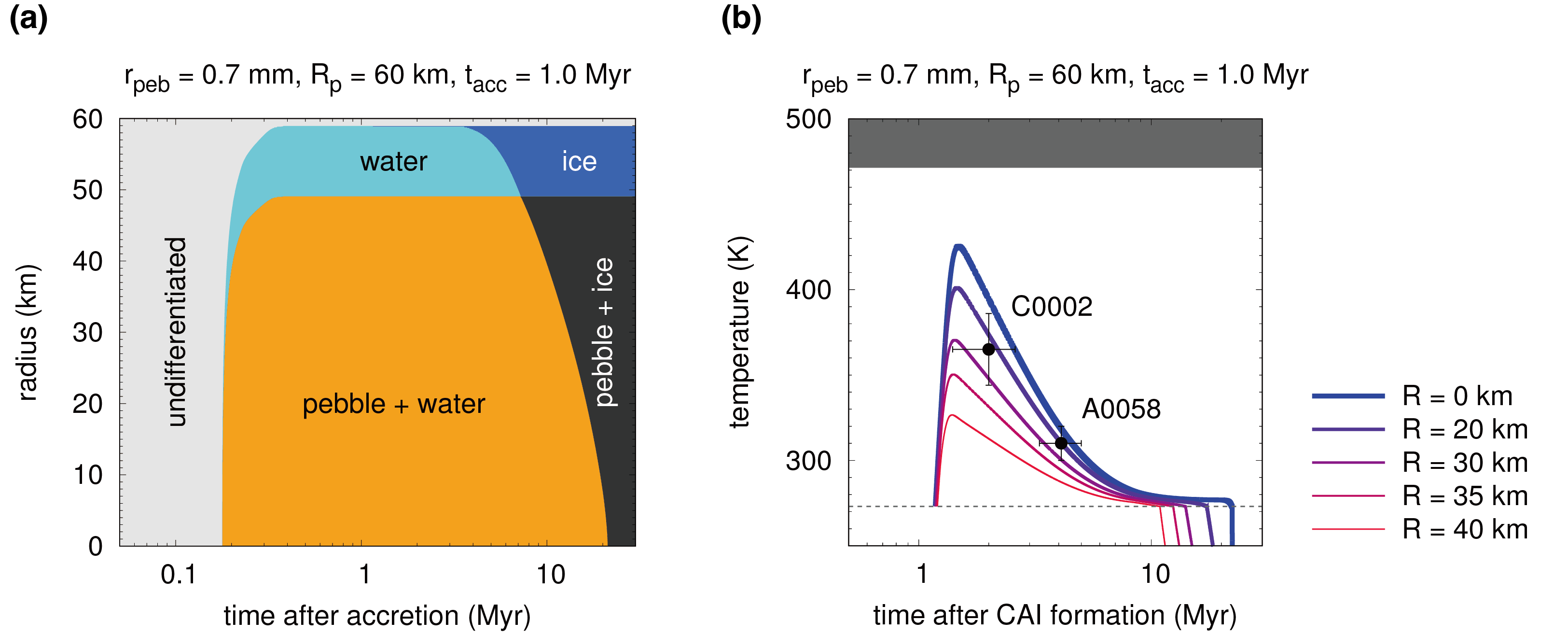}
\caption{
Thermal evolution of a parent planetesimal with $r_{\rm peb} = 0.7~{\rm mm}$, $R_{\rm p} = 60~{\rm km}$, and $t_{\rm acc} = 1.0~{\rm Myr}$.
(a) Structure evolution.
The horizontal axis indicates the time after accretion of the planetesimal ($t - t_{\rm acc}$).
(b) Temperature evolution within the pebble-pile core.
The horizontal axis indicates the time after CAI formation ($t$).
The points with error bars represent the precipitation ages and temperatures of dolomite in the A0058 and C0002 grains.
The gray shaded region indicates the temperature range in which liquid water is thermodynamically unstable at the center of the core.
}
\label{fig:2}
\end{figure*}

\begin{figure}[]
\centering
\includegraphics[width = 0.4\textwidth]{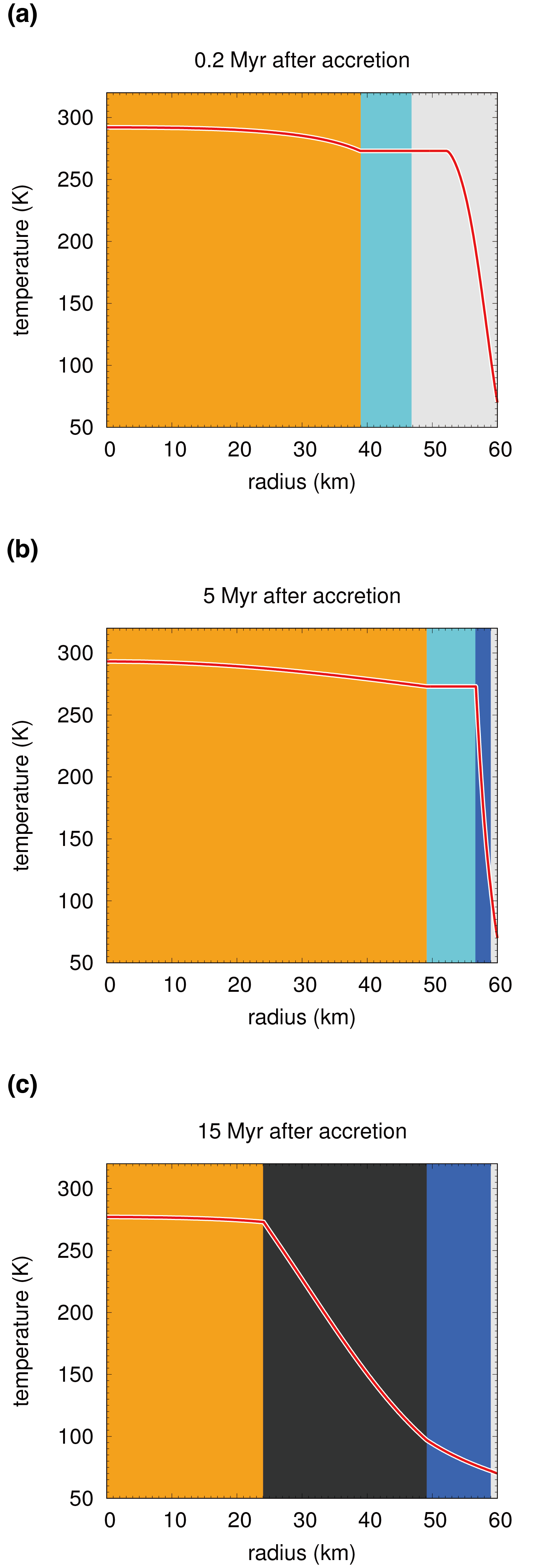}
\caption{
Snapshots of the radial temperature structure for the fiducial case: $r_{\rm peb} = 0.7~{\rm mm}$, $R_{\rm p} = 60~{\rm km}$, and $t_{\rm acc} = 1.0~{\rm Myr}$.
(a) At $0.2~{\rm Myr}$ after accretion.
(b) At $5~{\rm Myr}$ after accretion.
(c) At $15~{\rm Myr}$ after accretion.
The red line shows the radial temperature profile.
The background colors indicate the internal structure, consistent with the structural evolution shown in Figure \ref{fig:2}(a).
}
\label{fig:3}
\end{figure}

\begin{figure*}[]
\centering
\includegraphics[width = 0.9\textwidth]{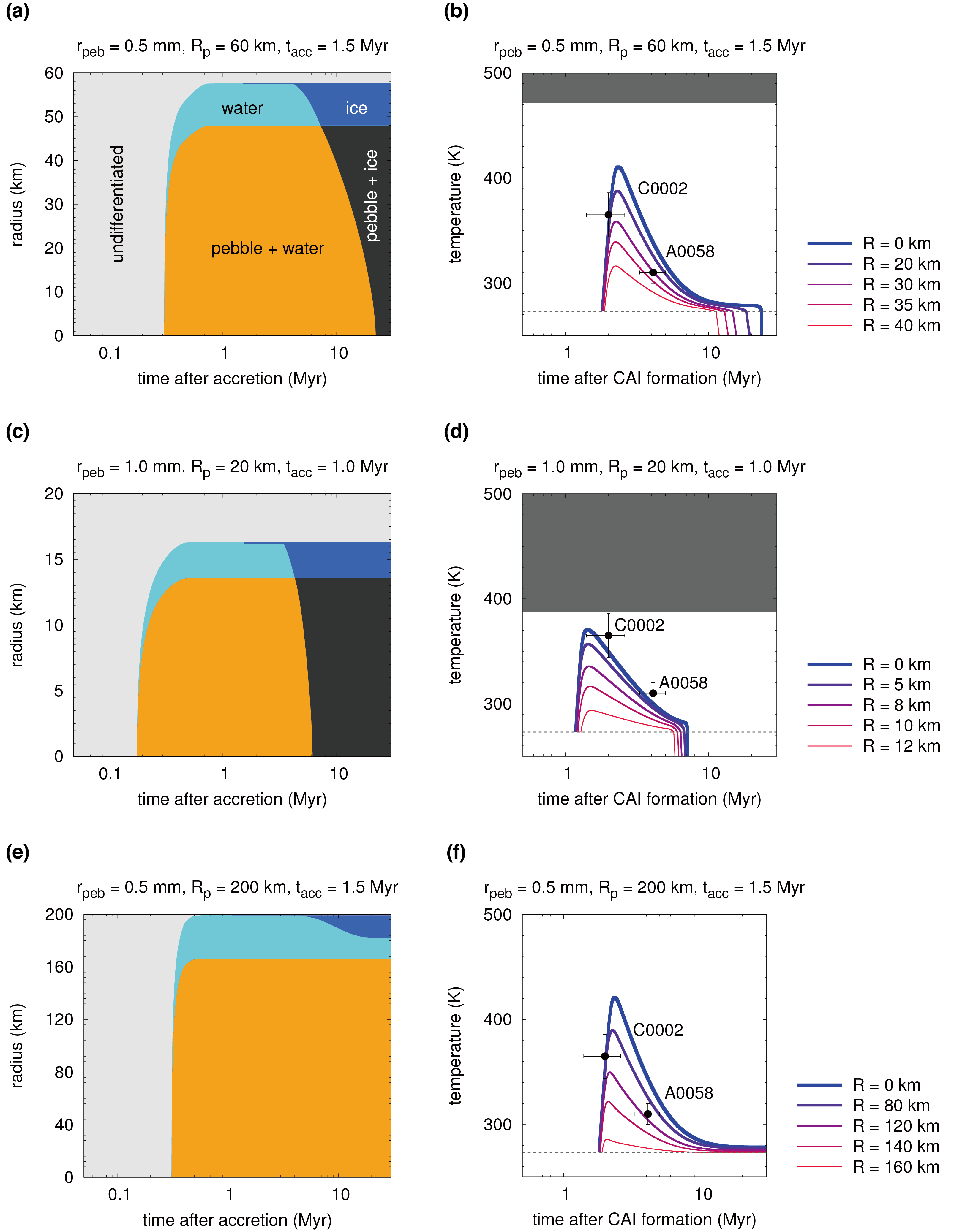}
\caption{
Examples of numerical results that reproduce the precipitation ages and formation temperatures of dolomite in the A0058 and C0002 grains.
(a, b): Same format as Figure \ref{fig:2}, but for $r_{\rm peb} = 0.5~{\rm mm}$, $R_{\rm p} = 60~{\rm km}$, and $t_{\rm acc} = 1.5~{\rm Myr}$.
(c, d): Results for $r_{\rm peb} = 1.0~{\rm mm}$, $R_{\rm p} = 20~{\rm km}$, and $t_{\rm acc} = 1.0~{\rm Myr}$.
(e, f): Results for $r_{\rm peb} = 0.5~{\rm mm}$, $R_{\rm p} = 200~{\rm km}$, and $t_{\rm acc} = 1.5~{\rm Myr}$.
}
\label{fig:4}
\end{figure*}

\begin{figure*}[]
\centering
\includegraphics[width = 0.9\textwidth]{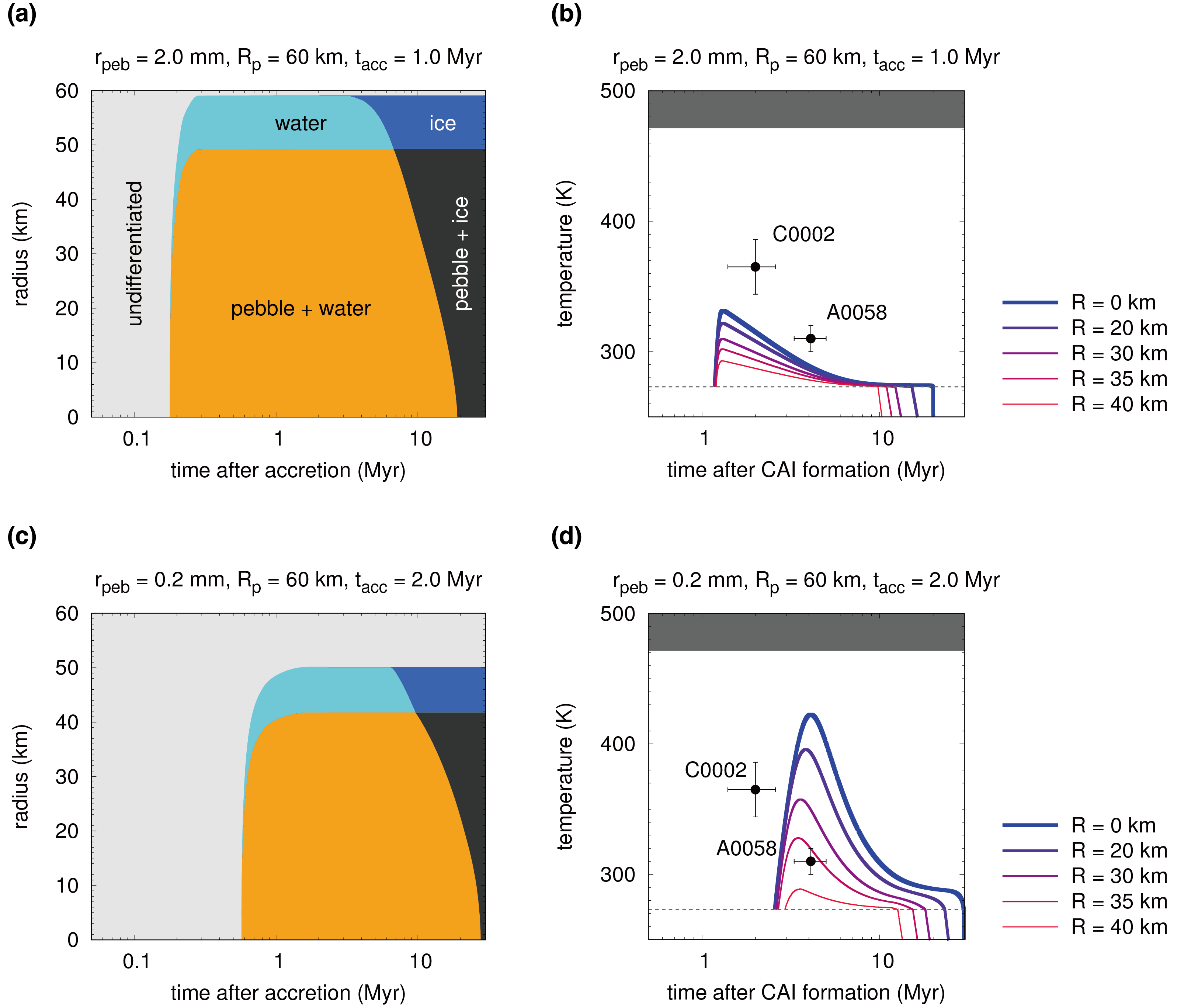}
\caption{
Examples of numerical results that fail to reproduce the precipitation ages and formation temperatures of dolomite in the A0058 and C0002 grains.
(a, b): Same format as Figure \ref{fig:2}, but for $r_{\rm peb} = 2.0~{\rm mm}$, $R_{\rm p} = 60~{\rm km}$, and $t_{\rm acc} = 1.0~{\rm Myr}$.
(c, d): Results for $r_{\rm peb} = 0.2~{\rm mm}$, $R_{\rm p} = 60~{\rm km}$, and $t_{\rm acc} = 2.0~{\rm Myr}$.
}
\label{fig:5}
\end{figure*}

\begin{figure}[]
\centering
\includegraphics[width = 0.45\textwidth]{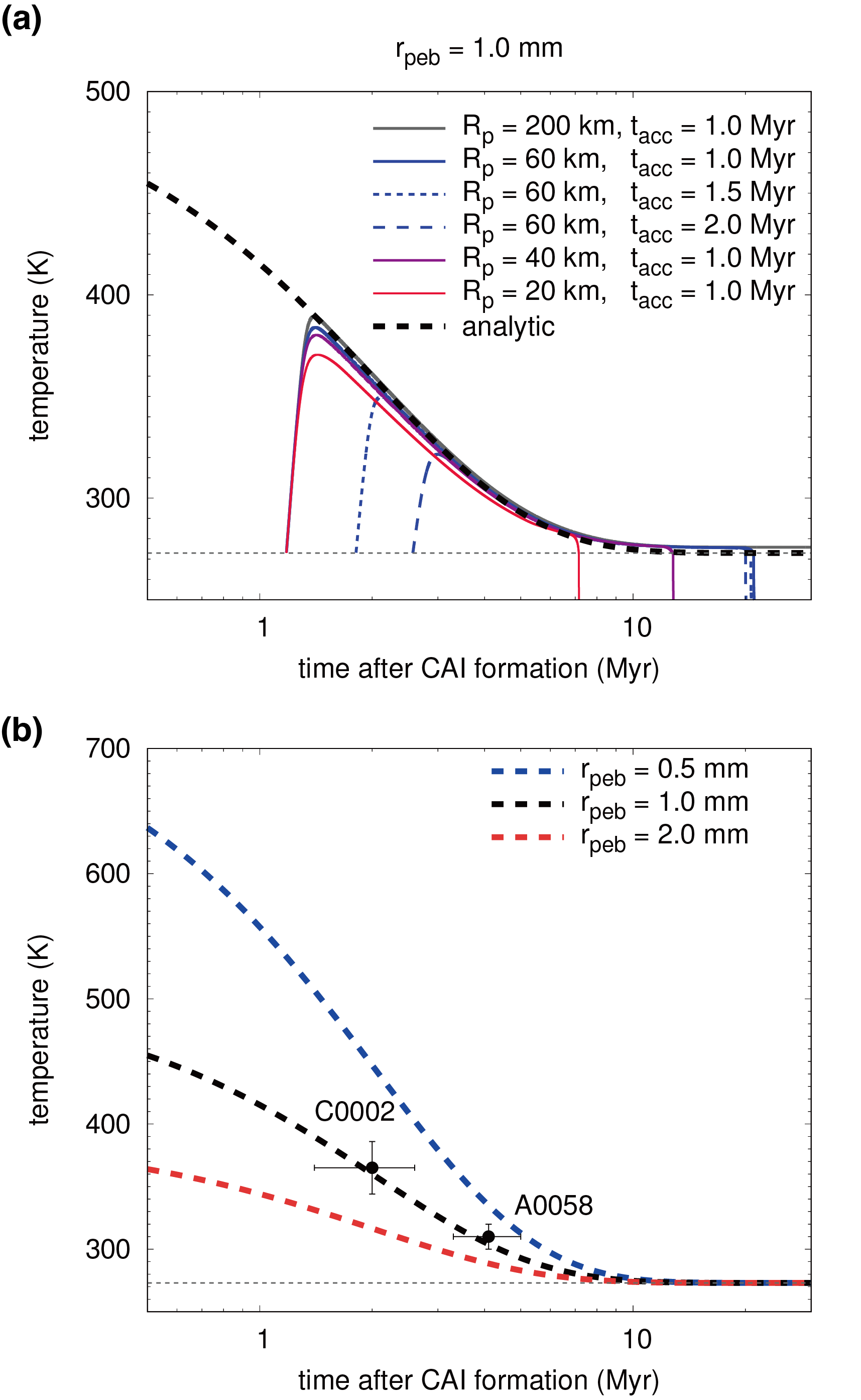}
\caption{
(a) Temperature evolution at the center of the pebble-pile core, $T_{\rm center}$, for various combinations of $R_{\rm p}$ and $t_{\rm acc}$.
The pebble radius is fixed at $r_{\rm peb} = 1.0~{\rm mm}$.
The dashed line shows the analytic solution (Equation (\ref{eq:T_center_analytic})).
(b) Analytic solution for the central temperature as a function of $r_{\rm peb}$.
The points with error bars represent the precipitation ages and temperatures of dolomite in the A0058 and C0002 grains.
}
\label{fig:6}
\end{figure}

\begin{figure*}[]
\centering
\includegraphics[width = 0.9\textwidth]{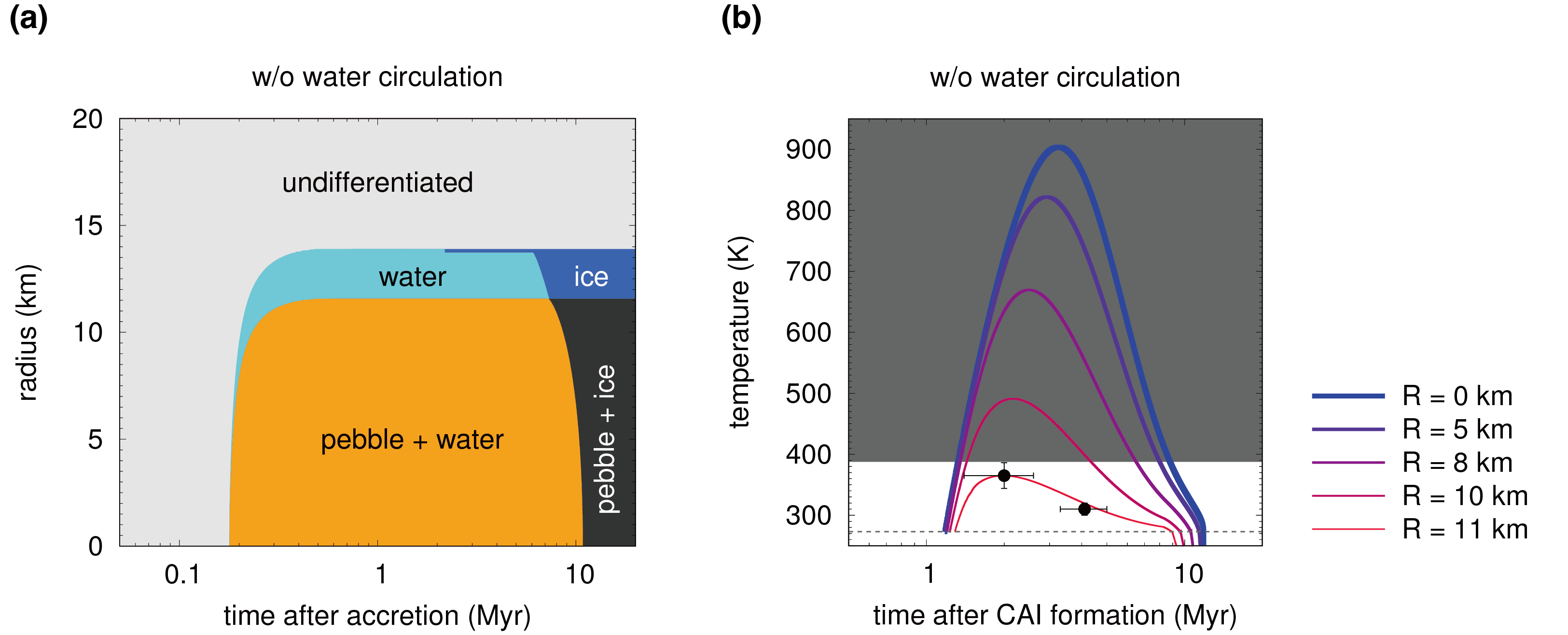}
\caption{
Same as Figures \ref{fig:4}(c) and \ref{fig:4}(d), but for the case without water circulation within the pebble-pile core.
Here, we set ${\rm Nu} = 1$ instead of using ${\rm Nu}$ calculated from Equation (\ref{eq:NuRa}).
}
\label{fig:7}
\end{figure*}

Here, we present several numerical results that either succeed or fail in explaining the precipitation ages and temperatures of dolomites in two Ryugu samples, namely, the C0002 and A0058 grains.
We treat $r_{\rm peb}$, $R_{\rm p}$, and $t_{\rm acc}$ as free parameters and find that the observed precipitation ages and temperatures can be reproduced across various combinations of ${( r_{\rm peb}, R_{\rm p}, t_{\rm acc} )}$.
Note that $f_{\rm ice} = 0.5$ and $\phi_{\rm str} = 80\%$ are fixed throughout Section \ref{sec:results}.
The dependence on these parameters is discussed in Supplementary Materials (Text S6, and Figures S2 and S3).

\subsection{Fiducial case}

We present our fiducial model, which reproduces the precipitation ages and temperatures of dolomite observed in two Ryugu samples (C0002 and A0058 grains; see Section \ref{sec:dolomite}) within a single parent planetesimal.
Figure \ref{fig:2} shows the thermal evolution of the parent planetesimal with the fiducial parameter set: $r_{\rm peb} = 0.7~{\rm mm}$, $R_{\rm p} = 60~{\rm km}$, and $t_{\rm acc} = 1.0~{\rm Myr}$.
Differentiation begins at $0.18~{\rm Myr}$ after accretion, and a pebble-pile core with a radius of $R_{\rm c} = 49.1~{\rm km}$ has formed (Figure \ref{fig:2}(a)).
The temperature at the center of the core reaches $T_{\rm center} = 425.4~{\rm K}$ at $t = 1.49~{\rm Myr}$ (i.e., $0.49~{\rm Myr}$ after accretion; Figure \ref{fig:2}(b)).
The saturation pressure of ${\rm H}_{2}{\rm O}$ at $425.4~{\rm K}$ is lower than the hydrostatic pressure at the center for $R_{\rm p} = 60~{\rm km}$, allowing water to remain in the liquid phase within the core.
Given that the boiling temperature within the core is nearly uniform (Figure S4 in Supplementary Materials), the presence of liquid water at the center implies its thermodynamic stability throughout the entire core.

The precipitation ages and temperatures of dolomites in the C0002 and A0058 grains are also plotted in Figure \ref{fig:2}(b).
We find that the temperature profiles for $0~{\rm km} < R < 35~{\rm km}$ can reproduce the precipitation ages and temperatures of both C0002 and A0058 grains.
Dolomites in both grains precipitated during retrograde cooling in our fiducial model, consistent with the previous finding that aqueous fluids and silicates were in oxygen isotopic equilibrium at the time of dolomite formation \citep{2023NatGe..16..675F}.

Our numerical result suggests that the C0002 and A0058 grains could have originated from the same parent planetesimal and from similar radial positions $R$.
\citet{2015P&SS..107...24M} performed numerical simulations of the catastrophic disruption of 100-km-sized asteroids followed by reaccumulation processes.
They found that most fragments within each reaccumulated body tend to cluster from the same original region, which is consistent with our findings.

Snapshots of the radial structure are shown in Figure \ref{fig:3}.
At $t - t_{\rm acc} = 0.2~{\rm Myr}$ (Figure \ref{fig:3}(a)), the interior of the planetesimal consists of three layers: a water-saturated pebble-pile core (orange), a water mantle (cyan), and an undifferentiated region (gray).
The temperature in the deep part of the undifferentiated region reaches $T_{\rm melt}$, and water--rock differentiation progresses.
An ice mantle (blue) appears at $t - t_{\rm acc} = 5~{\rm Myr}$ (Figure \ref{fig:3}(b)), and its thickness increases over time.
As freezing proceeds, the water mantle disappears and an ice-saturated pebble-pile core (black) emerges at $t - t_{\rm acc} = 15~{\rm Myr}$ (Figure \ref{fig:3}(c)).

\subsection{Parameter dependence}

We present other parameter sets that reproduce the precipitation ages and temperatures of dolomite in the A0058 and C0002 grains.
Figures \ref{fig:4}(a) and \ref{fig:4}(b) present the numerical results for the case with $r_{\rm peb} = 0.5~{\rm mm}$, $R_{\rm p} = 60~{\rm km}$, and $t_{\rm acc} = 1.5~{\rm Myr}$.
The duration during which the entire planetesimal remains undifferentiated is longer than that in Figure \ref{fig:2}(a), reflecting the lower value of $Q_{\rm decay}$ at $t = t_{\rm acc}$; the abundance of $^{26}{\rm Al}$ at $t = t_{\rm acc}$ is approximately 40\% lower than that for Figure \ref{fig:2} ($t_{\rm acc} = 1.0~{\rm Myr}$).
As shown in Figure \ref{fig:4}(b), this model also reproduces the precipitation ages and temperatures of dolomites in both the C0002 and A0058 grains.

Figures \ref{fig:4}(c) and \ref{fig:4}(d) show the results for the case with $r_{\rm peb} = 1.0~{\rm mm}$, $R_{\rm p} = 20~{\rm km}$, and $t_{\rm acc} = 1.0~{\rm Myr}$.
When $R_{\rm p} = 20~{\rm km}$, the condition for liquid water to be present at the center is $T_{\rm center} < 387.7~{\rm K}$.
As the temperature at the center increases with decreasing $r_{\rm peb}$ (see Section \ref{sec:analytic}), the key assumption of our model, water retention within the core, may not hold for $r_{\rm peb} \ll 1~{\rm mm}$ when $R_{\rm p} = 20~{\rm km}$ and $t_{\rm acc} = 1.0~{\rm Myr}$.
In this parameter set, the entire pebble-pile core freezes at $t = 7.2~{\rm Myr}$ (i.e., $6.2~{\rm Myr}$ after accretion), yet the model still reproduces the observed precipitation ages and temperatures.

Figures \ref{fig:4}(e) and \ref{fig:4}(f) show the results for the case with $r_{\rm peb} = 0.5~{\rm mm}$, $R_{\rm p} = 200~{\rm km}$, and $t_{\rm acc} = 1.5~{\rm Myr}$.
Since this case has approximately three times larger radius than our fiducial case, the entire core retains liquid water even at the end of our simulation ($30~{\rm Myr}$ after accretion).
The temperature evolution at the center closely resembles that in Figure \ref{fig:4}(b), except for the timing of freezing.
We note that for a body with $R_{\rm p} = 200~{\rm km}$, the boiling temperature at the center is higher than $500~{\rm K}$.

We also present some numerical results that fail to explain the precipitation ages or temperatures.
Figures \ref{fig:5}(a) and \ref{fig:5}(b) show the results for the case with $r_{\rm peb} = 2.0~{\rm mm}$, $R_{\rm p} = 60~{\rm km}$, and $t_{\rm acc} = 1.0~{\rm Myr}$.
For this case, the temperature curves do not reproduce the data for either the C0002 or A0058 grains.
The effective thermal conductivity within the water-saturated core ($k_{\rm c(w)}$) is approximately proportional to the permeability $K$ (Equation (\ref{eq:k_cw_analytic})), and $K$ is proportional to ${r_{\rm peb}}^{2}$ (Equation (\ref{eq:K_pmb})).
Thus, $k_{\rm c(w)}$ is roughly proportional to ${r_{\rm peb}}^{2}$, and the high effective thermal conductivity induced by permeable flow consequently suppresses the rise in core temperature.

Figures \ref{fig:5}(c) and \ref{fig:5}(d) show the results for the case with $r_{\rm peb} = 0.2~{\rm mm}$, $R_{\rm p} = 60~{\rm km}$, and $t_{\rm acc} = 2.0~{\rm Myr}$.
For this case, core formation begins at $t = 2.57~{\rm Myr}$ (i.e., $0.57~{\rm Myr}$ after accretion; Figure \ref{fig:5}(c)).
The precipitation age of dolomite in the C0002 grain is $2.0_{-0.6}^{+0.6}~{\rm Myr}$ \citep{kawasaki2025dolomite}, and thus core formation must precede dolomite precipitation.
Indeed, the temperature curves in Figure \ref{fig:5}(d) cannot account for the precipitation age and temperature of the C0002 grain, indicating that the parent body of the Ryugu C0002 grain must have accreted earlier than $2.0~{\rm Myr}$ after CAI formation.
We note, however, that the precipitation age and temperature of the A0058 grain are reproduced under this parameter setting.
Therefore, if Ryugu is a rubble pile composed of fragments from multiple parent planetesimals, it remains possible that the parent planetesimal of the A0058 grain accreted later than $2.0~{\rm Myr}$ after CAI formation.

\subsection{Analytic solution of the central temperature}
\label{sec:analytic}

The temperature $T (R)$ is maximum at the center of the core.
Figure \ref{fig:6}(a) presents the temporal evolution of $T_{\rm center}$ for various values of $R_{\rm p}$ and $t_{\rm acc}$.
We find that the evolution of $T_{\rm center}$ is nearly independent of $R_{\rm p}$ when $R_{\rm p} \gtrsim 20~{\rm km}$.
In this section, we provide an analytical explanation for this independence.

Here, we derive an analytic solution for $T_{\rm center}$ using a quasi-steady-state heat balance approximation.
Assuming quasi-steady-state conditions (i.e., ${\partial}T/{\partial}t = 0$), Equation (\ref{eq:conduction}) can be rewritten as:
\begin{equation}
\frac{k_{\rm c(w)}}{R^{2}} \frac{{\partial}}{{\partial}R} {\left( R^{2} \frac{{\partial}T}{{\partial}R} \right)} + Q_{\rm decay} = 0.
\label{eq:steady}
\end{equation}
Solving Equation (\ref{eq:steady}) with the boundary condition $T = T_{\rm melt}$ at $R = R_{\rm c(w)}$, we obtain the central temperature as:
\begin{equation}
T_{\rm center} = T_{\rm melt} + \frac{Q_{\rm decay}}{6 k_{\rm c(w)}} {R_{\rm c(w)}}^{2}.
\label{eq:T_center}
\end{equation}

We also provide approximate expressions for $k_{\rm c(w)}$ and $Q_{\rm decay}$.
The effective thermal conductivity $k_{\rm c(w)}$ depends on ${\rm Nu}$, and for ${\rm Ra} \gg 200$, ${\rm Nu}$ asymptotically approaches ${\rm Nu} = 0.0081 {\rm Ra}$ (see Equation (\ref{eq:NuRa})).
By combining Equations (\ref{eq:Nu}--\ref{eq:Ra}), $k_{\rm c(w)}$ is approximately given by
\begin{equation}
k_{\rm c(w)} = 194~{\left( \frac{K}{2.8 \times 10^{-10}~{\rm m}^{2}} \right)} {\left( \frac{{\Delta T}}{100~{\rm K}} \right)} {\left( \frac{R_{\rm c(w)}}{50~{\rm km}} \right)}^{2}~{\rm W}~{\rm m}^{-1}~{\rm K}^{-1}.
\label{eq:k_cw_analytic}
\end{equation}
The decay heat $Q_{\rm decay}$ in the pebble-pile core is given in Equation (S18) of Supplementary Text S2.
Since the decay of $^{26}$Al is the dominant heat source during the first 10 Myr of the solar system, $Q_{\rm decay}$ can be approximated by
\begin{equation}
Q_{\rm decay} = 2.5 \times 10^{-4}~\exp{\left( - \frac{t}{1.0~{\rm Myr}} \right)}~{\rm W}~{\rm m}^{-3}.
\label{eq:Q}
\end{equation}

Using Equations (\ref{eq:T_center}--\ref{eq:Q}), we finally derive an analytical expression for $T_{\rm center}$ as follows:
\begin{equation}
T_{\rm center} = T_{\rm melt} + 230~{\left( \frac{r_{\rm peb}}{1~{\rm mm}} \right)}^{-1} \exp{\left( - \frac{t}{2.0~{\rm Myr}} \right)}~{\rm K}.
\label{eq:T_center_analytic}
\end{equation}
The dashed line in Figure \ref{fig:6}(a) represents the analytical formula, which successfully reproduces the numerical results. 
Equation (\ref{eq:T_center_analytic}) also reveals that $T_{\rm center}$ is independent of $R_{\rm p}$.

Figure \ref{fig:6}(b) shows the analytical solutions of $T_{\rm center}$ for various values of $r_{\rm peb}$.
We find that the precipitation temperatures of dolomite in neither the C0002 nor the A0058 grains can be reproduced if the parent planetesimal had $r_{\rm peb} \ge 2~{\rm mm}$.
This constraint is consistent with that obtained from the numerical results in Figure \ref{fig:5}(b), supporting the validity of the analytical solution.
Although the threshold value of $r_{\rm peb}$ depends on model assumptions such as $\phi_{\rm str}$ (see Figure S3 in Supplementary Materials), our results suggest that the pebbles forming Ryugu's parent planetesimal were typically $\ll 1~{\rm cm}$ in radius.

\subsection{Thermal evolution for the case without water circulation}

A key physical process considered in the present study is the enhancement of heat transfer within the core due to efficient water circulation.
If water circulation is neglected, the peak temperature at the center exceeds the threshold required to maintain liquid water.
Figure \ref{fig:7} shows the thermal evolution of planetesimal with $R_{\rm p} = 20~{\rm km}$ and $t_{\rm acc} = 1.0~{\rm Myr}$.
In this case, we set ${\rm Nu} = 1$ within the pebble-pile core, instead of using the value of ${\rm Nu}$ derived from Equation (\ref{eq:NuRa}).
Taking water circulation into account and assuming $r_{\rm peb} = 1~{\rm mm}$, the temporal evolution of $T_{\rm center}$ satisfies the condition for the presence of liquid water (Figure \ref{fig:4}(d)).
In contrast, without water circulation, $T_{\rm center}$ exceeds the threshold, and liquid water can no longer exist at the core center.
Given that the accretion age of the parent planetesimal of the C0002 grain is $t_{\rm acc} < 2.0~{\rm Myr}$ and its radius is $R_{\rm p} \gg 10~{\rm km}$, water circulation within the pebble-pile core likely played a critical role in suppressing the temperature rise and sustaining liquid water.

\section{Discussion}

\subsection{Accretion age and pebble radius of the Ryugu's parent planetesimal}

\begin{figure*}[]
\centering
\includegraphics[width = 0.9\textwidth]{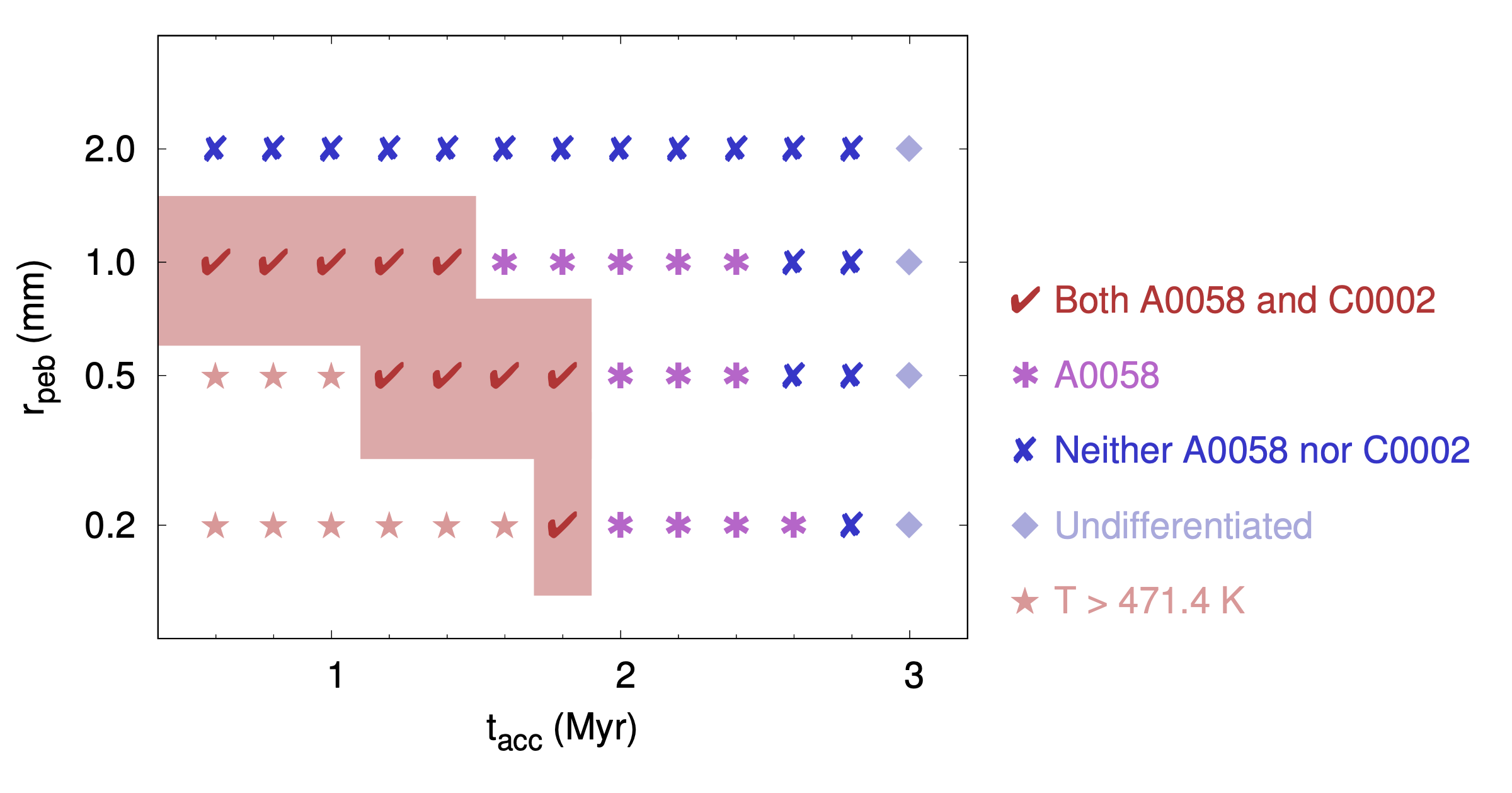}
\caption{
Summary of the thermal history of planetesimals for various combinations of $t_{\rm acc}$ and $r_{\rm peb}$.
Here, we fix the planetesimal radius at $R_{\rm p} = 60~{\rm km}$.
Checkmarks (\ding{52}) indicate cases where the precipitation conditions for both the A0058 and the C0002 grains were reproduced within a certain region of the pebble-pile core.
Asterisks (\ding{81}) indicate cases where only the conditions for the A0058 grain were reproduced, while those for the C0002 grain were not.
Crosses (\ding{56}) indicate cases where the conditions for neither the A0058 nor the C0002 grain were reproduced.
Diamonds (\ding{117}) denote cases where the entire planetesimal remained undifferentiated and the pebble-pile core did not form.
Stars (\ding{72}) indicate cases where the central temperature exceeded $471.4~{\rm K}$, the threshold for the thermophysical stability of liquid water.
The shaded region represents the model-derived range of $t_{\rm acc}$ and $r_{\rm peb}$ consistent with the parent planetesimal of Ryugu.
}
\label{fig:8}
\end{figure*}

We summarize the model-derived constraints on the accretion age and pebble radius of Ryugu's parent planetesimal.
Figure \ref{fig:8} presents the results for various combinations of $t_{\rm acc}$ and $r_{\rm peb}$, assuming a fixed planetesimal radius of $R_{\rm p} = 60~{\rm km}$.
The shaded region indicates the range of $t_{\rm acc}$ and $r_{\rm peb}$ values consistent with the thermal evolution of Ryugu's parent planetesimal.
To reproduce the precipitation ages and temperatures of both the C0002 and A0058 grains, the parent planetesimal must have accreted earlier than $2.0~{\rm Myr}$ after CAI formation, and the radius of the constituent pebbles should have been smaller than $2~{\rm mm}$.

In addition to the constraints on the precipitation ages and temperatures, dolomite in Ryugu samples is thought to have formed during the retrograde cooling stage of the parent planetesimal \citep{2023NatGe..16..675F, 2024M&PS...59.2097K}.
To satisfy this additional constraint, the accretion age of the parent planetesimal would need to be earlier than $1.5~{\rm Myr}$ after CAI formation.

These constraints on the accretion age are consistent with those reported by \citet{kawasaki2025dolomite}, regardless of differences in the thermal evolution modeling.
They also performed thermal evolution calculations using a model developed by \citet{2012NatCo...3..627F, 2013E&PSL.362..130F}.
Although their model does not account for the differentiation process, they concluded that Ryugu's parent planetesimal must have accreted earlier than $2~{\rm Myr}$ after CAI formation.

\citet{kawasaki2025dolomite} also suggested that the estimated accretion age of the parent planetesimal of Ryugu (and likely that of CI chondrites) is significantly earlier than those of other carbonaceous chondrite groups.
Unlike CI chondrites and Ryugu, most carbonaceous chondrites contain chondrules.
The formation ages of chondrules in the majority of carbonaceous chondrites, including CO (Ornans-type), CM (Mighei-type), and CV groups, range from $2.2$ to $2.8~{\rm Myr}$ after CAI formation \citep[e.g.,][]{2017GeCoA.201..303N, 2022GeCoA.322..194F}, while those in CR (Renazzo-type) chondrites formed at $\gtrsim 3.5~{\rm Myr}$ \citep[e.g.,][]{2017GeCoA.201..275S, 2019GeCoA.260..133T}.
Since chondrules must have formed before the accretion of their parent bodies, the accretion ages of these chondrite parent planetesimals must postdate $2.8~{\rm Myr}$ after CAI formation.
In contrast, Ryugu's parent planetesimal accreted within $2.0~{\rm Myr}$, and likely before $1.5~{\rm Myr}$ after CAI formation.
This substantial age difference may explain the near-absence of chondrules in Ryugu samples \citep[e.g.,][]{2022SciA....8E2067K, 2023Sci...379.8671N, 2023NatCo..14..532N, 2023NatAs...7..398Y, kawasaki2025solar}.

We acknowledge that the range of $r_{\rm peb}$ values consistent with the thermal evolution is sensitive to the parameters assumed in our calculations.
For example, in our fiducial model, we fixed the volume packing fraction of pebbles within the pebble-pile structure at $\phi_{\rm str} = 80\%$; however, the suitable range of $r_{\rm peb}$ shifts to smaller values when a lower $\phi_{\rm str}$ is adopted (see Supplementary Section S6).
Taking into account the uncertainty in $\phi_{\rm str}$, we provide a conservative constraint that the pebble radius within the parent body was likely smaller than $5~{\rm mm}$.

\subsection{Size and birthplace of Ryugu's parent planetesimal}

Ryugu's parent planetesimal is thought to have accreted in an orbit different from Ryugu's current near-Earth orbit, and its original size should have been larger than the current size of Ryugu.
Dynamical and spectral properties of Ryugu indicate that it likely originated from the Eulalia or New Polana asteroid families in the inner main asteroid belt \citep[e.g.,][]{2013Icar..225..283W, 2021NatAs...5...39T}.
The radii of the largest asteroids in the Eulalia or New Polana families are approximately $20~{\rm km}$ and $30~{\rm km}$, respectively, suggesting that the radius of the original parent planetesimal was at least larger than $20~{\rm km}$ \citep[e.g.,][]{2023Sci...379.8671N}.
This size constraint is consistent with the necessary condition for the presence of liquid water within the core at $T \approx 365~{\rm K}$ (Figure \ref{fig:1}(g)).

The birthplace of the parent planetesimal is thought to be beyond the ${\rm H}_{2}{\rm O}$ and ${\rm C}{\rm O}_{2}$ snow lines in the solar protoplanetary disk, possibly beyond the orbit of Jupiter.
\citet{2023Sci...379.8671N} performed synchrotron nano-computed tomography on a large pyrrhotite crystal in the Ryugu C0002 sample and discovered fluid inclusions.
These inclusions contain ${\rm C}{\rm O}_{2}$-bearing water with sulfur species and nitrogen- and chlorine-bearing organic compounds, indicating that Ryugu's parent planetesimal accreted beyond the ${\rm H}_{2}{\rm O}$ and ${\rm C}{\rm O}_{2}$ snow lines, similar to the case of CM chondrite \citep{2021SciA....7.9707T}.
In addition, the relative abundances of CAI-derived and chondrule-derived anhydrous primary minerals in Ryugu and CI chondrites differ from those in other carbonaceous chondrites, but closely resemble those observed in cometary dust grains, indicating that their parent planetesimal(s) likely accreted in the outer solar system far beyond the Jupiter's orbit \citep{2022SciA....8E2067K}.
This hypothesis is also supported by Fe nucleosynthetic isotopic anomalies observed in Ryugu and CI chondrites, which are distinguishable from those of other carbonaceous chondrites \citep{2022SciA....8D8141H}.
If the parent planetesimal formed beyond the ${\rm C}{\rm O}_{2}$ snow line, dust grains would have been coated with ${\rm C}{\rm O}_{2}$ ice, which is less sticky than ${\rm H}_{2}{\rm O}$ ice \citep[e.g.,][]{2021ApJ...910..130A, 2021ApJ...923..134F}.
The fragile nature of ${\rm C}{\rm O}_{2}$-mantled grains has been proposed as a reason why dust aggregates in the outer region of extrasolar protoplanetary disk are limited to sizes of $0.1$--$1~{\rm mm}$ \citep[e.g.,][]{2017ApJ...845...68P, 2019ApJ...878..132O}.
Therefore, the small pebble size of $r_{\rm peb} \ll 1~{\rm cm}$ is consistent with the scenario in which Ryugu's parent body accreted beyond the ${\rm C}{\rm O}_{2}$ snow line.

We assume that the pebble size within the core is approximately equivalent to that in the solar protoplanetary disk.
Although it remains uncertain whether pebbles can survive dissociation during ice melting in the undifferentiated region, physicochemical processes akin to cementation may promote their solidification as ice melting and aqueous alteration proceed \citep[e.g.,][]{2009JGRE..114.9006P}.
We also expect that the yield strengths of pebbles are greater than the hydrostatic pressure in planetesimals (see Section \ref{sec:structure}); however, the strengths depend on the details of mineral assemblages and vary considerably in reality.
Thus, future studies on the modification of pebble sizes within planetesimals are of great importance.

\subsection{Mud convection as an alternative scenario}

In our numerical model, the peak temperature readily exceeds the threshold for retaining liquid water when a small value of $r_{\rm peb}$ is chosen.
The internal structure and temperature evolution for the parameter set $r_{\rm peb} = 0.2~{\rm mm}$, $R_{\rm p} = 60~{\rm km}$ and $t_{\rm acc} = 1.0~{\rm Myr}$ are shown in Figure S4 in Supplementary Materials, where the peak temperature reaches approximately $700~{\rm K}$.
However, this does not imply that the building blocks of the parent planetesimal must have had $r_{\rm peb} > 0.2~{\rm mm}$.

In our model, we assume that pebbles settle and form a porous pebble-pile core upon ice melting.
This assumption is valid for cases where $r_{\rm peb} \gg 0.1~{\rm mm}$ \citep{2011EPS...63.1193W}.
In contrast, if $r_{\rm peb} \lesssim 0.1~{\rm mm}$, ``mud convection'' is expected to occur within the planetesimal \citep[e.g.,][]{2017SciA....3E2514B, 2018M&PS...53.2008T}.
Mud convection refers to efficient heat transfer driven by fluid convection dynamically coupled with fine dust grains.
The threshold value of $r_{\rm peb}$ separating pebble-pile core formation and mud convection likely depends on the balance between gravitational settling and stirring by convecting fluid.
Based on the numerical results of \citet{2017SciA....3E2514B}, we assume this threshold to be on the order of $0.1~{\rm mm}$.
Therefore, the model presented in this study is no longer applicable when $r_{\rm peb} \lesssim 0.1~{\rm mm}$, and in such cases, the temperature increase within the planetesimal would be suppressed by mud convection.

\subsection{Initial ${^{53}{\rm Mn} / ^{55}{\rm Mn}}$ ratio at the time of CAI formation}

We acknowledge that the initial ${^{53}{\rm Mn} / ^{55}{\rm Mn}}$ ratio at the time of CAI formation remains a subject of debate, and the calculated precipitation ages directly reflect the associated uncertainty.
In this study, we adopted a canonical initial ratio of $6.8 \times 10^{-6}$, which was derived from Pb--Pb ages and the initial ${^{53}{\rm Mn} / ^{55}{\rm Mn}}$ ratios of CAIs in CV chondrites and the D'Orbigny angrite \citep{2004M&PS...39..693G, 2012PNAS..109.9299B, 2012Sci...338..651C}.
However, a recent study by \citet{2023Icar..40215611D} proposed a higher initial ratio of $8.1 \times 10^{-6}$ using the initial ${^{26}{\rm Al} / ^{27}{\rm Al}}$ ratios of CAIs and angrites.
If this higher value is adopted, the precipitation ages of dolomite in the C0002 and A0058 grains are recalculated to be $2.9^{+0.6}_{-0.6}~{\rm Myr}$ and $5.0^{+0.9}_{-0.8}~{\rm Myr}$ after CAI formation, respectively \citep[see][]{kawasaki2025dolomite}.

Given this uncertainty, we evaluated the impact of adopting the alternative initial ${^{53}{\rm Mn} / ^{55}{\rm Mn}}$ ratio on the model-derived constraints for $t_{\rm acc}$ and $r_{\rm peb}$.
Figure S6 in Supplementary Materials presents the model results for the higher initial ratio, showing the parameter space consistent with the parent planetesimal of Ryugu.
While the overall trend remains similar to that obtained using the canonical ratio of $6.8 \times 10^{-6}$, the threshold value of $t_{\rm acc}$ required to reproduce the dolomite precipitation conditions are slightly shifted.
Nevertheless, the key conclusion of this study, that is, Ryugu's parent planetesimal formed within the first 2 Myr of the solar protoplanetary disk from pebbles smaller than a few millimeters in radius, remains robust and is consistent with the findings of \citet{kawasaki2025dolomite}.

\subsection{${\rm H}_{2}$ gas and unsaturated pores within the core}

In this study, we neglect the production of gaseous ${\rm H}_{2}$ during phyllosilicate formation.
However, in reality, ${\rm H}_{2}$ gas is produced through aqueous alteration \citep{2022AGUA....300568K, 2023Sci...379.8671N}, and a portion of it would be retained within the planetesimal.
The retention of ${\rm H}_{2}$ gas is essential to reproduce the oxidation state of iron: if all ${\rm H}_{2}$ gas had escaped from the system, the oxidation product of metallic ${\rm Fe}$ would not be magnetite (${\rm Fe}_{3}{\rm O}_{4}$) but hematite (${\rm Fe}_{2}{\rm O}_{3}$), which is not found in the major lithology \citep[e.g.,][]{2023Sci...379.8671N}.
Some of the produced ${\rm H}_{2}$ gas would also be transported outward within the planetesimal \citep{2024GeCoA.374..264S}, and the amount of retained ${\rm H}_{2}$ gas affects the assemblage of precipitating minerals.
This assemblage also depends on the local water-to-rock mass ratio, a point we discuss further in Supplementary Text S7.

We speculate that ${\rm H}_{2}$ gas could be retained within nm- or $\upmu$m-sized pores in pebbles.
Using X-ray computed nanotomography, \citet{2024GeCoA.375..146T} identified numerous $\upmu$m-scale pores and cracks in the Ryugu samples.
\citet{2023GeCoA.346...65D} also observed abundant nanopores with facets and dislocations in spherulitic magnetite using transmission electron microscopy.
The retention of ${\rm H}_{2}$ gas within pebbles indicates the presence of intra-pebble pores that are unsaturated with water.
Water could evaporate in these unsaturated pores, leading to the formation of evaporite minerals, such as chlorides and sulfates, observed in Ryugu samples \citep{2024NatAs...8.1536M}.

\section{Summary}

Recent astronomical observations of planet-forming circumstellar disks have revealed that planetesimals formed from 0.1-mm to 1-cm-sized dust aggregates, commonly referred to as "pebbles."
Icy planetesimals that formed beyond the water snow line were composed of both rock and ice.
 A rocky core would form through water--rock differentiation once the internal temperature exceeds $273~{\rm K}$.
Given the presence of pebbles, the resulting rocky core after differentiation is expected to have a porous pebble-pile structure with voids saturated by liquid water (see Figure \ref{fig:1}).
Water circulation through these voids enhances heat transfer within the core, thereby suppressing the temperature rise caused by the decay of radioactive nuclides.

Since the efficiency of water circulation depends on the radius of the constituent pebbles, the pebble size can be constrained by analyzing extraterrestrial materials that records the aqueous alteration history of its parent planetesimal.
Indeed, aqueously formed minerals, particularly dolomite, found in samples returned from asteroid Ryugu provide valuable insights into the formation and evolutionary history of its parent planetesimal (see Section \ref{sec:dolomite}).
\citet{kawasaki2025dolomite} determined the precipitation ages and temperatures of dolomite in two Ryugu samples, C0002 and A0058.
Using these constraints, we derived plausible estimates for the accretion age and pebble radius of Ryugu's parent planetesimal.

We numerically investigated the thermal evolution of icy planetesimals while accounting for their structural evolution (see Section \ref{sec:results}).
Our model includes the formation of a pebble-pile core following ice melting and incorporates enhanced heat transfer within the core due to water circulation (see Section \ref{sec:k_cw}).
We found that the precipitation ages and temperatures of dolomite in both the C0002 and A0058 grains can be reproduced if the parent planetesimal accreted earlier than $2.0~{\rm Myr}$ after CAI formation and had a pebble radius smaller than $2~{\rm mm}$ (see Figure \ref{fig:8}).
Regardless of differences in the details of thermal evolution modeling, our estimated accretion age aligns with the findings of \citet{kawasaki2025dolomite}, suggesting that Ryugu's parent planetesimal accreted earlier than the majority of chondrule-bearing carbonaceous chondrite parent bodies \citep[e.g.,][]{2022GeCoA.322..194F}.
Furthermore, a pebble radius smaller than $2~{\rm mm}$ is consistent with theoretical predictions for planetesimals that accreted beyond the ${\rm C}{\rm O}_{2}$ snow line \citep[e.g.,][]{2023Sci...379.8671N}, where pebbles are less adhesive upon collision \citep[e.g.,][]{2019ApJ...878..132O}.







\section*{Acknowledgments}

We appreciate Sei-ichiro Watanabe, Akira Tsuchiyama, and Hiroshi Kobayashi for constructive comments.
This work was supported by JSPS KAKENHI Grant (JP24K17118, JP24KK0072, JP25H00678).
Numerical computations were partly carried out on the general-purpose PC cluster at the Center for Computational Astrophysics, National Astronomical Observatory of Japan.




\bibliographystyle{elsarticle-harv} 

\begin{thebibliography}{0}
\expandafter\ifx\csname natexlab\endcsname\relax\def\natexlab#1{#1}\fi
\providecommand{\url}[1]{\texttt{#1}}
\providecommand{\href}[2]{#2}
\providecommand{\path}[1]{#1}
\providecommand{\DOIprefix}{doi:}
\providecommand{\ArXivprefix}{arXiv:}
\providecommand{\URLprefix}{URL: }
\providecommand{\Pubmedprefix}{pmid:}
\providecommand{\doi}[1]{\href{http://dx.doi.org/#1}{\path{#1}}}
\providecommand{\Pubmed}[1]{\href{pmid:#1}{\path{#1}}}
\providecommand{\bibinfo}[2]{#2}
\ifx\xfnm\relax \def\xfnm[#1]{\unskip,\space#1}\fi

\end{thebibliography}


\begin{thebibliography}{71}
\expandafter\ifx\csname natexlab\endcsname\relax\def\natexlab#1{#1}\fi
\providecommand{\url}[1]{\texttt{#1}}
\providecommand{\href}[2]{#2}
\providecommand{\path}[1]{#1}
\providecommand{\DOIprefix}{doi:}
\providecommand{\ArXivprefix}{arXiv:}
\providecommand{\URLprefix}{URL: }
\providecommand{\Pubmedprefix}{pmid:}
\providecommand{\doi}[1]{\href{http://dx.doi.org/#1}{\path{#1}}}
\providecommand{\Pubmed}[1]{\href{pmid:#1}{\path{#1}}}
\providecommand{\bibinfo}[2]{#2}
\ifx\xfnm\relax \def\xfnm[#1]{\unskip,\space#1}\fi
\bibitem[{{Arakawa} and {Krijt}(2021)}]{2021ApJ...910..130A}
\bibinfo{author}{{Arakawa}, S.}, \bibinfo{author}{{Krijt}, S.},
  \bibinfo{year}{2021}.
\newblock \bibinfo{title}{{On the Stickiness of CO$_{2}$ and H$_{2}$O Ice
  Particles}}.
\newblock \bibinfo{journal}{The Astrophysical Journal} \bibinfo{volume}{910},
  \bibinfo{pages}{130}.
\newblock \DOIprefix\doi{10.3847/1538-4357/abe61d}.
\bibitem[{{Arakawa} et~al.(2023){Arakawa}, {Okuzumi}, {Tatsuuma}, {Tanaka},
  {Kokubo}, {Nishiura}, {Furuichi} and {Nakamoto}}]{2023ApJ...951L..16A}
\bibinfo{author}{{Arakawa}, S.}, \bibinfo{author}{{Okuzumi}, S.},
  \bibinfo{author}{{Tatsuuma}, M.}, \bibinfo{author}{{Tanaka}, H.},
  \bibinfo{author}{{Kokubo}, E.}, \bibinfo{author}{{Nishiura}, D.},
  \bibinfo{author}{{Furuichi}, M.}, \bibinfo{author}{{Nakamoto}, T.},
  \bibinfo{year}{2023}.
\newblock \bibinfo{title}{{Size Dependence of the Bouncing Barrier in
  Protoplanetary Dust Growth}}.
\newblock \bibinfo{journal}{The Astrophysical Journal Letters}
  \bibinfo{volume}{951}, \bibinfo{pages}{L16}.
\newblock \DOIprefix\doi{10.3847/2041-8213/acdb5f}.
\bibitem[{{Arakawa} and {Wakita}(2024)}]{2024PASJ...76..130A}
\bibinfo{author}{{Arakawa}, S.}, \bibinfo{author}{{Wakita}, S.},
  \bibinfo{year}{2024}.
\newblock \bibinfo{title}{{Survivability of amorphous ice in comets depends on
  the latent heat of crystallization of impure water ice}}.
\newblock \bibinfo{journal}{Publications of the Astronomical Society of Japan}
  \bibinfo{volume}{76}, \bibinfo{pages}{130--141}.
\newblock \DOIprefix\doi{10.1093/pasj/psad086}.
\bibitem[{{Bland} et~al.(2009){Bland}, {Jackson}, {Coker}, {Cohen}, {Webber},
  {Lee}, {Duffy}, {Chater}, {Ardakani}, {McPhail}, {McComb} and
  {Benedix}}]{2009E&PSL.287..559B}
\bibinfo{author}{{Bland}, P.A.}, \bibinfo{author}{{Jackson}, M.D.},
  \bibinfo{author}{{Coker}, R.F.}, \bibinfo{author}{{Cohen}, B.A.},
  \bibinfo{author}{{Webber}, J.B.W.}, \bibinfo{author}{{Lee}, M.R.},
  \bibinfo{author}{{Duffy}, C.M.}, \bibinfo{author}{{Chater}, R.J.},
  \bibinfo{author}{{Ardakani}, M.G.}, \bibinfo{author}{{McPhail}, D.S.},
  \bibinfo{author}{{McComb}, D.W.}, \bibinfo{author}{{Benedix}, G.K.},
  \bibinfo{year}{2009}.
\newblock \bibinfo{title}{{Why aqueous alteration in asteroids was isochemical:
  High porosity {\ensuremath{\neq}} high permeability}}.
\newblock \bibinfo{journal}{Earth and Planetary Science Letters}
  \bibinfo{volume}{287}, \bibinfo{pages}{559--568}.
\newblock \DOIprefix\doi{10.1016/j.epsl.2009.09.004}.
\bibitem[{{Bland} and {Travis}(2017)}]{2017SciA....3E2514B}
\bibinfo{author}{{Bland}, P.A.}, \bibinfo{author}{{Travis}, B.J.},
  \bibinfo{year}{2017}.
\newblock \bibinfo{title}{{Giant convecting mud balls of the early solar
  system}}.
\newblock \bibinfo{journal}{Science Advances} \bibinfo{volume}{3},
  \bibinfo{pages}{e1602514}.
\newblock \DOIprefix\doi{10.1126/sciadv.1602514}.
\bibitem[{{Brennecka} and {Wadhwa}(2012)}]{2012PNAS..109.9299B}
\bibinfo{author}{{Brennecka}, G.A.}, \bibinfo{author}{{Wadhwa}, M.},
  \bibinfo{year}{2012}.
\newblock \bibinfo{title}{{Uranium isotope compositions of the basaltic angrite
  meteorites and the chronological implications for the early Solar System}}.
\newblock \bibinfo{journal}{Proceedings of the National Academy of Science}
  \bibinfo{volume}{109}, \bibinfo{pages}{9299--9303}.
\newblock \DOIprefix\doi{10.1073/pnas.1114043109}.
\bibitem[{{Connelly} et~al.(2012){Connelly}, {Bizzarro}, {Krot}, {Nordlund},
  {Wielandt} and {Ivanova}}]{2012Sci...338..651C}
\bibinfo{author}{{Connelly}, J.N.}, \bibinfo{author}{{Bizzarro}, M.},
  \bibinfo{author}{{Krot}, A.N.}, \bibinfo{author}{{Nordlund}, {\r{A}}.},
  \bibinfo{author}{{Wielandt}, D.}, \bibinfo{author}{{Ivanova}, M.A.},
  \bibinfo{year}{2012}.
\newblock \bibinfo{title}{{The Absolute Chronology and Thermal Processing of
  Solids in the Solar Protoplanetary Disk}}.
\newblock \bibinfo{journal}{Science} \bibinfo{volume}{338},
  \bibinfo{pages}{651}.
\newblock \DOIprefix\doi{10.1126/science.1226919}.
\bibitem[{{Desch} et~al.(2023){Desch}, {Dunlap}, {Williams}, {Mane} and
  {Dunham}}]{2023Icar..40215611D}
\bibinfo{author}{{Desch}, S.J.}, \bibinfo{author}{{Dunlap}, D.R.},
  \bibinfo{author}{{Williams}, C.D.}, \bibinfo{author}{{Mane}, P.},
  \bibinfo{author}{{Dunham}, E.T.}, \bibinfo{year}{2023}.
\newblock \bibinfo{title}{{Statistical chronometry of Meteorites: II. Initial
  abundances and homogeneity of short-lived radionuclides}}.
\newblock \bibinfo{journal}{Icarus} \bibinfo{volume}{402},
  \bibinfo{pages}{115611}.
\newblock \DOIprefix\doi{10.1016/j.icarus.2023.115611}.
\bibitem[{{Dobric{\u{a}}} et~al.(2023){Dobric{\u{a}}}, {Ishii}, {Bradley},
  {Ohtaki}, {Brearley}, {Noguchi}, {Matsumoto}, {Miyake}, {Igami}, {Haruta},
  {Saito}, {Hata}, {Seto}, {Miyahara}, {Tomioka}, {Leroux}, {Le Guillou},
  {Jacob}, {de la Pe{\~n}a}, {Laforet}, {Marinova}, {Langenhorst}, {Harries},
  {Beck}, {Phan}, {Rebois}, {Abreu}, {Gray}, {Zega}, {Zanetta}, {Thompson},
  {Stroud}, {Burgess}, {Cymes}, {Bridges}, {Hicks}, {Lee}, {Daly}, {Bland},
  {Zolensky}, {Frank}, {Martinez}, {Tsuchiyama}, {Yasutake}, {Matsuno},
  {Okumura}, {Mitsukawa}, {Uesugi}, {Uesugi}, {Takeuchi}, {Sun}, {Enju},
  {Takigawa}, {Michikami}, {Nakamura}, {Matsumoto}, {Nakauchi}, {Yurimoto},
  {Okazaki}, {Yabuta}, {Naraoka}, {Sakamoto}, {Tachibana}, {Yada}, {Nishimura},
  {Nakato}, {Miyazaki}, {Yogata}, {Abe}, {Okada}, {Usui}, {Yoshikawa}, {Saiki},
  {Tanaka}, {Terui}, {Nakazawa}, {Watanabe} and {Tsuda}}]{2023GeCoA.346...65D}
\bibinfo{author}{{Dobric{\u{a}}}, E.}, \bibinfo{author}{{Ishii}, H.A.},
  \bibinfo{author}{{Bradley}, J.P.}, \bibinfo{author}{{Ohtaki}, K.},
  \bibinfo{author}{{Brearley}, A.J.}, \bibinfo{author}{{Noguchi}, T.},
  \bibinfo{author}{{Matsumoto}, T.}, \bibinfo{author}{{Miyake}, A.},
  \bibinfo{author}{{Igami}, Y.}, \bibinfo{author}{{Haruta}, M.},
  \bibinfo{author}{{Saito}, H.}, \bibinfo{author}{{Hata}, S.},
  \bibinfo{author}{{Seto}, Y.}, \bibinfo{author}{{Miyahara}, M.},
  \bibinfo{author}{{Tomioka}, N.}, \bibinfo{author}{{Leroux}, H.},
  \bibinfo{author}{{Le Guillou}, C.}, \bibinfo{author}{{Jacob}, D.},
  \bibinfo{author}{{de la Pe{\~n}a}, F.}, \bibinfo{author}{{Laforet}, S.},
  \bibinfo{author}{{Marinova}, M.}, \bibinfo{author}{{Langenhorst}, F.},
  \bibinfo{author}{{Harries}, D.}, \bibinfo{author}{{Beck}, P.},
  \bibinfo{author}{{Phan}, T.H.V.}, \bibinfo{author}{{Rebois}, R.},
  \bibinfo{author}{{Abreu}, N.M.}, \bibinfo{author}{{Gray}, J.},
  \bibinfo{author}{{Zega}, T.}, \bibinfo{author}{{Zanetta}, P.M.},
  \bibinfo{author}{{Thompson}, M.S.}, \bibinfo{author}{{Stroud}, R.},
  \bibinfo{author}{{Burgess}, K.}, \bibinfo{author}{{Cymes}, B.A.},
  \bibinfo{author}{{Bridges}, J.C.}, \bibinfo{author}{{Hicks}, L.},
  \bibinfo{author}{{Lee}, M.R.}, \bibinfo{author}{{Daly}, L.},
  \bibinfo{author}{{Bland}, P.A.}, \bibinfo{author}{{Zolensky}, M.E.},
  \bibinfo{author}{{Frank}, D.R.}, \bibinfo{author}{{Martinez}, J.},
  \bibinfo{author}{{Tsuchiyama}, A.}, \bibinfo{author}{{Yasutake}, M.},
  \bibinfo{author}{{Matsuno}, J.}, \bibinfo{author}{{Okumura}, S.},
  \bibinfo{author}{{Mitsukawa}, I.}, \bibinfo{author}{{Uesugi}, K.},
  \bibinfo{author}{{Uesugi}, M.}, \bibinfo{author}{{Takeuchi}, A.},
  \bibinfo{author}{{Sun}, M.}, \bibinfo{author}{{Enju}, S.},
  \bibinfo{author}{{Takigawa}, A.}, \bibinfo{author}{{Michikami}, T.},
  \bibinfo{author}{{Nakamura}, T.}, \bibinfo{author}{{Matsumoto}, M.},
  \bibinfo{author}{{Nakauchi}, Y.}, \bibinfo{author}{{Yurimoto}, H.},
  \bibinfo{author}{{Okazaki}, R.}, \bibinfo{author}{{Yabuta}, H.},
  \bibinfo{author}{{Naraoka}, H.}, \bibinfo{author}{{Sakamoto}, K.},
  \bibinfo{author}{{Tachibana}, S.}, \bibinfo{author}{{Yada}, T.},
  \bibinfo{author}{{Nishimura}, M.}, \bibinfo{author}{{Nakato}, A.},
  \bibinfo{author}{{Miyazaki}, A.}, \bibinfo{author}{{Yogata}, K.},
  \bibinfo{author}{{Abe}, M.}, \bibinfo{author}{{Okada}, T.},
  \bibinfo{author}{{Usui}, T.}, \bibinfo{author}{{Yoshikawa}, M.},
  \bibinfo{author}{{Saiki}, T.}, \bibinfo{author}{{Tanaka}, S.},
  \bibinfo{author}{{Terui}, F.}, \bibinfo{author}{{Nakazawa}, S.},
  \bibinfo{author}{{Watanabe}, S.i.}, \bibinfo{author}{{Tsuda}, Y.},
  \bibinfo{year}{2023}.
\newblock \bibinfo{title}{{Nonequilibrium spherulitic magnetite in the Ryugu
  samples}}.
\newblock \bibinfo{journal}{Geochimica et Cosmochimica Acta}
  \bibinfo{volume}{346}, \bibinfo{pages}{65--75}.
\newblock \DOIprefix\doi{10.1016/j.gca.2023.02.003}.
\bibitem[{{Dr{\k{a}}{\.z}kowska} et~al.(2023){Dr{\k{a}}{\.z}kowska}, {Bitsch},
  {Lambrechts}, {Mulders}, {Harsono}, {Vazan}, {Liu}, {Ormel}, {Kretke} and
  {Morbidelli}}]{2023ASPC..534..717D}
\bibinfo{author}{{Dr{\k{a}}{\.z}kowska}, J.}, \bibinfo{author}{{Bitsch}, B.},
  \bibinfo{author}{{Lambrechts}, M.}, \bibinfo{author}{{Mulders}, G.D.},
  \bibinfo{author}{{Harsono}, D.}, \bibinfo{author}{{Vazan}, A.},
  \bibinfo{author}{{Liu}, B.}, \bibinfo{author}{{Ormel}, C.W.},
  \bibinfo{author}{{Kretke}, K.}, \bibinfo{author}{{Morbidelli}, A.},
  \bibinfo{year}{2023}.
\newblock \bibinfo{title}{{Planet Formation Theory in the Era of ALMA and
  Kepler: from Pebbles to Exoplanets}}, in: \bibinfo{editor}{{Inutsuka}, S.},
  \bibinfo{editor}{{Aikawa}, Y.}, \bibinfo{editor}{{Muto}, T.},
  \bibinfo{editor}{{Tomida}, K.}, \bibinfo{editor}{{Tamura}, M.} (Eds.),
  \bibinfo{booktitle}{Protostars and Planets VII}, p. \bibinfo{pages}{717}.
\newblock \DOIprefix\doi{10.48550/arXiv.2203.09759}.
\bibitem[{{Dullien}(2012)}]{dullien2012porous}
\bibinfo{author}{{Dullien}, F.A.}, \bibinfo{year}{2012}.
\newblock \bibinfo{title}{{Porous media: fluid transport and pore structure}}.
\newblock \bibinfo{publisher}{Academic press}.
\bibitem[{{Fritscher} and {Teiser}(2021)}]{2021ApJ...923..134F}
\bibinfo{author}{{Fritscher}, M.}, \bibinfo{author}{{Teiser}, J.},
  \bibinfo{year}{2021}.
\newblock \bibinfo{title}{{CO$_{2}$-ice Collisions: A New Experimental
  Approach}}.
\newblock \bibinfo{journal}{The Astrophysical Journal} \bibinfo{volume}{923},
  \bibinfo{pages}{134}.
\newblock \DOIprefix\doi{10.3847/1538-4357/ac2df4}.
\bibitem[{{Fujiya} et~al.(2023){Fujiya}, {Kawasaki}, {Nagashima}, {Sakamoto},
  {O'D. Alexander}, {Kita}, {Kitajima}, {Abe}, {Al{\'e}on}, {Amari}, {Amelin},
  {Bajo}, {Bizzarro}, {Bouvier}, {Carlson}, {Chaussidon}, {Choi}, {Dauphas},
  {Davis}, {Di Rocco}, {Fukai}, {Gautam}, {Haba}, {Hibiya}, {Hidaka}, {Homma},
  {Hoppe}, {Huss}, {Ichida}, {Iizuka}, {Ireland}, {Ishikawa}, {Itoh}, {Kleine},
  {Komatani}, {Krot}, {Liu}, {Masuda}, {McKeegan}, {Morita}, {Motomura},
  {Moynier}, {Nakai}, {Nguyen}, {Nittler}, {Onose}, {Pack}, {Park}, {Piani},
  {Qin}, {Russell}, {Sch{\"o}nb{\"a}chler}, {Tafla}, {Tang}, {Terada},
  {Terada}, {Usui}, {Wada}, {Wadhwa}, {Walker}, {Yamashita}, {Yin}, {Yokoyama},
  {Yoneda}, {Young}, {Yui}, {Zhang}, {Nakamura}, {Naraoka}, {Noguchi},
  {Okazaki}, {Sakamoto}, {Yabuta}, {Abe}, {Miyazaki}, {Nakato}, {Nishimura},
  {Okada}, {Yada}, {Yogata}, {Nakazawa}, {Saiki}, {Tanaka}, {Terui}, {Tsuda},
  {Watanabe}, {Yoshikawa}, {Tachibana} and {Yurimoto}}]{2023NatGe..16..675F}
\bibinfo{author}{{Fujiya}, W.}, \bibinfo{author}{{Kawasaki}, N.},
  \bibinfo{author}{{Nagashima}, K.}, \bibinfo{author}{{Sakamoto}, N.},
  \bibinfo{author}{{O'D. Alexander}, C.M.}, \bibinfo{author}{{Kita}, N.T.},
  \bibinfo{author}{{Kitajima}, K.}, \bibinfo{author}{{Abe}, Y.},
  \bibinfo{author}{{Al{\'e}on}, J.}, \bibinfo{author}{{Amari}, S.},
  \bibinfo{author}{{Amelin}, Y.}, \bibinfo{author}{{Bajo}, K.i.},
  \bibinfo{author}{{Bizzarro}, M.}, \bibinfo{author}{{Bouvier}, A.},
  \bibinfo{author}{{Carlson}, R.W.}, \bibinfo{author}{{Chaussidon}, M.},
  \bibinfo{author}{{Choi}, B.G.}, \bibinfo{author}{{Dauphas}, N.},
  \bibinfo{author}{{Davis}, A.M.}, \bibinfo{author}{{Di Rocco}, T.},
  \bibinfo{author}{{Fukai}, R.}, \bibinfo{author}{{Gautam}, I.},
  \bibinfo{author}{{Haba}, M.K.}, \bibinfo{author}{{Hibiya}, Y.},
  \bibinfo{author}{{Hidaka}, H.}, \bibinfo{author}{{Homma}, H.},
  \bibinfo{author}{{Hoppe}, P.}, \bibinfo{author}{{Huss}, G.R.},
  \bibinfo{author}{{Ichida}, K.}, \bibinfo{author}{{Iizuka}, T.},
  \bibinfo{author}{{Ireland}, T.R.}, \bibinfo{author}{{Ishikawa}, A.},
  \bibinfo{author}{{Itoh}, S.}, \bibinfo{author}{{Kleine}, T.},
  \bibinfo{author}{{Komatani}, S.}, \bibinfo{author}{{Krot}, A.N.},
  \bibinfo{author}{{Liu}, M.C.}, \bibinfo{author}{{Masuda}, Y.},
  \bibinfo{author}{{McKeegan}, K.D.}, \bibinfo{author}{{Morita}, M.},
  \bibinfo{author}{{Motomura}, K.}, \bibinfo{author}{{Moynier}, F.},
  \bibinfo{author}{{Nakai}, I.}, \bibinfo{author}{{Nguyen}, A.},
  \bibinfo{author}{{Nittler}, L.}, \bibinfo{author}{{Onose}, M.},
  \bibinfo{author}{{Pack}, A.}, \bibinfo{author}{{Park}, C.},
  \bibinfo{author}{{Piani}, L.}, \bibinfo{author}{{Qin}, L.},
  \bibinfo{author}{{Russell}, S.S.}, \bibinfo{author}{{Sch{\"o}nb{\"a}chler},
  M.}, \bibinfo{author}{{Tafla}, L.}, \bibinfo{author}{{Tang}, H.},
  \bibinfo{author}{{Terada}, K.}, \bibinfo{author}{{Terada}, Y.},
  \bibinfo{author}{{Usui}, T.}, \bibinfo{author}{{Wada}, S.},
  \bibinfo{author}{{Wadhwa}, M.}, \bibinfo{author}{{Walker}, R.J.},
  \bibinfo{author}{{Yamashita}, K.}, \bibinfo{author}{{Yin}, Q.Z.},
  \bibinfo{author}{{Yokoyama}, T.}, \bibinfo{author}{{Yoneda}, S.},
  \bibinfo{author}{{Young}, E.D.}, \bibinfo{author}{{Yui}, H.},
  \bibinfo{author}{{Zhang}, A.C.}, \bibinfo{author}{{Nakamura}, T.},
  \bibinfo{author}{{Naraoka}, H.}, \bibinfo{author}{{Noguchi}, T.},
  \bibinfo{author}{{Okazaki}, R.}, \bibinfo{author}{{Sakamoto}, K.},
  \bibinfo{author}{{Yabuta}, H.}, \bibinfo{author}{{Abe}, M.},
  \bibinfo{author}{{Miyazaki}, A.}, \bibinfo{author}{{Nakato}, A.},
  \bibinfo{author}{{Nishimura}, M.}, \bibinfo{author}{{Okada}, T.},
  \bibinfo{author}{{Yada}, T.}, \bibinfo{author}{{Yogata}, K.},
  \bibinfo{author}{{Nakazawa}, S.}, \bibinfo{author}{{Saiki}, T.},
  \bibinfo{author}{{Tanaka}, S.}, \bibinfo{author}{{Terui}, F.},
  \bibinfo{author}{{Tsuda}, Y.}, \bibinfo{author}{{Watanabe}, S.i.},
  \bibinfo{author}{{Yoshikawa}, M.}, \bibinfo{author}{{Tachibana}, S.},
  \bibinfo{author}{{Yurimoto}, H.}, \bibinfo{year}{2023}.
\newblock \bibinfo{title}{{Carbonate record of temporal change in oxygen
  fugacity and gaseous species in asteroid Ryugu}}.
\newblock \bibinfo{journal}{Nature Geoscience} \bibinfo{volume}{16},
  \bibinfo{pages}{675--682}.
\newblock \DOIprefix\doi{10.1038/s41561-023-01226-y}.
\bibitem[{{Fujiya} et~al.(2012){Fujiya}, {Sugiura}, {Hotta}, {Ichimura} and
  {Sano}}]{2012NatCo...3..627F}
\bibinfo{author}{{Fujiya}, W.}, \bibinfo{author}{{Sugiura}, N.},
  \bibinfo{author}{{Hotta}, H.}, \bibinfo{author}{{Ichimura}, K.},
  \bibinfo{author}{{Sano}, Y.}, \bibinfo{year}{2012}.
\newblock \bibinfo{title}{{Evidence for the late formation of hydrous asteroids
  from young meteoritic carbonates}}.
\newblock \bibinfo{journal}{Nature Communications} \bibinfo{volume}{3},
  \bibinfo{pages}{627}.
\newblock \DOIprefix\doi{10.1038/ncomms1635}.
\bibitem[{{Fujiya} et~al.(2013){Fujiya}, {Sugiura}, {Sano} and
  {Hiyagon}}]{2013E&PSL.362..130F}
\bibinfo{author}{{Fujiya}, W.}, \bibinfo{author}{{Sugiura}, N.},
  \bibinfo{author}{{Sano}, Y.}, \bibinfo{author}{{Hiyagon}, H.},
  \bibinfo{year}{2013}.
\newblock \bibinfo{title}{{Mn-Cr ages of dolomites in CI chondrites and the
  Tagish Lake ungrouped carbonaceous chondrite}}.
\newblock \bibinfo{journal}{Earth and Planetary Science Letters}
  \bibinfo{volume}{362}, \bibinfo{pages}{130--142}.
\newblock \DOIprefix\doi{10.1016/j.epsl.2012.11.057}.
\bibitem[{{Fukuda} et~al.(2022){Fukuda}, {Tenner}, {Kimura}, {Tomioka},
  {Siron}, {Ushikubo}, {Chaumard}, {Hertwig} and {Kita}}]{2022GeCoA.322..194F}
\bibinfo{author}{{Fukuda}, K.}, \bibinfo{author}{{Tenner}, T.J.},
  \bibinfo{author}{{Kimura}, M.}, \bibinfo{author}{{Tomioka}, N.},
  \bibinfo{author}{{Siron}, G.}, \bibinfo{author}{{Ushikubo}, T.},
  \bibinfo{author}{{Chaumard}, N.}, \bibinfo{author}{{Hertwig}, A.T.},
  \bibinfo{author}{{Kita}, N.T.}, \bibinfo{year}{2022}.
\newblock \bibinfo{title}{{A temporal shift of chondrule generation from the
  inner to outer Solar System inferred from oxygen isotopes and Al-Mg
  chronology of chondrules from primitive CM and CO chondrites}}.
\newblock \bibinfo{journal}{Geochimica et Cosmochimica Acta}
  \bibinfo{volume}{322}, \bibinfo{pages}{194--226}.
\newblock \DOIprefix\doi{10.1016/j.gca.2021.12.027}.
\bibitem[{{Genda}(2016)}]{2016GeocJ..50...27G}
\bibinfo{author}{{Genda}, H.}, \bibinfo{year}{2016}.
\newblock \bibinfo{title}{{Origin of Earth's oceans: An assessment of the total
  amount, history and supply of water}}.
\newblock \bibinfo{journal}{Geochemical Journal} \bibinfo{volume}{50},
  \bibinfo{pages}{27--42}.
\newblock \DOIprefix\doi{10.2343/geochemj.2.0398}.
\bibitem[{{Glavin} et~al.(2004){Glavin}, {Kubny}, {Jagoutz} and
  {Lugmair}}]{2004M&PS...39..693G}
\bibinfo{author}{{Glavin}, D.P.}, \bibinfo{author}{{Kubny}, A.},
  \bibinfo{author}{{Jagoutz}, E.}, \bibinfo{author}{{Lugmair}, G.W.},
  \bibinfo{year}{2004}.
\newblock \bibinfo{title}{{Mn-Cr isotope systematics of the D'Orbigny
  angrite}}.
\newblock \bibinfo{journal}{Meteoritics \& Planetary Science}
  \bibinfo{volume}{39}, \bibinfo{pages}{693--700}.
\newblock \DOIprefix\doi{10.1111/j.1945-5100.2004.tb00112.x}.
\bibitem[{{Grimm} and {McSween}(1989)}]{1989Icar...82..244G}
\bibinfo{author}{{Grimm}, R.E.}, \bibinfo{author}{{McSween}, Jr., H.Y.},
  \bibinfo{year}{1989}.
\newblock \bibinfo{title}{{Water and the thermal evolution of carbonaceous
  chondrite parent bodies}}.
\newblock \bibinfo{journal}{Icarus} \bibinfo{volume}{82},
  \bibinfo{pages}{244--280}.
\newblock \DOIprefix\doi{10.1016/0019-1035(89)90038-9}.
\bibitem[{{Grott} et~al.(2020){Grott}, {Biele}, {Michel}, {Sugita},
  {Schr{\"o}der}, {Sakatani}, {Neumann}, {Kameda}, {Michikami} and
  {Honda}}]{2020JGRE..12506519G}
\bibinfo{author}{{Grott}, M.}, \bibinfo{author}{{Biele}, J.},
  \bibinfo{author}{{Michel}, P.}, \bibinfo{author}{{Sugita}, S.},
  \bibinfo{author}{{Schr{\"o}der}, S.}, \bibinfo{author}{{Sakatani}, N.},
  \bibinfo{author}{{Neumann}, W.}, \bibinfo{author}{{Kameda}, S.},
  \bibinfo{author}{{Michikami}, T.}, \bibinfo{author}{{Honda}, C.},
  \bibinfo{year}{2020}.
\newblock \bibinfo{title}{{Macroporosity and Grain Density of Rubble Pile
  Asteroid (162173) Ryugu}}.
\newblock \bibinfo{journal}{Journal of Geophysical Research (Planets)}
  \bibinfo{volume}{125}, \bibinfo{pages}{e06519}.
\newblock \DOIprefix\doi{10.1029/2020JE00651910.1002/essoar.10503201.2}.
\bibitem[{{Hopp} et~al.(2022){Hopp}, {Dauphas}, {Abe}, {Al{\'e}on}, {O'D.
  Alexander}, {Amari}, {Amelin}, {Bajo}, {Bizzarro}, {Bouvier}, {Carlson},
  {Chaussidon}, {Choi}, {Davis}, {Di Rocco}, {Fujiya}, {Fukai}, {Gautam},
  {Haba}, {Hibiya}, {Hidaka}, {Homma}, {Hoppe}, {Huss}, {Ichida}, {Iizuka},
  {Ireland}, {Ishikawa}, {Ito}, {Itoh}, {Kawasaki}, {Kita}, {Kitajima},
  {Kleine}, {Komatani}, {Krot}, {Liu}, {Masuda}, {McKeegan}, {Morita},
  {Motomura}, {Moynier}, {Nakai}, {Nagashima}, {Nesvorn{\'y}}, {Nguyen},
  {Nittler}, {Onose}, {Pack}, {Park}, {Piani}, {Qin}, {Russell}, {Sakamoto},
  {Sch{\"o}nb{\"a}chler}, {Tafla}, {Tang}, {Terada}, {Terada}, {Usui}, {Wada},
  {Wadhwa}, {Walker}, {Yamashita}, {Yin}, {Yokoyama}, {Yoneda}, {Young}, {Yui},
  {Zhang}, {Nakamura}, {Naraoka}, {Noguchi}, {Okazaki}, {Sakamoto}, {Yabuta},
  {Abe}, {Miyazaki}, {Nakato}, {Nishimura}, {Okada}, {Yada}, {Yogata},
  {Nakazawa}, {Saiki}, {Tanaka}, {Terui}, {Tsuda}, {Watanabe}, {Yoshikawa},
  {Tachibana} and {Yurimoto}}]{2022SciA....8D8141H}
\bibinfo{author}{{Hopp}, T.}, \bibinfo{author}{{Dauphas}, N.},
  \bibinfo{author}{{Abe}, Y.}, \bibinfo{author}{{Al{\'e}on}, J.},
  \bibinfo{author}{{O'D. Alexander}, C.M.}, \bibinfo{author}{{Amari}, S.},
  \bibinfo{author}{{Amelin}, Y.}, \bibinfo{author}{{Bajo}, K.i.},
  \bibinfo{author}{{Bizzarro}, M.}, \bibinfo{author}{{Bouvier}, A.},
  \bibinfo{author}{{Carlson}, R.W.}, \bibinfo{author}{{Chaussidon}, M.},
  \bibinfo{author}{{Choi}, B.G.}, \bibinfo{author}{{Davis}, A.M.},
  \bibinfo{author}{{Di Rocco}, T.}, \bibinfo{author}{{Fujiya}, W.},
  \bibinfo{author}{{Fukai}, R.}, \bibinfo{author}{{Gautam}, I.},
  \bibinfo{author}{{Haba}, M.K.}, \bibinfo{author}{{Hibiya}, Y.},
  \bibinfo{author}{{Hidaka}, H.}, \bibinfo{author}{{Homma}, H.},
  \bibinfo{author}{{Hoppe}, P.}, \bibinfo{author}{{Huss}, G.R.},
  \bibinfo{author}{{Ichida}, K.}, \bibinfo{author}{{Iizuka}, T.},
  \bibinfo{author}{{Ireland}, T.R.}, \bibinfo{author}{{Ishikawa}, A.},
  \bibinfo{author}{{Ito}, M.}, \bibinfo{author}{{Itoh}, S.},
  \bibinfo{author}{{Kawasaki}, N.}, \bibinfo{author}{{Kita}, N.T.},
  \bibinfo{author}{{Kitajima}, K.}, \bibinfo{author}{{Kleine}, T.},
  \bibinfo{author}{{Komatani}, S.}, \bibinfo{author}{{Krot}, A.N.},
  \bibinfo{author}{{Liu}, M.C.}, \bibinfo{author}{{Masuda}, Y.},
  \bibinfo{author}{{McKeegan}, K.D.}, \bibinfo{author}{{Morita}, M.},
  \bibinfo{author}{{Motomura}, K.}, \bibinfo{author}{{Moynier}, F.},
  \bibinfo{author}{{Nakai}, I.}, \bibinfo{author}{{Nagashima}, K.},
  \bibinfo{author}{{Nesvorn{\'y}}, D.}, \bibinfo{author}{{Nguyen}, A.},
  \bibinfo{author}{{Nittler}, L.}, \bibinfo{author}{{Onose}, M.},
  \bibinfo{author}{{Pack}, A.}, \bibinfo{author}{{Park}, C.},
  \bibinfo{author}{{Piani}, L.}, \bibinfo{author}{{Qin}, L.},
  \bibinfo{author}{{Russell}, S.S.}, \bibinfo{author}{{Sakamoto}, N.},
  \bibinfo{author}{{Sch{\"o}nb{\"a}chler}, M.}, \bibinfo{author}{{Tafla}, L.},
  \bibinfo{author}{{Tang}, H.}, \bibinfo{author}{{Terada}, K.},
  \bibinfo{author}{{Terada}, Y.}, \bibinfo{author}{{Usui}, T.},
  \bibinfo{author}{{Wada}, S.}, \bibinfo{author}{{Wadhwa}, M.},
  \bibinfo{author}{{Walker}, R.J.}, \bibinfo{author}{{Yamashita}, K.},
  \bibinfo{author}{{Yin}, Q.Z.}, \bibinfo{author}{{Yokoyama}, T.},
  \bibinfo{author}{{Yoneda}, S.}, \bibinfo{author}{{Young}, E.D.},
  \bibinfo{author}{{Yui}, H.}, \bibinfo{author}{{Zhang}, A.C.},
  \bibinfo{author}{{Nakamura}, T.}, \bibinfo{author}{{Naraoka}, H.},
  \bibinfo{author}{{Noguchi}, T.}, \bibinfo{author}{{Okazaki}, R.},
  \bibinfo{author}{{Sakamoto}, K.}, \bibinfo{author}{{Yabuta}, H.},
  \bibinfo{author}{{Abe}, M.}, \bibinfo{author}{{Miyazaki}, A.},
  \bibinfo{author}{{Nakato}, A.}, \bibinfo{author}{{Nishimura}, M.},
  \bibinfo{author}{{Okada}, T.}, \bibinfo{author}{{Yada}, T.},
  \bibinfo{author}{{Yogata}, K.}, \bibinfo{author}{{Nakazawa}, S.},
  \bibinfo{author}{{Saiki}, T.}, \bibinfo{author}{{Tanaka}, S.},
  \bibinfo{author}{{Terui}, F.}, \bibinfo{author}{{Tsuda}, Y.},
  \bibinfo{author}{{Watanabe}, S.i.}, \bibinfo{author}{{Yoshikawa}, M.},
  \bibinfo{author}{{Tachibana}, S.}, \bibinfo{author}{{Yurimoto}, H.},
  \bibinfo{year}{2022}.
\newblock \bibinfo{title}{{Ryugu's nucleosynthetic heritage from the outskirts
  of the Solar System}}.
\newblock \bibinfo{journal}{Science Advances} \bibinfo{volume}{8},
  \bibinfo{pages}{eadd8141}.
\newblock \DOIprefix\doi{10.1126/sciadv.add8141}.
\bibitem[{{Ito} et~al.(2022){Ito}, {Tomioka}, {Uesugi}, {Yamaguchi}, {Shirai},
  {Ohigashi}, {Liu}, {Greenwood}, {Kimura}, {Imae}, {Uesugi}, {Nakato},
  {Yogata}, {Yuzawa}, {Kodama}, {Tsuchiyama}, {Yasutake}, {Findlay}, {Franchi},
  {Malley}, {McCain}, {Matsuda}, {McKeegan}, {Hirahara}, {Takeuchi},
  {Sekimoto}, {Sakurai}, {Okada}, {Karouji}, {Arakawa}, {Fujii}, {Fujimoto},
  {Hayakawa}, {Hirata}, {Hirata}, {Honda}, {Honda}, {Hosoda}, {Iijima},
  {Ikeda}, {Ishiguro}, {Ishihara}, {Iwata}, {Kawahara}, {Kikuchi}, {Kitazato},
  {Matsumoto}, {Matsuoka}, {Michikami}, {Mimasu}, {Miura}, {Mori}, {Morota},
  {Nakazawa}, {Namiki}, {Noda}, {Noguchi}, {Ogawa}, {Ogawa}, {Okada},
  {Okamoto}, {Ono}, {Ozaki}, {Saiki}, {Sakatani}, {Sawada}, {Senshu},
  {Shimaki}, {Shirai}, {Sugita}, {Takei}, {Takeuchi}, {Tanaka}, {Tatsumi},
  {Terui}, {Tsukizaki}, {Wada}, {Yamada}, {Yamada}, {Yamamoto}, {Yano},
  {Yokota}, {Yoshihara}, {Yoshikawa}, {Yoshikawa}, {Fukai}, {Furuya},
  {Hatakeda}, {Hayashi}, {Hitomi}, {Kumagai}, {Miyazaki}, {Nishimura},
  {Soejima}, {Iwamae}, {Yamamoto}, {Yoshitake}, {Yada}, {Abe}, {Usui},
  {Watanabe} and {Tsuda}}]{2022NatAs...6.1163I}
\bibinfo{author}{{Ito}, M.}, \bibinfo{author}{{Tomioka}, N.},
  \bibinfo{author}{{Uesugi}, M.}, \bibinfo{author}{{Yamaguchi}, A.},
  \bibinfo{author}{{Shirai}, N.}, \bibinfo{author}{{Ohigashi}, T.},
  \bibinfo{author}{{Liu}, M.C.}, \bibinfo{author}{{Greenwood}, R.C.},
  \bibinfo{author}{{Kimura}, M.}, \bibinfo{author}{{Imae}, N.},
  \bibinfo{author}{{Uesugi}, K.}, \bibinfo{author}{{Nakato}, A.},
  \bibinfo{author}{{Yogata}, K.}, \bibinfo{author}{{Yuzawa}, H.},
  \bibinfo{author}{{Kodama}, Y.}, \bibinfo{author}{{Tsuchiyama}, A.},
  \bibinfo{author}{{Yasutake}, M.}, \bibinfo{author}{{Findlay}, R.},
  \bibinfo{author}{{Franchi}, I.A.}, \bibinfo{author}{{Malley}, J.A.},
  \bibinfo{author}{{McCain}, K.A.}, \bibinfo{author}{{Matsuda}, N.},
  \bibinfo{author}{{McKeegan}, K.D.}, \bibinfo{author}{{Hirahara}, K.},
  \bibinfo{author}{{Takeuchi}, A.}, \bibinfo{author}{{Sekimoto}, S.},
  \bibinfo{author}{{Sakurai}, I.}, \bibinfo{author}{{Okada}, I.},
  \bibinfo{author}{{Karouji}, Y.}, \bibinfo{author}{{Arakawa}, M.},
  \bibinfo{author}{{Fujii}, A.}, \bibinfo{author}{{Fujimoto}, M.},
  \bibinfo{author}{{Hayakawa}, M.}, \bibinfo{author}{{Hirata}, N.},
  \bibinfo{author}{{Hirata}, N.}, \bibinfo{author}{{Honda}, R.},
  \bibinfo{author}{{Honda}, C.}, \bibinfo{author}{{Hosoda}, S.},
  \bibinfo{author}{{Iijima}, Y.i.}, \bibinfo{author}{{Ikeda}, H.},
  \bibinfo{author}{{Ishiguro}, M.}, \bibinfo{author}{{Ishihara}, Y.},
  \bibinfo{author}{{Iwata}, T.}, \bibinfo{author}{{Kawahara}, K.},
  \bibinfo{author}{{Kikuchi}, S.}, \bibinfo{author}{{Kitazato}, K.},
  \bibinfo{author}{{Matsumoto}, K.}, \bibinfo{author}{{Matsuoka}, M.},
  \bibinfo{author}{{Michikami}, T.}, \bibinfo{author}{{Mimasu}, Y.},
  \bibinfo{author}{{Miura}, A.}, \bibinfo{author}{{Mori}, O.},
  \bibinfo{author}{{Morota}, T.}, \bibinfo{author}{{Nakazawa}, S.},
  \bibinfo{author}{{Namiki}, N.}, \bibinfo{author}{{Noda}, H.},
  \bibinfo{author}{{Noguchi}, R.}, \bibinfo{author}{{Ogawa}, N.},
  \bibinfo{author}{{Ogawa}, K.}, \bibinfo{author}{{Okada}, T.},
  \bibinfo{author}{{Okamoto}, C.}, \bibinfo{author}{{Ono}, G.},
  \bibinfo{author}{{Ozaki}, M.}, \bibinfo{author}{{Saiki}, T.},
  \bibinfo{author}{{Sakatani}, N.}, \bibinfo{author}{{Sawada}, H.},
  \bibinfo{author}{{Senshu}, H.}, \bibinfo{author}{{Shimaki}, Y.},
  \bibinfo{author}{{Shirai}, K.}, \bibinfo{author}{{Sugita}, S.},
  \bibinfo{author}{{Takei}, Y.}, \bibinfo{author}{{Takeuchi}, H.},
  \bibinfo{author}{{Tanaka}, S.}, \bibinfo{author}{{Tatsumi}, E.},
  \bibinfo{author}{{Terui}, F.}, \bibinfo{author}{{Tsukizaki}, R.},
  \bibinfo{author}{{Wada}, K.}, \bibinfo{author}{{Yamada}, M.},
  \bibinfo{author}{{Yamada}, T.}, \bibinfo{author}{{Yamamoto}, Y.},
  \bibinfo{author}{{Yano}, H.}, \bibinfo{author}{{Yokota}, Y.},
  \bibinfo{author}{{Yoshihara}, K.}, \bibinfo{author}{{Yoshikawa}, M.},
  \bibinfo{author}{{Yoshikawa}, K.}, \bibinfo{author}{{Fukai}, R.},
  \bibinfo{author}{{Furuya}, S.}, \bibinfo{author}{{Hatakeda}, K.},
  \bibinfo{author}{{Hayashi}, T.}, \bibinfo{author}{{Hitomi}, Y.},
  \bibinfo{author}{{Kumagai}, K.}, \bibinfo{author}{{Miyazaki}, A.},
  \bibinfo{author}{{Nishimura}, M.}, \bibinfo{author}{{Soejima}, H.},
  \bibinfo{author}{{Iwamae}, A.}, \bibinfo{author}{{Yamamoto}, D.},
  \bibinfo{author}{{Yoshitake}, M.}, \bibinfo{author}{{Yada}, T.},
  \bibinfo{author}{{Abe}, M.}, \bibinfo{author}{{Usui}, T.},
  \bibinfo{author}{{Watanabe}, S.i.}, \bibinfo{author}{{Tsuda}, Y.},
  \bibinfo{year}{2022}.
\newblock \bibinfo{title}{{A pristine record of outer Solar System materials
  from asteroid Ryugu's returned sample}}.
\newblock \bibinfo{journal}{Nature Astronomy} \bibinfo{volume}{6},
  \bibinfo{pages}{1163--1171}.
\newblock \DOIprefix\doi{10.1038/s41550-022-01745-5}.
\bibitem[{{Johansen} et~al.(2014){Johansen}, {Blum}, {Tanaka}, {Ormel},
  {Bizzarro} and {Rickman}}]{2014prpl.conf..547J}
\bibinfo{author}{{Johansen}, A.}, \bibinfo{author}{{Blum}, J.},
  \bibinfo{author}{{Tanaka}, H.}, \bibinfo{author}{{Ormel}, C.},
  \bibinfo{author}{{Bizzarro}, M.}, \bibinfo{author}{{Rickman}, H.},
  \bibinfo{year}{2014}.
\newblock \bibinfo{title}{{The Multifaceted Planetesimal Formation Process}},
  in: \bibinfo{editor}{{Beuther}, H.}, \bibinfo{editor}{{Klessen}, R.S.},
  \bibinfo{editor}{{Dullemond}, C.P.}, \bibinfo{editor}{{Henning}, T.} (Eds.),
  \bibinfo{booktitle}{Protostars and Planets VI}, pp.
  \bibinfo{pages}{547--570}.
\newblock \DOIprefix\doi{10.2458/azu\_uapress\_9780816531240-ch024}.
\bibitem[{{Kawasaki} et~al.(2025){Kawasaki}, {Arakawa}, {Miyamoto},
  {Sakamoto}, {Yamamoto}, {Russell} and {Yurimoto}}]{kawasaki2025solar}
\bibinfo{author}{{Kawasaki}, N.}, \bibinfo{author}{{Arakawa}, S.},
  \bibinfo{author}{{Miyamoto}, Y.}, \bibinfo{author}{{Sakamoto}, N.},
  \bibinfo{author}{{Yamamoto}, D.}, \bibinfo{author}{{Russell}, S.S.},
  \bibinfo{author}{{Yurimoto}, H.}, \bibinfo{year}{2025}.
\newblock \bibinfo{title}{{Solar System’s earliest solids as tracers of the
  accretion region of Ryugu and Ivuna-type carbonaceous chondrites}}.
\newblock \bibinfo{journal}{Communications Earth \& Environment}
  \bibinfo{volume}{6}, \bibinfo{pages}{537}.
\newblock \DOIprefix\doi{10.1038/s43247-025-02511-x}.
\bibitem[{{Kawasaki} et~al.(2026){Kawasaki}, {Nagashima}, {Sakamoto},
  {Fujiya}, {Arakawa}, {Bajo}, {Kita}, {Kitajima}, {Krot}, {Huss}, {Abe},
  {Al^^c3^^a9on}, {Alexander}, {Amari}, {Amelin}, {Bizzarro}, {Bouvier},
  {Carlson}, {Chaussidon}, {Choi}, {Dauphas}, {Davis}, {Di Rocco}, {Fukai},
  {Gautam}, {Haba}, {Hibiya}, {Hidaka}, {Homma}, {Iizuka}, {Ireland},
  {Ishikawa}, {Itoh}, {Kleine}, {Komatani}, {Liu}, {Masuda}, {Motomura},
  {Moynier}, {Nakai}, {Nguyen}, {Nittler}, Andreas, {Park}, {Piani}, {Qin},
  {Russell}, {Sch^^c3^^b6nb^^c3^^a4chler}, {Terada}, {Terada}, {Usui}, {Wada},
  {Wadhwa}, {Walker}, {Yamashita}, {Yin}, {Yokoyama}, {Yoneda}, {Young}, {Yui},
  {Zhang}, {Nakamura}, {Naraoka}, {Noguchi}, {Okazaki}, {Sakamoto}, {Yabuta},
  {Abe}, {Miyazaki}, {Nakato}, {Nishimura}, {Okada}, {Yada}, {Yogata},
  {Nakazawa}, {Saiki}, {Tanaka}, {Terui}, {Tsuda}, {Watanabe}, {Yoshikawa},
  {Tachibana} and {Yurimoto}}]{kawasaki2025dolomite}
\bibinfo{author}{{Kawasaki}, N.}, \bibinfo{author}{{Nagashima}, K.},
  \bibinfo{author}{{Sakamoto}, N.}, \bibinfo{author}{{Fujiya}, W.},
  \bibinfo{author}{{Arakawa}, S.}, \bibinfo{author}{{Bajo}, K.i.},
  \bibinfo{author}{{Kita}, N.T.}, \bibinfo{author}{{Kitajima}, K.},
  \bibinfo{author}{{Krot}, A.N.}, \bibinfo{author}{{Huss}, G.R.},
  \bibinfo{author}{{Abe}, Y.}, \bibinfo{author}{{Al^^c3^^a9on}, J.},
  \bibinfo{author}{{Alexander}, C.M.O.}, \bibinfo{author}{{Amari}, S.},
  \bibinfo{author}{{Amelin}, Y.}, \bibinfo{author}{{Bizzarro}, M.},
  \bibinfo{author}{{Bouvier}, A.}, \bibinfo{author}{{Carlson}, R.W.},
  \bibinfo{author}{{Chaussidon}, M.}, \bibinfo{author}{{Choi}, B.G.},
  \bibinfo{author}{{Dauphas}, N.}, \bibinfo{author}{{Davis}, A.M.},
  \bibinfo{author}{{Di Rocco}, T.}, \bibinfo{author}{{Fukai}, R.},
  \bibinfo{author}{{Gautam}, I.}, \bibinfo{author}{{Haba}, M.K.},
  \bibinfo{author}{{Hibiya}, Y.}, \bibinfo{author}{{Hidaka}, H.},
  \bibinfo{author}{{Homma}, H.}, \bibinfo{author}{{Iizuka}, T.},
  \bibinfo{author}{{Ireland}, T.R.}, \bibinfo{author}{{Ishikawa}, A.},
  \bibinfo{author}{{Itoh}, S.}, \bibinfo{author}{{Kleine}, T.},
  \bibinfo{author}{{Komatani}, S.}, \bibinfo{author}{{Liu}, M.C.},
  \bibinfo{author}{{Masuda}, Y.}, \bibinfo{author}{{Motomura}, K.},
  \bibinfo{author}{{Moynier}, F.}, \bibinfo{author}{{Nakai}, I.},
  \bibinfo{author}{{Nguyen}, A.}, \bibinfo{author}{{Nittler}, L.},
  \bibinfo{author}{Andreas, P.}, \bibinfo{author}{{Park}, C.},
  \bibinfo{author}{{Piani}, L.}, \bibinfo{author}{{Qin}, L.},
  \bibinfo{author}{{Russell}, S.S.},
  \bibinfo{author}{{Sch^^c3^^b6nb^^c3^^a4chler}, M.},
  \bibinfo{author}{{Terada}, K.}, \bibinfo{author}{{Terada}, Y.},
  \bibinfo{author}{{Usui}, T.}, \bibinfo{author}{{Wada}, S.},
  \bibinfo{author}{{Wadhwa}, M.}, \bibinfo{author}{{Walker}, R.J.},
  \bibinfo{author}{{Yamashita}, K.}, \bibinfo{author}{{Yin}, Q.Z.},
  \bibinfo{author}{{Yokoyama}, T.}, \bibinfo{author}{{Yoneda}, S.},
  \bibinfo{author}{{Young}, E.D.}, \bibinfo{author}{{Yui}, H.},
  \bibinfo{author}{{Zhang}, A.C.}, \bibinfo{author}{{Nakamura}, T.},
  \bibinfo{author}{{Naraoka}, H.}, \bibinfo{author}{{Noguchi}, T.},
  \bibinfo{author}{{Okazaki}, R.}, \bibinfo{author}{{Sakamoto}, K.},
  \bibinfo{author}{{Yabuta}, H.}, \bibinfo{author}{{Abe}, M.},
  \bibinfo{author}{{Miyazaki}, A.}, \bibinfo{author}{{Nakato}, A.},
  \bibinfo{author}{{Nishimura}, M.}, \bibinfo{author}{{Okada}, T.},
  \bibinfo{author}{{Yada}, T.}, \bibinfo{author}{{Yogata}, K.},
  \bibinfo{author}{{Nakazawa}, S.}, \bibinfo{author}{{Saiki}, T.},
  \bibinfo{author}{{Tanaka}, S.}, \bibinfo{author}{{Terui}, F.},
  \bibinfo{author}{{Tsuda}, Y.}, \bibinfo{author}{{Watanabe}, S.i.},
  \bibinfo{author}{{Yoshikawa}, M.}, \bibinfo{author}{{Tachibana}, S.},
  \bibinfo{author}{{Yurimoto}, H.}, \bibinfo{year}{2026}.
\newblock \bibinfo{title}{{The parent bodies of Ryugu and Ivuna formed before those of other carbonaceous chondrites}}.
\newblock \bibinfo{journal}{Science} \bibinfo{volume}{393},
  \bibinfo{pages}{6816}.
\newblock \DOIprefix\doi{10.1126/science.adw4498}.

\bibitem[{{Kawasaki} et~al.(2022){Kawasaki}, {Nagashima}, {Sakamoto},
  {Matsumoto}, {Bajo}, {Wada}, {Igami}, {Miyake}, {Noguchi}, {Yamamoto},
  {Russell}, {Abe}, {Al{\'e}on}, {Alexander}, {Amari}, {Amelin}, {Bizzarro},
  {Bouvier}, {Carlson}, {Chaussidon}, {Choi}, {Dauphas}, {Davis}, {Di Rocco},
  {Fujiya}, {Fukai}, {Gautam}, {Haba}, {Hibiya}, {Hidaka}, {Homma}, {Hoppe},
  {Huss}, {Ichida}, {Iizuka}, {Ireland}, {Ishikawa}, {Ito}, {Itoh}, {Kita},
  {Kitajima}, {Kleine}, {Komatani}, {Krot}, {Liu}, {Masuda}, {McKeegan},
  {Morita}, {Motomura}, {Moynier}, {Nakai}, {Nguyen}, {Nittler}, {Onose},
  {Pack}, {Park}, {Piani}, {Qin}, {Sch{\"o}nb{\"a}chler}, {Tafla}, {Tang},
  {Terada}, {Terada}, {Usui}, {Wadhwa}, {Walker}, {Yamashita}, {Yin},
  {Yokoyama}, {Yoneda}, {Young}, {Yui}, {Zhang}, {Nakamura}, {Naraoka},
  {Okazaki}, {Sakamoto}, {Yabuta}, {Abe}, {Miyazaki}, {Nakato}, {Nishimura},
  {Okada}, {Yada}, {Yogata}, {Nakazawa}, {Saiki}, {Tanaka}, {Terui}, {Tsuda},
  {Watanabe}, {Yoshikawa}, {Tachibana} and {Yurimoto}}]{2022SciA....8E2067K}
\bibinfo{author}{{Kawasaki}, N.}, \bibinfo{author}{{Nagashima}, K.},
  \bibinfo{author}{{Sakamoto}, N.}, \bibinfo{author}{{Matsumoto}, T.},
  \bibinfo{author}{{Bajo}, K.i.}, \bibinfo{author}{{Wada}, S.},
  \bibinfo{author}{{Igami}, Y.}, \bibinfo{author}{{Miyake}, A.},
  \bibinfo{author}{{Noguchi}, T.}, \bibinfo{author}{{Yamamoto}, D.},
  \bibinfo{author}{{Russell}, S.S.}, \bibinfo{author}{{Abe}, Y.},
  \bibinfo{author}{{Al{\'e}on}, J.}, \bibinfo{author}{{Alexander}, C.M.O.D.},
  \bibinfo{author}{{Amari}, S.}, \bibinfo{author}{{Amelin}, Y.},
  \bibinfo{author}{{Bizzarro}, M.}, \bibinfo{author}{{Bouvier}, A.},
  \bibinfo{author}{{Carlson}, R.W.}, \bibinfo{author}{{Chaussidon}, M.},
  \bibinfo{author}{{Choi}, B.G.}, \bibinfo{author}{{Dauphas}, N.},
  \bibinfo{author}{{Davis}, A.M.}, \bibinfo{author}{{Di Rocco}, T.},
  \bibinfo{author}{{Fujiya}, W.}, \bibinfo{author}{{Fukai}, R.},
  \bibinfo{author}{{Gautam}, I.}, \bibinfo{author}{{Haba}, M.K.},
  \bibinfo{author}{{Hibiya}, Y.}, \bibinfo{author}{{Hidaka}, H.},
  \bibinfo{author}{{Homma}, H.}, \bibinfo{author}{{Hoppe}, P.},
  \bibinfo{author}{{Huss}, G.R.}, \bibinfo{author}{{Ichida}, K.},
  \bibinfo{author}{{Iizuka}, T.}, \bibinfo{author}{{Ireland}, T.R.},
  \bibinfo{author}{{Ishikawa}, A.}, \bibinfo{author}{{Ito}, M.},
  \bibinfo{author}{{Itoh}, S.}, \bibinfo{author}{{Kita}, N.T.},
  \bibinfo{author}{{Kitajima}, K.}, \bibinfo{author}{{Kleine}, T.},
  \bibinfo{author}{{Komatani}, S.}, \bibinfo{author}{{Krot}, A.N.},
  \bibinfo{author}{{Liu}, M.C.}, \bibinfo{author}{{Masuda}, Y.},
  \bibinfo{author}{{McKeegan}, K.D.}, \bibinfo{author}{{Morita}, M.},
  \bibinfo{author}{{Motomura}, K.}, \bibinfo{author}{{Moynier}, F.},
  \bibinfo{author}{{Nakai}, I.}, \bibinfo{author}{{Nguyen}, A.},
  \bibinfo{author}{{Nittler}, L.}, \bibinfo{author}{{Onose}, M.},
  \bibinfo{author}{{Pack}, A.}, \bibinfo{author}{{Park}, C.},
  \bibinfo{author}{{Piani}, L.}, \bibinfo{author}{{Qin}, L.},
  \bibinfo{author}{{Sch{\"o}nb{\"a}chler}, M.}, \bibinfo{author}{{Tafla}, L.},
  \bibinfo{author}{{Tang}, H.}, \bibinfo{author}{{Terada}, K.},
  \bibinfo{author}{{Terada}, Y.}, \bibinfo{author}{{Usui}, T.},
  \bibinfo{author}{{Wadhwa}, M.}, \bibinfo{author}{{Walker}, R.J.},
  \bibinfo{author}{{Yamashita}, K.}, \bibinfo{author}{{Yin}, Q.Z.},
  \bibinfo{author}{{Yokoyama}, T.}, \bibinfo{author}{{Yoneda}, S.},
  \bibinfo{author}{{Young}, E.D.}, \bibinfo{author}{{Yui}, H.},
  \bibinfo{author}{{Zhang}, A.C.}, \bibinfo{author}{{Nakamura}, T.},
  \bibinfo{author}{{Naraoka}, H.}, \bibinfo{author}{{Okazaki}, R.},
  \bibinfo{author}{{Sakamoto}, K.}, \bibinfo{author}{{Yabuta}, H.},
  \bibinfo{author}{{Abe}, M.}, \bibinfo{author}{{Miyazaki}, A.},
  \bibinfo{author}{{Nakato}, A.}, \bibinfo{author}{{Nishimura}, M.},
  \bibinfo{author}{{Okada}, T.}, \bibinfo{author}{{Yada}, T.},
  \bibinfo{author}{{Yogata}, K.}, \bibinfo{author}{{Nakazawa}, S.},
  \bibinfo{author}{{Saiki}, T.}, \bibinfo{author}{{Tanaka}, S.},
  \bibinfo{author}{{Terui}, F.}, \bibinfo{author}{{Tsuda}, Y.},
  \bibinfo{author}{{Watanabe}, S.i.}, \bibinfo{author}{{Yoshikawa}, M.},
  \bibinfo{author}{{Tachibana}, S.}, \bibinfo{author}{{Yurimoto}, H.},
  \bibinfo{year}{2022}.
\newblock \bibinfo{title}{{Oxygen isotopes of anhydrous primary minerals show
  kinship between asteroid Ryugu and comet 81P/Wild2}}.
\newblock \bibinfo{journal}{Science Advances} \bibinfo{volume}{8},
  \bibinfo{pages}{eade2067}.
\newblock \DOIprefix\doi{10.1126/sciadv.ade2067}.
\bibitem[{{Kimura} et~al.(2021){Kimura}, {Yamamoto} and
  {Wakita}}]{2021ApJ...917L...5K}
\bibinfo{author}{{Kimura}, Y.}, \bibinfo{author}{{Yamamoto}, K.},
  \bibinfo{author}{{Wakita}, S.}, \bibinfo{year}{2021}.
\newblock \bibinfo{title}{{Electron Holography Details the Tagish Lake Parent
  Body and Implies Early Planetary Dynamics of the Solar System}}.
\newblock \bibinfo{journal}{The Astrophysical Journal Letters}
  \bibinfo{volume}{917}, \bibinfo{pages}{L5}.
\newblock \DOIprefix\doi{10.3847/2041-8213/ac13a8}.
\bibitem[{{Kita} et~al.(2024){Kita}, {Kitajima}, {Nagashima}, {Kawasaki},
  {Sakamoto}, {Fujiya}, {Abe}, {Al{\'e}on}, {Alexander}, {Amari}, {Amelin},
  {Bajo}, {Bizzarro}, {Bouvier}, {Carlson}, {Chaussidon}, {Choi}, {Dauphas},
  {Davis}, {Di Rocco}, {Fukai}, {Gautam}, {Haba}, {Hibiya}, {Hidaka}, {Homma},
  {Hoppe}, {Huss}, {Ichida}, {Iizuka}, {Ireland}, {Ishikawa}, {Itoh}, {Kleine},
  {Komatani}, {Krot}, {Liu}, {Masuda}, {McKeegan}, {Morita}, {Motomura},
  {Moynier}, {Nakai}, {Nguyen}, {Nittler}, {Onose}, {Pack}, {Park}, {Piani},
  {Qin}, {Russell}, {Sch{\"o}nb{\"a}chler}, {Tafla}, {Tang}, {Terada},
  {Terada}, {Usui}, {Wada}, {Wadhwa}, {Walker}, {Yamashita}, {Yin}, {Yokoyama},
  {Yoneda}, {Young}, {Yui}, {Zhang}, {Nakamura}, {Naraoka}, {Noguchi},
  {Okazaki}, {Sakamoto}, {Yabuta}, {Abe}, {Miyazaki}, {Nakato}, {Nishimura},
  {Okada}, {Yada}, {Yogata}, {Nakazawa}, {Saiki}, {Tanaka}, {Terui}, {Tsuda},
  {Watanabe}, {Yoshikawa}, {Tachibana} and {Yurimoto}}]{2024M&PS...59.2097K}
\bibinfo{author}{{Kita}, N.T.}, \bibinfo{author}{{Kitajima}, K.},
  \bibinfo{author}{{Nagashima}, K.}, \bibinfo{author}{{Kawasaki}, N.},
  \bibinfo{author}{{Sakamoto}, N.}, \bibinfo{author}{{Fujiya}, W.},
  \bibinfo{author}{{Abe}, Y.}, \bibinfo{author}{{Al{\'e}on}, J.},
  \bibinfo{author}{{Alexander}, C.M.O.}, \bibinfo{author}{{Amari}, S.},
  \bibinfo{author}{{Amelin}, Y.}, \bibinfo{author}{{Bajo}, K.i.},
  \bibinfo{author}{{Bizzarro}, M.}, \bibinfo{author}{{Bouvier}, A.},
  \bibinfo{author}{{Carlson}, R.W.}, \bibinfo{author}{{Chaussidon}, M.},
  \bibinfo{author}{{Choi}, B.G.}, \bibinfo{author}{{Dauphas}, N.},
  \bibinfo{author}{{Davis}, A.M.}, \bibinfo{author}{{Di Rocco}, T.},
  \bibinfo{author}{{Fukai}, R.}, \bibinfo{author}{{Gautam}, I.},
  \bibinfo{author}{{Haba}, M.K.}, \bibinfo{author}{{Hibiya}, Y.},
  \bibinfo{author}{{Hidaka}, H.}, \bibinfo{author}{{Homma}, H.},
  \bibinfo{author}{{Hoppe}, P.}, \bibinfo{author}{{Huss}, G.R.},
  \bibinfo{author}{{Ichida}, K.}, \bibinfo{author}{{Iizuka}, T.},
  \bibinfo{author}{{Ireland}, T.R.}, \bibinfo{author}{{Ishikawa}, A.},
  \bibinfo{author}{{Itoh}, S.}, \bibinfo{author}{{Kleine}, T.},
  \bibinfo{author}{{Komatani}, S.}, \bibinfo{author}{{Krot}, A.N.},
  \bibinfo{author}{{Liu}, M.C.}, \bibinfo{author}{{Masuda}, Y.},
  \bibinfo{author}{{McKeegan}, K.D.}, \bibinfo{author}{{Morita}, M.},
  \bibinfo{author}{{Motomura}, K.}, \bibinfo{author}{{Moynier}, F.},
  \bibinfo{author}{{Nakai}, I.}, \bibinfo{author}{{Nguyen}, A.},
  \bibinfo{author}{{Nittler}, L.}, \bibinfo{author}{{Onose}, M.},
  \bibinfo{author}{{Pack}, A.}, \bibinfo{author}{{Park}, C.},
  \bibinfo{author}{{Piani}, L.}, \bibinfo{author}{{Qin}, L.},
  \bibinfo{author}{{Russell}, S.S.}, \bibinfo{author}{{Sch{\"o}nb{\"a}chler},
  M.}, \bibinfo{author}{{Tafla}, L.}, \bibinfo{author}{{Tang}, H.},
  \bibinfo{author}{{Terada}, K.}, \bibinfo{author}{{Terada}, Y.},
  \bibinfo{author}{{Usui}, T.}, \bibinfo{author}{{Wada}, S.},
  \bibinfo{author}{{Wadhwa}, M.}, \bibinfo{author}{{Walker}, R.J.},
  \bibinfo{author}{{Yamashita}, K.}, \bibinfo{author}{{Yin}, Q.Z.},
  \bibinfo{author}{{Yokoyama}, T.}, \bibinfo{author}{{Yoneda}, S.},
  \bibinfo{author}{{Young}, E.D.}, \bibinfo{author}{{Yui}, H.},
  \bibinfo{author}{{Zhang}, A.C.}, \bibinfo{author}{{Nakamura}, T.},
  \bibinfo{author}{{Naraoka}, H.}, \bibinfo{author}{{Noguchi}, T.},
  \bibinfo{author}{{Okazaki}, R.}, \bibinfo{author}{{Sakamoto}, K.},
  \bibinfo{author}{{Yabuta}, H.}, \bibinfo{author}{{Abe}, M.},
  \bibinfo{author}{{Miyazaki}, A.}, \bibinfo{author}{{Nakato}, A.},
  \bibinfo{author}{{Nishimura}, M.}, \bibinfo{author}{{Okada}, T.},
  \bibinfo{author}{{Yada}, T.}, \bibinfo{author}{{Yogata}, K.},
  \bibinfo{author}{{Nakazawa}, S.}, \bibinfo{author}{{Saiki}, T.},
  \bibinfo{author}{{Tanaka}, S.}, \bibinfo{author}{{Terui}, F.},
  \bibinfo{author}{{Tsuda}, Y.}, \bibinfo{author}{{Watanabe}, S.i.},
  \bibinfo{author}{{Yoshikawa}, M.}, \bibinfo{author}{{Tachibana}, S.},
  \bibinfo{author}{{Yurimoto}, H.}, \bibinfo{year}{2024}.
\newblock \bibinfo{title}{{Disequilibrium oxygen isotope distribution among
  aqueously altered minerals in Ryugu asteroid returned samples}}.
\newblock \bibinfo{journal}{Meteoritics \& Planetary Science}
  \bibinfo{volume}{59}, \bibinfo{pages}{2097--2116}.
\newblock \DOIprefix\doi{10.1111/maps.14163}.
\bibitem[{{Krijt} et~al.(2023){Krijt}, {Kama}, {McClure}, {Teske}, {Bergin},
  {Shorttle}, {Walsh} and {Raymond}}]{2023ASPC..534.1031K}
\bibinfo{author}{{Krijt}, S.}, \bibinfo{author}{{Kama}, M.},
  \bibinfo{author}{{McClure}, M.}, \bibinfo{author}{{Teske}, J.},
  \bibinfo{author}{{Bergin}, E.A.}, \bibinfo{author}{{Shorttle}, O.},
  \bibinfo{author}{{Walsh}, K.J.}, \bibinfo{author}{{Raymond}, S.N.},
  \bibinfo{year}{2023}.
\newblock \bibinfo{title}{{Chemical Habitability: Supply and Retention of
  Life's Essential Elements During Planet Formation}}, in:
  \bibinfo{editor}{{Inutsuka}, S.}, \bibinfo{editor}{{Aikawa}, Y.},
  \bibinfo{editor}{{Muto}, T.}, \bibinfo{editor}{{Tomida}, K.},
  \bibinfo{editor}{{Tamura}, M.} (Eds.), \bibinfo{booktitle}{Protostars and
  Planets VII}, p. \bibinfo{pages}{1031}.
\newblock \DOIprefix\doi{10.48550/arXiv.2203.10056}.
\bibitem[{{Kurokawa} et~al.(2022){Kurokawa}, {Shibuya}, {Sekine}, {Ehlmann},
  {Usui}, {Kikuchi} and {Yoda}}]{2022AGUA....300568K}
\bibinfo{author}{{Kurokawa}, H.}, \bibinfo{author}{{Shibuya}, T.},
  \bibinfo{author}{{Sekine}, Y.}, \bibinfo{author}{{Ehlmann}, B.L.},
  \bibinfo{author}{{Usui}, F.}, \bibinfo{author}{{Kikuchi}, S.},
  \bibinfo{author}{{Yoda}, M.}, \bibinfo{year}{2022}.
\newblock \bibinfo{title}{{Distant Formation and Differentiation of Outer Main
  Belt Asteroids and Carbonaceous Chondrite Parent Bodies}}.
\newblock \bibinfo{journal}{AGU Advances} \bibinfo{volume}{3},
  \bibinfo{pages}{e2021AV000568}.
\newblock \DOIprefix\doi{10.1029/2021AV000568}.
\bibitem[{{Lauretta} et~al.(2024){Lauretta}, {Connolly}, {Aebersold},
  {Alexander}, {Ballouz}, {Barnes}, {Bates}, {Bennett}, {Blanche},
  {Blumenfeld}, {Clemett}, {Cody}, {DellaGiustina}, {Dworkin}, {Eckley},
  {Foustoukos}, {Franchi}, {Glavin}, {Greenwood}, {Haenecour}, {Hamilton},
  {Hill}, {Hiroi}, {Ishimaru}, {Jourdan}, {Kaplan}, {Keller}, {King},
  {Koefoed}, {Kontogiannis}, {Le}, {Macke}, {McCoy}, {Milliken}, {Najorka},
  {Nguyen}, {Pajola}, {Polit}, {Righter}, {Roper}, {Russell}, {Ryan},
  {Sandford}, {Schofield}, {Schultz}, {Seifert}, {Tachibana}, {Thomas-Keprta},
  {Thompson}, {Tu}, {Tusberti}, {Wang}, {Zega} and
  {Wolner}}]{2024M&PS...59.2453L}
\bibinfo{author}{{Lauretta}, D.S.}, \bibinfo{author}{{Connolly}, H.C.},
  \bibinfo{author}{{Aebersold}, J.E.}, \bibinfo{author}{{Alexander}, C.M.O.},
  \bibinfo{author}{{Ballouz}, R.L.}, \bibinfo{author}{{Barnes}, J.J.},
  \bibinfo{author}{{Bates}, H.C.}, \bibinfo{author}{{Bennett}, C.A.},
  \bibinfo{author}{{Blanche}, L.}, \bibinfo{author}{{Blumenfeld}, E.H.},
  \bibinfo{author}{{Clemett}, S.J.}, \bibinfo{author}{{Cody}, G.D.},
  \bibinfo{author}{{DellaGiustina}, D.N.}, \bibinfo{author}{{Dworkin}, J.P.},
  \bibinfo{author}{{Eckley}, S.A.}, \bibinfo{author}{{Foustoukos}, D.I.},
  \bibinfo{author}{{Franchi}, I.A.}, \bibinfo{author}{{Glavin}, D.P.},
  \bibinfo{author}{{Greenwood}, R.C.}, \bibinfo{author}{{Haenecour}, P.},
  \bibinfo{author}{{Hamilton}, V.E.}, \bibinfo{author}{{Hill}, D.H.},
  \bibinfo{author}{{Hiroi}, T.}, \bibinfo{author}{{Ishimaru}, K.},
  \bibinfo{author}{{Jourdan}, F.}, \bibinfo{author}{{Kaplan}, H.H.},
  \bibinfo{author}{{Keller}, L.P.}, \bibinfo{author}{{King}, A.J.},
  \bibinfo{author}{{Koefoed}, P.}, \bibinfo{author}{{Kontogiannis}, M.K.},
  \bibinfo{author}{{Le}, L.}, \bibinfo{author}{{Macke}, R.J.},
  \bibinfo{author}{{McCoy}, T.J.}, \bibinfo{author}{{Milliken}, R.E.},
  \bibinfo{author}{{Najorka}, J.}, \bibinfo{author}{{Nguyen}, A.N.},
  \bibinfo{author}{{Pajola}, M.}, \bibinfo{author}{{Polit}, A.T.},
  \bibinfo{author}{{Righter}, K.}, \bibinfo{author}{{Roper}, H.L.},
  \bibinfo{author}{{Russell}, S.S.}, \bibinfo{author}{{Ryan}, A.J.},
  \bibinfo{author}{{Sandford}, S.A.}, \bibinfo{author}{{Schofield}, P.F.},
  \bibinfo{author}{{Schultz}, C.D.}, \bibinfo{author}{{Seifert}, L.B.},
  \bibinfo{author}{{Tachibana}, S.}, \bibinfo{author}{{Thomas-Keprta}, K.L.},
  \bibinfo{author}{{Thompson}, M.S.}, \bibinfo{author}{{Tu}, V.},
  \bibinfo{author}{{Tusberti}, F.}, \bibinfo{author}{{Wang}, K.},
  \bibinfo{author}{{Zega}, T.J.}, \bibinfo{author}{{Wolner}, C.W.V.},
  \bibinfo{year}{2024}.
\newblock \bibinfo{title}{{Asteroid (101955) Bennu in the laboratory:
  Properties of the sample collected by OSIRIS-REx}}.
\newblock \bibinfo{journal}{Meteoritics \& Planetary Science}
  \bibinfo{volume}{59}, \bibinfo{pages}{2453--2486}.
\newblock \DOIprefix\doi{10.1111/maps.14227}.
\bibitem[{{Lesur} et~al.(2023){Lesur}, {Flock}, {Ercolano}, {Lin}, {Yang},
  {Barranco}, {Benitez-Llambay}, {Goodman}, {Johansen}, {Klahr}, {Laibe},
  {Lyra}, {Marcus}, {Nelson}, {Squire}, {Simon}, {Turner}, {Umurhan} and
  {Youdin}}]{2023ASPC..534..465L}
\bibinfo{author}{{Lesur}, G.}, \bibinfo{author}{{Flock}, M.},
  \bibinfo{author}{{Ercolano}, B.}, \bibinfo{author}{{Lin}, M.K.},
  \bibinfo{author}{{Yang}, C.}, \bibinfo{author}{{Barranco}, J.A.},
  \bibinfo{author}{{Benitez-Llambay}, P.}, \bibinfo{author}{{Goodman}, J.},
  \bibinfo{author}{{Johansen}, A.}, \bibinfo{author}{{Klahr}, H.},
  \bibinfo{author}{{Laibe}, G.}, \bibinfo{author}{{Lyra}, W.},
  \bibinfo{author}{{Marcus}, P.S.}, \bibinfo{author}{{Nelson}, R.P.},
  \bibinfo{author}{{Squire}, J.}, \bibinfo{author}{{Simon}, J.B.},
  \bibinfo{author}{{Turner}, N.J.}, \bibinfo{author}{{Umurhan}, O.M.},
  \bibinfo{author}{{Youdin}, A.N.}, \bibinfo{year}{2023}.
\newblock \bibinfo{title}{{Hydro-, Magnetohydro-, and Dust-Gas Dynamics of
  Protoplanetary Disks}}, in: \bibinfo{editor}{{Inutsuka}, S.},
  \bibinfo{editor}{{Aikawa}, Y.}, \bibinfo{editor}{{Muto}, T.},
  \bibinfo{editor}{{Tomida}, K.}, \bibinfo{editor}{{Tamura}, M.} (Eds.),
  \bibinfo{booktitle}{Protostars and Planets VII}, p. \bibinfo{pages}{465}.
\newblock \DOIprefix\doi{10.48550/arXiv.2203.09821}.
\bibitem[{{Malamud} et~al.(2024){Malamud}, {Sch{\"a}fer}, {Luciana San
  Sebasti{\'a}n}, {Timpe}, {Alexander Essink}, {Kreuzig}, {Meier}, {Blum},
  {Perets} and {Burger}}]{2024ApJ...974...76M}
\bibinfo{author}{{Malamud}, U.}, \bibinfo{author}{{Sch{\"a}fer}, C.M.},
  \bibinfo{author}{{Luciana San Sebasti{\'a}n}, I.}, \bibinfo{author}{{Timpe},
  M.}, \bibinfo{author}{{Alexander Essink}, K.}, \bibinfo{author}{{Kreuzig},
  C.}, \bibinfo{author}{{Meier}, G.}, \bibinfo{author}{{Blum}, J.},
  \bibinfo{author}{{Perets}, H.B.}, \bibinfo{author}{{Burger}, C.},
  \bibinfo{year}{2024}.
\newblock \bibinfo{title}{{New Versus Past Silica Crush Curve Experiments:
  Application to Dimorphos Benchmarking Impact Simulations}}.
\newblock \bibinfo{journal}{The Astrophysical Journal} \bibinfo{volume}{974},
  \bibinfo{pages}{76}.
\newblock \DOIprefix\doi{10.3847/1538-4357/ad6c4a}.
\bibitem[{{Matsumoto} et~al.(2024){Matsumoto}, {Noguchi}, {Miyake}, {Igami},
  {Matsumoto}, {Yada}, {Uesugi}, {Yasutake}, {Uesugi}, {Takeuchi}, {Yuzawa},
  {Ohigashi} and {Araki}}]{2024NatAs...8.1536M}
\bibinfo{author}{{Matsumoto}, T.}, \bibinfo{author}{{Noguchi}, T.},
  \bibinfo{author}{{Miyake}, A.}, \bibinfo{author}{{Igami}, Y.},
  \bibinfo{author}{{Matsumoto}, M.}, \bibinfo{author}{{Yada}, T.},
  \bibinfo{author}{{Uesugi}, M.}, \bibinfo{author}{{Yasutake}, M.},
  \bibinfo{author}{{Uesugi}, K.}, \bibinfo{author}{{Takeuchi}, A.},
  \bibinfo{author}{{Yuzawa}, H.}, \bibinfo{author}{{Ohigashi}, T.},
  \bibinfo{author}{{Araki}, T.}, \bibinfo{year}{2024}.
\newblock \bibinfo{title}{{Sodium carbonates on Ryugu as evidence of highly
  saline water in the outer Solar System}}.
\newblock \bibinfo{journal}{Nature Astronomy} \bibinfo{volume}{8},
  \bibinfo{pages}{1536--1543}.
\newblock \DOIprefix\doi{10.1038/s41550-024-02418-1}.
\bibitem[{{McCain} et~al.(2023){McCain}, {Matsuda}, {Liu}, {McKeegan},
  {Yamaguchi}, {Kimura}, {Tomioka}, {Ito}, {Imae}, {Uesugi}, {Shirai},
  {Ohigashi}, {Greenwood}, {Uesugi}, {Nakato}, {Yogata}, {Yuzawa}, {Kodama},
  {Hirahara}, {Sakurai}, {Okada}, {Karouji}, {Nakazawa}, {Okada}, {Saiki},
  {Tanaka}, {Terui}, {Yoshikawa}, {Miyazaki}, {Nishimura}, {Yada}, {Abe},
  {Usui}, {Watanabe} and {Tsuda}}]{2023NatAs...7..309M}
\bibinfo{author}{{McCain}, K.A.}, \bibinfo{author}{{Matsuda}, N.},
  \bibinfo{author}{{Liu}, M.C.}, \bibinfo{author}{{McKeegan}, K.D.},
  \bibinfo{author}{{Yamaguchi}, A.}, \bibinfo{author}{{Kimura}, M.},
  \bibinfo{author}{{Tomioka}, N.}, \bibinfo{author}{{Ito}, M.},
  \bibinfo{author}{{Imae}, N.}, \bibinfo{author}{{Uesugi}, M.},
  \bibinfo{author}{{Shirai}, N.}, \bibinfo{author}{{Ohigashi}, T.},
  \bibinfo{author}{{Greenwood}, R.C.}, \bibinfo{author}{{Uesugi}, K.},
  \bibinfo{author}{{Nakato}, A.}, \bibinfo{author}{{Yogata}, K.},
  \bibinfo{author}{{Yuzawa}, H.}, \bibinfo{author}{{Kodama}, Y.},
  \bibinfo{author}{{Hirahara}, K.}, \bibinfo{author}{{Sakurai}, I.},
  \bibinfo{author}{{Okada}, I.}, \bibinfo{author}{{Karouji}, Y.},
  \bibinfo{author}{{Nakazawa}, S.}, \bibinfo{author}{{Okada}, T.},
  \bibinfo{author}{{Saiki}, T.}, \bibinfo{author}{{Tanaka}, S.},
  \bibinfo{author}{{Terui}, F.}, \bibinfo{author}{{Yoshikawa}, M.},
  \bibinfo{author}{{Miyazaki}, A.}, \bibinfo{author}{{Nishimura}, M.},
  \bibinfo{author}{{Yada}, T.}, \bibinfo{author}{{Abe}, M.},
  \bibinfo{author}{{Usui}, T.}, \bibinfo{author}{{Watanabe}, S.i.},
  \bibinfo{author}{{Tsuda}, Y.}, \bibinfo{year}{2023}.
\newblock \bibinfo{title}{{Early fluid activity on Ryugu inferred by isotopic
  analyses of carbonates and magnetite}}.
\newblock \bibinfo{journal}{Nature Astronomy} \bibinfo{volume}{7},
  \bibinfo{pages}{309--317}.
\newblock \DOIprefix\doi{10.1038/s41550-022-01863-0}.
\bibitem[{{Michel} et~al.(2015){Michel}, {Jutzi}, {Richardson}, {Goodrich},
  {Hartmann} and {O`Brien}}]{2015P&SS..107...24M}
\bibinfo{author}{{Michel}, P.}, \bibinfo{author}{{Jutzi}, M.},
  \bibinfo{author}{{Richardson}, D.C.}, \bibinfo{author}{{Goodrich}, C.A.},
  \bibinfo{author}{{Hartmann}, W.K.}, \bibinfo{author}{{O`Brien}, D.P.},
  \bibinfo{year}{2015}.
\newblock \bibinfo{title}{{Selective sampling during catastrophic disruption:
  Mapping the location of reaccumulated fragments in the original parent
  body}}.
\newblock \bibinfo{journal}{Planetary and Space Science} \bibinfo{volume}{107},
  \bibinfo{pages}{24--28}.
\newblock \DOIprefix\doi{10.1016/j.pss.2014.08.005}.
\bibitem[{{Miotello} et~al.(2023){Miotello}, {Kamp}, {Birnstiel}, {Cleeves} and
  {Kataoka}}]{2023ASPC..534..501M}
\bibinfo{author}{{Miotello}, A.}, \bibinfo{author}{{Kamp}, I.},
  \bibinfo{author}{{Birnstiel}, T.}, \bibinfo{author}{{Cleeves}, L.C.},
  \bibinfo{author}{{Kataoka}, A.}, \bibinfo{year}{2023}.
\newblock \bibinfo{title}{{Setting the Stage for Planet Formation: Measurements
  and Implications of the Fundamental Disk Properties}}, in:
  \bibinfo{editor}{{Inutsuka}, S.}, \bibinfo{editor}{{Aikawa}, Y.},
  \bibinfo{editor}{{Muto}, T.}, \bibinfo{editor}{{Tomida}, K.},
  \bibinfo{editor}{{Tamura}, M.} (Eds.), \bibinfo{booktitle}{Protostars and
  Planets VII}, p. \bibinfo{pages}{501}.
\newblock \DOIprefix\doi{10.48550/arXiv.2203.09818}.
\bibitem[{{Morbidelli} et~al.(2024){Morbidelli}, {Marrocchi}, {Ahmad},
  {Bhandare}, {Charnoz}, {Commer{\c{c}}on}, {Dullemond}, {Guillot},
  {Hennebelle}, {Lee}, {Lovascio}, {Marschall}, {Marty}, {Maury} and
  {Tamami}}]{2024A&A...691A.147M}
\bibinfo{author}{{Morbidelli}, A.}, \bibinfo{author}{{Marrocchi}, Y.},
  \bibinfo{author}{{Ahmad}, A.A.}, \bibinfo{author}{{Bhandare}, A.},
  \bibinfo{author}{{Charnoz}, S.}, \bibinfo{author}{{Commer{\c{c}}on}, B.},
  \bibinfo{author}{{Dullemond}, C.P.}, \bibinfo{author}{{Guillot}, T.},
  \bibinfo{author}{{Hennebelle}, P.}, \bibinfo{author}{{Lee}, Y.N.},
  \bibinfo{author}{{Lovascio}, F.}, \bibinfo{author}{{Marschall}, R.},
  \bibinfo{author}{{Marty}, B.}, \bibinfo{author}{{Maury}, A.},
  \bibinfo{author}{{Tamami}, O.}, \bibinfo{year}{2024}.
\newblock \bibinfo{title}{{Formation and evolution of a protoplanetary disk:
  Combining observations, simulations, and cosmochemical constraints}}.
\newblock \bibinfo{journal}{Astronomy \& Astrophysics} \bibinfo{volume}{691},
  \bibinfo{pages}{A147}.
\newblock \DOIprefix\doi{10.1051/0004-6361/202451388}.
\bibitem[{{Nagashima} et~al.(2017){Nagashima}, {Krot} and
  {Komatsu}}]{2017GeCoA.201..303N}
\bibinfo{author}{{Nagashima}, K.}, \bibinfo{author}{{Krot}, A.N.},
  \bibinfo{author}{{Komatsu}, M.}, \bibinfo{year}{2017}.
\newblock \bibinfo{title}{{$^{26}$Al-$^{26}$Mg systematics in chondrules from
  Kaba and Yamato 980145 CV3 carbonaceous chondrites}}.
\newblock \bibinfo{journal}{Geochimica et Cosmochimica Acta}
  \bibinfo{volume}{201}, \bibinfo{pages}{303--319}.
\newblock \DOIprefix\doi{10.1016/j.gca.2016.10.030}.
\bibitem[{{Nakamura, E.} et~al.(2022){Nakamura, E.}, {Kobayashi}, {Tanaka},
  {Kunihiro}, {Kitagawa}, {Potiszil}, {Ota}, {Sakaguchi}, {Yamanaka},
  {Ratnayake}, {Tripathi}, {Kumar}, {Avramescu}, {Tsuchida}, {Yachi}, {Miura},
  {Abe}, {Fukai}, {Furuya}, {Hatakeda}, {Hayashi}, {Hitomi}, {Kumagai},
  {Miyazaki}, {Nakato}, {Nishimura}, {Okada}, {Soejima}, {Sugita}, {Suzuki},
  {Usui}, {Yada}, {Yamamoto}, {Yogata}, {Yoshitake}, {Arakawa}, {Fujii},
  {Hayakawa}, {Hirata}, {Hirata}, {Honda}, {Honda}, {Hosoda}, {Iijima},
  {Ikeda}, {Ishiguro}, {Ishihara}, {Iwata}, {Kawahara}, {Kikuchi}, {Kitazato},
  {Matsumoto}, {Matsuoka}, {Michikami}, {Mimasu}, {Miura}, {Morota},
  {Nakazawa}, {Namiki}, {Noda}, {Noguchi}, {Ogawa}, {Ogawa}, {Okamoto}, {Ono},
  {Ozaki}, {Saiki}, {Sakatani}, {Sawada}, {Senshu}, {Shimaki}, {Shirai},
  {Takei}, {Takeuchi}, {Tanaka}, {Tatsumi}, {Terui}, {Tsukizaki}, {Wada},
  {Yamada}, {Yamada}, {Yamamoto}, {Yano}, {Yokota}, {Yoshihara}, {Yoshikawa},
  {Yoshikawa}, {Fujimoto}, {Watanabe} and {Tsuda}}]{2022PJAB...98..227N}
\bibinfo{author}{{Nakamura, E.}}, \bibinfo{author}{{Kobayashi}, K.},
  \bibinfo{author}{{Tanaka}, R.}, \bibinfo{author}{{Kunihiro}, T.},
  \bibinfo{author}{{Kitagawa}, .H.}, \bibinfo{author}{{Potiszil}, C.},
  \bibinfo{author}{{Ota}, T.}, \bibinfo{author}{{Sakaguchi}, C.},
  \bibinfo{author}{{Yamanaka}, M.}, \bibinfo{author}{{Ratnayake}, D.M..},
  \bibinfo{author}{{Tripathi}, H.}, \bibinfo{author}{{Kumar}, R.},
  \bibinfo{author}{{Avramescu}, M.L.}, \bibinfo{author}{{Tsuchida}, H.},
  \bibinfo{author}{{Yachi}, Y.}, \bibinfo{author}{{Miura}, H.},
  \bibinfo{author}{{Abe}, M.}, \bibinfo{author}{{Fukai}, R.},
  \bibinfo{author}{{Furuya}, S.}, \bibinfo{author}{{Hatakeda}, K.},
  \bibinfo{author}{{Hayashi}, T.}, \bibinfo{author}{{Hitomi}, Y.},
  \bibinfo{author}{{Kumagai}, K.}, \bibinfo{author}{{Miyazaki}, A.},
  \bibinfo{author}{{Nakato}, A.}, \bibinfo{author}{{Nishimura}, M.},
  \bibinfo{author}{{Okada}, T.}, \bibinfo{author}{{Soejima}, H.},
  \bibinfo{author}{{Sugita}, S.}, \bibinfo{author}{{Suzuki}, A.},
  \bibinfo{author}{{Usui}, T.}, \bibinfo{author}{{Yada}, T.},
  \bibinfo{author}{{Yamamoto}, D.}, \bibinfo{author}{{Yogata}, K.},
  \bibinfo{author}{{Yoshitake}, M.}, \bibinfo{author}{{Arakawa}, M.},
  \bibinfo{author}{{Fujii}, A.}, \bibinfo{author}{{Hayakawa}, M.},
  \bibinfo{author}{{Hirata}, N.}, \bibinfo{author}{{Hirata}, N.},
  \bibinfo{author}{{Honda}, R.}, \bibinfo{author}{{Honda}, C.},
  \bibinfo{author}{{Hosoda}, S.}, \bibinfo{author}{{Iijima}, Y.},
  \bibinfo{author}{{Ikeda}, H.}, \bibinfo{author}{{Ishiguro}, M.},
  \bibinfo{author}{{Ishihara}, Y.}, \bibinfo{author}{{Iwata}, T.},
  \bibinfo{author}{{Kawahara}, K.}, \bibinfo{author}{{Kikuchi}, S.},
  \bibinfo{author}{{Kitazato}, K.}, \bibinfo{author}{{Matsumoto}, K.},
  \bibinfo{author}{{Matsuoka}, M.}, \bibinfo{author}{{Michikami}, T.},
  \bibinfo{author}{{Mimasu}, Y.}, \bibinfo{author}{{Miura}, A.},
  \bibinfo{author}{{Morota}, T.}, \bibinfo{author}{{Nakazawa}, S.},
  \bibinfo{author}{{Namiki}, N.}, \bibinfo{author}{{Noda}, H.},
  \bibinfo{author}{{Noguchi}, R.}, \bibinfo{author}{{Ogawa}, N.},
  \bibinfo{author}{{Ogawa}, K.}, \bibinfo{author}{{Okamoto}, C.},
  \bibinfo{author}{{Ono}, G.}, \bibinfo{author}{{Ozaki}, M.},
  \bibinfo{author}{{Saiki}, T.}, \bibinfo{author}{{Sakatani}, N.},
  \bibinfo{author}{{Sawada}, H.}, \bibinfo{author}{{Senshu}, H.},
  \bibinfo{author}{{Shimaki}, Y.}, \bibinfo{author}{{Shirai}, K.},
  \bibinfo{author}{{Takei}, Y.}, \bibinfo{author}{{Takeuchi}, H.},
  \bibinfo{author}{{Tanaka}, S.}, \bibinfo{author}{{Tatsumi}, E.},
  \bibinfo{author}{{Terui}, F.}, \bibinfo{author}{{Tsukizaki}, R.},
  \bibinfo{author}{{Wada}, K.}, \bibinfo{author}{{Yamada}, M.},
  \bibinfo{author}{{Yamada}, T.}, \bibinfo{author}{{Yamamoto}, Y.},
  \bibinfo{author}{{Yano}, H.}, \bibinfo{author}{{Yokota}, Y.},
  \bibinfo{author}{{Yoshihara}, K.}, \bibinfo{author}{{Yoshikawa}, M.},
  \bibinfo{author}{{Yoshikawa}, K.}, \bibinfo{author}{{Fujimoto}, M.},
  \bibinfo{author}{{Watanabe}, S.}, \bibinfo{author}{{Tsuda}, Y.},
  \bibinfo{year}{2022}.
\newblock \bibinfo{title}{{On the origin and evolution of the asteroid Ryugu: A
  comprehensive geochemical perspective}}.
\newblock \bibinfo{journal}{Proceedings of the Japan Academy, Series B}
  \bibinfo{volume}{98}, \bibinfo{pages}{227--282}.
\newblock \DOIprefix\doi{10.2183/pjab.98.015}.
\bibitem[{{Nakamura, T.} et~al.(2023){Nakamura, T.}, {Matsumoto}, {Amano},
  {Enokido}, {Zolensky}, {Mikouchi}, {Genda}, {Tanaka}, {Zolotov}, {Kurosawa},
  {Wakita}, {Hyodo}, {Nagano}, {Nakashima}, {Takahashi}, {Fujioka}, {Kikuiri},
  {Kagawa}, {Matsuoka}, {Brearley}, {Tsuchiyama}, {Uesugi}, {Matsuno},
  {Kimura}, {Sato}, {Milliken}, {Tatsumi}, {Sugita}, {Hiroi}, {Kitazato},
  {Brownlee}, {Joswiak}, {Takahashi}, {Ninomiya}, {Takahashi}, {Osawa},
  {Terada}, {Brenker}, {Tkalcec}, {Vincze}, {Brunetto}, {Al{\'e}on-Toppani},
  {Chan}, {Roskosz}, {Viennet}, {Beck}, {Alp}, {Michikami}, {Nagaashi},
  {Tsuji}, {Ino}, {Martinez}, {Han}, {Dolocan}, {Bodnar}, {Tanaka}, {Yoshida},
  {Sugiyama}, {King}, {Fukushi}, {Suga}, {Yamashita}, {Kawai}, {Inoue},
  {Nakato}, {Noguchi}, {Vilas}, {Hendrix}, {Jaramillo-Correa}, {Domingue},
  {Dominguez}, {Gainsforth}, {Engrand}, {Duprat}, {Russell}, {Bonato}, {Ma},
  {Kawamoto}, {Wada}, {Watanabe}, {Endo}, {Enju}, {Riu}, {Rubino}, {Tack},
  {Takeshita}, {Takeichi}, {Takeuchi}, {Takigawa}, {Takir}, {Tanigaki},
  {Taniguchi}, {Tsukamoto}, {Yagi}, {Yamada}, {Yamamoto}, {Yamashita},
  {Yasutake}, {Uesugi}, {Umegaki}, {Chiu}, {Ishizaki}, {Okumura}, {Palomba},
  {Pilorget}, {Potin}, {Alasli}, {Anada}, {Araki}, {Sakatani}, {Schultz},
  {Sekizawa}, {Sitzman}, {Sugiura}, {Sun}, {Dartois}, {De Pauw}, {Dionnet},
  {Djouadi}, {Falkenberg}, {Fujita}, {Fukuma}, {Gearba}, {Hagiya}, {Hu},
  {Kato}, {Kawamura}, {Kimura}, {Kubo}, {Langenhorst}, {Lantz}, {Lavina},
  {Lindner}, {Zhao}, {Vekemans}, {Baklouti}, {Bazi}, {Borondics}, {Nagasawa},
  {Nishiyama}, {Nitta}, {Mathurin}, {Matsumoto}, {Mitsukawa}, {Miura},
  {Miyake}, {Miyake}, {Yurimoto}, {Okazaki}, {Yabuta}, {Naraoka}, {Sakamoto},
  {Tachibana}, {Connolly}, {Lauretta}, {Yoshitake}, {Yoshikawa}, {Yoshikawa},
  {Yoshihara}, {Yokota}, {Yogata}, {Yano}, {Yamamoto}, {Yamamoto}, {Yamada},
  {Yamada}, {Yada}, {Wada}, {Usui}, {Tsukizaki}, {Terui}, {Takeuchi}, {Takei},
  {Iwamae}, {Soejima}, {Shirai}, {Shimaki}, {Senshu}, {Sawada}, {Saiki},
  {Ozaki}, {Ono}, {Okada}, {Ogawa}, {Ogawa}, {Noguchi}, {Noda}, {Nishimura},
  {Namiki}, {Nakazawa}, {Morota}, {Miyazaki}, {Miura}, {Mimasu}, {Matsumoto},
  {Kumagai}, {Kouyama}, {Kikuchi}, {Kawahara} and
  {Kameda}}]{2023Sci...379.8671N}
\bibinfo{author}{{Nakamura, T.}}, \bibinfo{author}{{Matsumoto}, M.},
  \bibinfo{author}{{Amano}, K.}, \bibinfo{author}{{Enokido}, Y.},
  \bibinfo{author}{{Zolensky}, M.E.}, \bibinfo{author}{{Mikouchi}, T.},
  \bibinfo{author}{{Genda}, H.}, \bibinfo{author}{{Tanaka}, S.},
  \bibinfo{author}{{Zolotov}, M.Y.}, \bibinfo{author}{{Kurosawa}, K.},
  \bibinfo{author}{{Wakita}, S.}, \bibinfo{author}{{Hyodo}, R.},
  \bibinfo{author}{{Nagano}, H.}, \bibinfo{author}{{Nakashima}, D.},
  \bibinfo{author}{{Takahashi}, Y.}, \bibinfo{author}{{Fujioka}, Y.},
  \bibinfo{author}{{Kikuiri}, M.}, \bibinfo{author}{{Kagawa}, E.},
  \bibinfo{author}{{Matsuoka}, M.}, \bibinfo{author}{{Brearley}, A.J.},
  \bibinfo{author}{{Tsuchiyama}, A.}, \bibinfo{author}{{Uesugi}, M.},
  \bibinfo{author}{{Matsuno}, J.}, \bibinfo{author}{{Kimura}, Y.},
  \bibinfo{author}{{Sato}, M.}, \bibinfo{author}{{Milliken}, R.E.},
  \bibinfo{author}{{Tatsumi}, E.}, \bibinfo{author}{{Sugita}, S.},
  \bibinfo{author}{{Hiroi}, T.}, \bibinfo{author}{{Kitazato}, K.},
  \bibinfo{author}{{Brownlee}, D.}, \bibinfo{author}{{Joswiak}, D.J.},
  \bibinfo{author}{{Takahashi}, M.}, \bibinfo{author}{{Ninomiya}, K.},
  \bibinfo{author}{{Takahashi}, T.}, \bibinfo{author}{{Osawa}, T.},
  \bibinfo{author}{{Terada}, K.}, \bibinfo{author}{{Brenker}, F.E.},
  \bibinfo{author}{{Tkalcec}, B.J.}, \bibinfo{author}{{Vincze}, L.},
  \bibinfo{author}{{Brunetto}, R.}, \bibinfo{author}{{Al{\'e}on-Toppani}, A.},
  \bibinfo{author}{{Chan}, Q.H.S.}, \bibinfo{author}{{Roskosz}, M.},
  \bibinfo{author}{{Viennet}, J.C.}, \bibinfo{author}{{Beck}, P.},
  \bibinfo{author}{{Alp}, E.E.}, \bibinfo{author}{{Michikami}, T.},
  \bibinfo{author}{{Nagaashi}, Y.}, \bibinfo{author}{{Tsuji}, T.},
  \bibinfo{author}{{Ino}, Y.}, \bibinfo{author}{{Martinez}, J.},
  \bibinfo{author}{{Han}, J.}, \bibinfo{author}{{Dolocan}, A.},
  \bibinfo{author}{{Bodnar}, R.J.}, \bibinfo{author}{{Tanaka}, M.},
  \bibinfo{author}{{Yoshida}, H.}, \bibinfo{author}{{Sugiyama}, K.},
  \bibinfo{author}{{King}, A.J.}, \bibinfo{author}{{Fukushi}, K.},
  \bibinfo{author}{{Suga}, H.}, \bibinfo{author}{{Yamashita}, S.},
  \bibinfo{author}{{Kawai}, T.}, \bibinfo{author}{{Inoue}, K.},
  \bibinfo{author}{{Nakato}, A.}, \bibinfo{author}{{Noguchi}, T.},
  \bibinfo{author}{{Vilas}, F.}, \bibinfo{author}{{Hendrix}, A.R.},
  \bibinfo{author}{{Jaramillo-Correa}, C.}, \bibinfo{author}{{Domingue}, D.L.},
  \bibinfo{author}{{Dominguez}, G.}, \bibinfo{author}{{Gainsforth}, Z.},
  \bibinfo{author}{{Engrand}, C.}, \bibinfo{author}{{Duprat}, J.},
  \bibinfo{author}{{Russell}, S.S.}, \bibinfo{author}{{Bonato}, E.},
  \bibinfo{author}{{Ma}, C.}, \bibinfo{author}{{Kawamoto}, T.},
  \bibinfo{author}{{Wada}, T.}, \bibinfo{author}{{Watanabe}, S.},
  \bibinfo{author}{{Endo}, R.}, \bibinfo{author}{{Enju}, S.},
  \bibinfo{author}{{Riu}, L.}, \bibinfo{author}{{Rubino}, S.},
  \bibinfo{author}{{Tack}, P.}, \bibinfo{author}{{Takeshita}, S.},
  \bibinfo{author}{{Takeichi}, Y.}, \bibinfo{author}{{Takeuchi}, A.},
  \bibinfo{author}{{Takigawa}, A.}, \bibinfo{author}{{Takir}, D.},
  \bibinfo{author}{{Tanigaki}, T.}, \bibinfo{author}{{Taniguchi}, A.},
  \bibinfo{author}{{Tsukamoto}, K.}, \bibinfo{author}{{Yagi}, T.},
  \bibinfo{author}{{Yamada}, S.}, \bibinfo{author}{{Yamamoto}, K.},
  \bibinfo{author}{{Yamashita}, Y.}, \bibinfo{author}{{Yasutake}, M.},
  \bibinfo{author}{{Uesugi}, K.}, \bibinfo{author}{{Umegaki}, I.},
  \bibinfo{author}{{Chiu}, I.}, \bibinfo{author}{{Ishizaki}, T.},
  \bibinfo{author}{{Okumura}, S.}, \bibinfo{author}{{Palomba}, E.},
  \bibinfo{author}{{Pilorget}, C.}, \bibinfo{author}{{Potin}, S.M.},
  \bibinfo{author}{{Alasli}, A.}, \bibinfo{author}{{Anada}, S.},
  \bibinfo{author}{{Araki}, Y.}, \bibinfo{author}{{Sakatani}, N.},
  \bibinfo{author}{{Schultz}, C.}, \bibinfo{author}{{Sekizawa}, O.},
  \bibinfo{author}{{Sitzman}, S.D.}, \bibinfo{author}{{Sugiura}, K.},
  \bibinfo{author}{{Sun}, M.}, \bibinfo{author}{{Dartois}, E.},
  \bibinfo{author}{{De Pauw}, E.}, \bibinfo{author}{{Dionnet}, Z.},
  \bibinfo{author}{{Djouadi}, Z.}, \bibinfo{author}{{Falkenberg}, G.},
  \bibinfo{author}{{Fujita}, R.}, \bibinfo{author}{{Fukuma}, T.},
  \bibinfo{author}{{Gearba}, I.R.}, \bibinfo{author}{{Hagiya}, K.},
  \bibinfo{author}{{Hu}, M.Y.}, \bibinfo{author}{{Kato}, T.},
  \bibinfo{author}{{Kawamura}, T.}, \bibinfo{author}{{Kimura}, M.},
  \bibinfo{author}{{Kubo}, M.K.}, \bibinfo{author}{{Langenhorst}, F.},
  \bibinfo{author}{{Lantz}, C.}, \bibinfo{author}{{Lavina}, B.},
  \bibinfo{author}{{Lindner}, M.}, \bibinfo{author}{{Zhao}, J.},
  \bibinfo{author}{{Vekemans}, B.}, \bibinfo{author}{{Baklouti}, D.},
  \bibinfo{author}{{Bazi}, B.}, \bibinfo{author}{{Borondics}, F.},
  \bibinfo{author}{{Nagasawa}, S.}, \bibinfo{author}{{Nishiyama}, G.},
  \bibinfo{author}{{Nitta}, K.}, \bibinfo{author}{{Mathurin}, J.},
  \bibinfo{author}{{Matsumoto}, T.}, \bibinfo{author}{{Mitsukawa}, I.},
  \bibinfo{author}{{Miura}, H.}, \bibinfo{author}{{Miyake}, A.},
  \bibinfo{author}{{Miyake}, Y.}, \bibinfo{author}{{Yurimoto}, H.},
  \bibinfo{author}{{Okazaki}, R.}, \bibinfo{author}{{Yabuta}, H.},
  \bibinfo{author}{{Naraoka}, H.}, \bibinfo{author}{{Sakamoto}, K.},
  \bibinfo{author}{{Tachibana}, S.}, \bibinfo{author}{{Connolly}, H.C.},
  \bibinfo{author}{{Lauretta}, D.S.}, \bibinfo{author}{{Yoshitake}, M.},
  \bibinfo{author}{{Yoshikawa}, M.}, \bibinfo{author}{{Yoshikawa}, K.},
  \bibinfo{author}{{Yoshihara}, K.}, \bibinfo{author}{{Yokota}, Y.},
  \bibinfo{author}{{Yogata}, K.}, \bibinfo{author}{{Yano}, H.},
  \bibinfo{author}{{Yamamoto}, Y.}, \bibinfo{author}{{Yamamoto}, D.},
  \bibinfo{author}{{Yamada}, M.}, \bibinfo{author}{{Yamada}, T.},
  \bibinfo{author}{{Yada}, T.}, \bibinfo{author}{{Wada}, K.},
  \bibinfo{author}{{Usui}, T.}, \bibinfo{author}{{Tsukizaki}, R.},
  \bibinfo{author}{{Terui}, F.}, \bibinfo{author}{{Takeuchi}, H.},
  \bibinfo{author}{{Takei}, Y.}, \bibinfo{author}{{Iwamae}, A.},
  \bibinfo{author}{{Soejima}, H.}, \bibinfo{author}{{Shirai}, K.},
  \bibinfo{author}{{Shimaki}, Y.}, \bibinfo{author}{{Senshu}, H.},
  \bibinfo{author}{{Sawada}, H.}, \bibinfo{author}{{Saiki}, T.},
  \bibinfo{author}{{Ozaki}, M.}, \bibinfo{author}{{Ono}, G.},
  \bibinfo{author}{{Okada}, T.}, \bibinfo{author}{{Ogawa}, N.},
  \bibinfo{author}{{Ogawa}, K.}, \bibinfo{author}{{Noguchi}, R.},
  \bibinfo{author}{{Noda}, H.}, \bibinfo{author}{{Nishimura}, M.},
  \bibinfo{author}{{Namiki}, N.}, \bibinfo{author}{{Nakazawa}, S.},
  \bibinfo{author}{{Morota}, T.}, \bibinfo{author}{{Miyazaki}, A.},
  \bibinfo{author}{{Miura}, A.}, \bibinfo{author}{{Mimasu}, Y.},
  \bibinfo{author}{{Matsumoto}, K.}, \bibinfo{author}{{Kumagai}, K.},
  \bibinfo{author}{{Kouyama}, T.}, \bibinfo{author}{{Kikuchi}, S.},
  \bibinfo{author}{{Kawahara}, K.}, \bibinfo{author}{{Kameda}, S.},
  \bibinfo{year}{2023}.
\newblock \bibinfo{title}{{Formation and evolution of carbonaceous asteroid
  Ryugu: Direct evidence from returned samples}}.
\newblock \bibinfo{journal}{Science} \bibinfo{volume}{379},
  \bibinfo{pages}{abn8671}.
\newblock \DOIprefix\doi{10.1126/science.abn8671}.
\bibitem[{{Nakashima} et~al.(2023){Nakashima}, {Nakamura}, {Zhang}, {Kita},
  {Mikouchi}, {Yoshida}, {Enokido}, {Morita}, {Kikuiri}, {Amano}, {Kagawa},
  {Yada}, {Nishimura}, {Nakato}, {Miyazaki}, {Yogata}, {Abe}, {Okada}, {Usui},
  {Yoshikawa}, {Saiki}, {Tanaka}, {Nakazawa}, {Terui}, {Yurimoto}, {Noguchi},
  {Yabuta}, {Naraoka}, {Okazaki}, {Sakamoto}, {Watanabe}, {Tachibana} and
  {Tsuda}}]{2023NatCo..14..532N}
\bibinfo{author}{{Nakashima}, D.}, \bibinfo{author}{{Nakamura}, T.},
  \bibinfo{author}{{Zhang}, M.}, \bibinfo{author}{{Kita}, N.T.},
  \bibinfo{author}{{Mikouchi}, T.}, \bibinfo{author}{{Yoshida}, H.},
  \bibinfo{author}{{Enokido}, Y.}, \bibinfo{author}{{Morita}, T.},
  \bibinfo{author}{{Kikuiri}, M.}, \bibinfo{author}{{Amano}, K.},
  \bibinfo{author}{{Kagawa}, E.}, \bibinfo{author}{{Yada}, T.},
  \bibinfo{author}{{Nishimura}, M.}, \bibinfo{author}{{Nakato}, A.},
  \bibinfo{author}{{Miyazaki}, A.}, \bibinfo{author}{{Yogata}, K.},
  \bibinfo{author}{{Abe}, M.}, \bibinfo{author}{{Okada}, T.},
  \bibinfo{author}{{Usui}, T.}, \bibinfo{author}{{Yoshikawa}, M.},
  \bibinfo{author}{{Saiki}, T.}, \bibinfo{author}{{Tanaka}, S.},
  \bibinfo{author}{{Nakazawa}, S.}, \bibinfo{author}{{Terui}, F.},
  \bibinfo{author}{{Yurimoto}, H.}, \bibinfo{author}{{Noguchi}, T.},
  \bibinfo{author}{{Yabuta}, H.}, \bibinfo{author}{{Naraoka}, H.},
  \bibinfo{author}{{Okazaki}, R.}, \bibinfo{author}{{Sakamoto}, K.},
  \bibinfo{author}{{Watanabe}, S.i.}, \bibinfo{author}{{Tachibana}, S.},
  \bibinfo{author}{{Tsuda}, Y.}, \bibinfo{year}{2023}.
\newblock \bibinfo{title}{{Chondrule-like objects and Ca-Al-rich inclusions in
  Ryugu may potentially be the oldest Solar System materials}}.
\newblock \bibinfo{journal}{Nature Communications} \bibinfo{volume}{14},
  \bibinfo{pages}{532}.
\newblock \DOIprefix\doi{10.1038/s41467-023-36268-8}.
\bibitem[{{Neumann} et~al.(2021){Neumann}, {Grott}, {Trieloff}, {Jaumann},
  {Biele}, {Hamm} and {K{\"u}hrt}}]{2021Icar..35814166N}
\bibinfo{author}{{Neumann}, W.}, \bibinfo{author}{{Grott}, M.},
  \bibinfo{author}{{Trieloff}, M.}, \bibinfo{author}{{Jaumann}, R.},
  \bibinfo{author}{{Biele}, J.}, \bibinfo{author}{{Hamm}, M.},
  \bibinfo{author}{{K{\"u}hrt}, E.}, \bibinfo{year}{2021}.
\newblock \bibinfo{title}{{Microporosity and parent body of the rubble-pile NEA
  (162173) Ryugu}}.
\newblock \bibinfo{journal}{Icarus} \bibinfo{volume}{358},
  \bibinfo{pages}{114166}.
\newblock \DOIprefix\doi{10.1016/j.icarus.2020.114166}.
\bibitem[{{Neveu} et~al.(2015){Neveu}, {Desch} and
  {Castillo-Rogez}}]{2015JGRE..120..123N}
\bibinfo{author}{{Neveu}, M.}, \bibinfo{author}{{Desch}, S.J.},
  \bibinfo{author}{{Castillo-Rogez}, J.C.}, \bibinfo{year}{2015}.
\newblock \bibinfo{title}{{Core cracking and hydrothermal circulation can
  profoundly affect Ceres' geophysical evolution}}.
\newblock \bibinfo{journal}{Journal of Geophysical Research (Planets)}
  \bibinfo{volume}{120}, \bibinfo{pages}{123--154}.
\newblock \DOIprefix\doi{10.1002/2014JE004714}.
\bibitem[{{Okada} et~al.(2020){Okada}, {Fukuhara}, {Tanaka}, {Taguchi}, {Arai},
  {Senshu}, {Sakatani}, {Shimaki}, {Demura}, {Ogawa}, {Suko}, {Sekiguchi},
  {Kouyama}, {Takita}, {Matsunaga}, {Imamura}, {Wada}, {Hasegawa}, {Helbert},
  {M{\"u}ller}, {Hagermann}, {Biele}, {Grott}, {Hamm}, {Delbo}, {Hirata},
  {Hirata}, {Yamamoto}, {Sugita}, {Namiki}, {Kitazato}, {Arakawa}, {Tachibana},
  {Ikeda}, {Ishiguro}, {Wada}, {Honda}, {Honda}, {Ishihara}, {Matsumoto},
  {Matsuoka}, {Michikami}, {Miura}, {Morota}, {Noda}, {Noguchi}, {Ogawa},
  {Shirai}, {Tatsumi}, {Yabuta}, {Yokota}, {Yamada}, {Abe}, {Hayakawa},
  {Iwata}, {Ozaki}, {Yano}, {Hosoda}, {Mori}, {Sawada}, {Shimada}, {Takeuchi},
  {Tsukizaki}, {Fujii}, {Hirose}, {Kikuchi}, {Mimasu}, {Ogawa}, {Ono},
  {Takahashi}, {Takei}, {Yamaguchi}, {Yoshikawa}, {Terui}, {Saiki}, {Nakazawa},
  {Yoshikawa}, {Watanabe} and {Tsuda}}]{2020Natur.579..518O}
\bibinfo{author}{{Okada}, T.}, \bibinfo{author}{{Fukuhara}, T.},
  \bibinfo{author}{{Tanaka}, S.}, \bibinfo{author}{{Taguchi}, M.},
  \bibinfo{author}{{Arai}, T.}, \bibinfo{author}{{Senshu}, H.},
  \bibinfo{author}{{Sakatani}, N.}, \bibinfo{author}{{Shimaki}, Y.},
  \bibinfo{author}{{Demura}, H.}, \bibinfo{author}{{Ogawa}, Y.},
  \bibinfo{author}{{Suko}, K.}, \bibinfo{author}{{Sekiguchi}, T.},
  \bibinfo{author}{{Kouyama}, T.}, \bibinfo{author}{{Takita}, J.},
  \bibinfo{author}{{Matsunaga}, T.}, \bibinfo{author}{{Imamura}, T.},
  \bibinfo{author}{{Wada}, T.}, \bibinfo{author}{{Hasegawa}, S.},
  \bibinfo{author}{{Helbert}, J.}, \bibinfo{author}{{M{\"u}ller}, T.G.},
  \bibinfo{author}{{Hagermann}, A.}, \bibinfo{author}{{Biele}, J.},
  \bibinfo{author}{{Grott}, M.}, \bibinfo{author}{{Hamm}, M.},
  \bibinfo{author}{{Delbo}, M.}, \bibinfo{author}{{Hirata}, N.},
  \bibinfo{author}{{Hirata}, N.}, \bibinfo{author}{{Yamamoto}, Y.},
  \bibinfo{author}{{Sugita}, S.}, \bibinfo{author}{{Namiki}, N.},
  \bibinfo{author}{{Kitazato}, K.}, \bibinfo{author}{{Arakawa}, M.},
  \bibinfo{author}{{Tachibana}, S.}, \bibinfo{author}{{Ikeda}, H.},
  \bibinfo{author}{{Ishiguro}, M.}, \bibinfo{author}{{Wada}, K.},
  \bibinfo{author}{{Honda}, C.}, \bibinfo{author}{{Honda}, R.},
  \bibinfo{author}{{Ishihara}, Y.}, \bibinfo{author}{{Matsumoto}, K.},
  \bibinfo{author}{{Matsuoka}, M.}, \bibinfo{author}{{Michikami}, T.},
  \bibinfo{author}{{Miura}, A.}, \bibinfo{author}{{Morota}, T.},
  \bibinfo{author}{{Noda}, H.}, \bibinfo{author}{{Noguchi}, R.},
  \bibinfo{author}{{Ogawa}, K.}, \bibinfo{author}{{Shirai}, K.},
  \bibinfo{author}{{Tatsumi}, E.}, \bibinfo{author}{{Yabuta}, H.},
  \bibinfo{author}{{Yokota}, Y.}, \bibinfo{author}{{Yamada}, M.},
  \bibinfo{author}{{Abe}, M.}, \bibinfo{author}{{Hayakawa}, M.},
  \bibinfo{author}{{Iwata}, T.}, \bibinfo{author}{{Ozaki}, M.},
  \bibinfo{author}{{Yano}, H.}, \bibinfo{author}{{Hosoda}, S.},
  \bibinfo{author}{{Mori}, O.}, \bibinfo{author}{{Sawada}, H.},
  \bibinfo{author}{{Shimada}, T.}, \bibinfo{author}{{Takeuchi}, H.},
  \bibinfo{author}{{Tsukizaki}, R.}, \bibinfo{author}{{Fujii}, A.},
  \bibinfo{author}{{Hirose}, C.}, \bibinfo{author}{{Kikuchi}, S.},
  \bibinfo{author}{{Mimasu}, Y.}, \bibinfo{author}{{Ogawa}, N.},
  \bibinfo{author}{{Ono}, G.}, \bibinfo{author}{{Takahashi}, T.},
  \bibinfo{author}{{Takei}, Y.}, \bibinfo{author}{{Yamaguchi}, T.},
  \bibinfo{author}{{Yoshikawa}, K.}, \bibinfo{author}{{Terui}, F.},
  \bibinfo{author}{{Saiki}, T.}, \bibinfo{author}{{Nakazawa}, S.},
  \bibinfo{author}{{Yoshikawa}, M.}, \bibinfo{author}{{Watanabe}, S.},
  \bibinfo{author}{{Tsuda}, Y.}, \bibinfo{year}{2020}.
\newblock \bibinfo{title}{{Highly porous nature of a primitive asteroid
  revealed by thermal imaging}}.
\newblock \bibinfo{journal}{Nature} \bibinfo{volume}{579},
  \bibinfo{pages}{518--522}.
\newblock \DOIprefix\doi{10.1038/s41586-020-2102-6}.
\bibitem[{{Okuzumi} and {Tazaki}(2019)}]{2019ApJ...878..132O}
\bibinfo{author}{{Okuzumi}, S.}, \bibinfo{author}{{Tazaki}, R.},
  \bibinfo{year}{2019}.
\newblock \bibinfo{title}{{Nonsticky Ice at the Origin of the Uniformly
  Polarized Submillimeter Emission from the HL Tau Disk}}.
\newblock \bibinfo{journal}{The Astrophysical Journal} \bibinfo{volume}{878},
  \bibinfo{pages}{132}.
\newblock \DOIprefix\doi{10.3847/1538-4357/ab204d}.
\bibitem[{{Pinilla} et~al.(2017){Pinilla}, {Pohl}, {Stammler} and
  {Birnstiel}}]{2017ApJ...845...68P}
\bibinfo{author}{{Pinilla}, P.}, \bibinfo{author}{{Pohl}, A.},
  \bibinfo{author}{{Stammler}, S.M.}, \bibinfo{author}{{Birnstiel}, T.},
  \bibinfo{year}{2017}.
\newblock \bibinfo{title}{{Dust Density Distribution and Imaging Analysis of
  Different Ice Lines in Protoplanetary Disks}}.
\newblock \bibinfo{journal}{The Astrophysical Journal} \bibinfo{volume}{845},
  \bibinfo{pages}{68}.
\newblock \DOIprefix\doi{10.3847/1538-4357/aa7edb}.
\bibitem[{{Piqueux} and {Christensen}(2009)}]{2009JGRE..114.9006P}
\bibinfo{author}{{Piqueux}, S.}, \bibinfo{author}{{Christensen}, P.R.},
  \bibinfo{year}{2009}.
\newblock \bibinfo{title}{{A model of thermal conductivity for planetary soils:
  2. Theory for cemented soils}}.
\newblock \bibinfo{journal}{Journal of Geophysical Research (Planets)}
  \bibinfo{volume}{114}, \bibinfo{pages}{E09006}.
\newblock \DOIprefix\doi{10.1029/2008JE003309}.
\bibitem[{{Pirozzoli} et~al.(2021){Pirozzoli}, {De Paoli}, {Zonta} and
  {Soldati}}]{2021JFM...911R...4P}
\bibinfo{author}{{Pirozzoli}, S.}, \bibinfo{author}{{De Paoli}, M.},
  \bibinfo{author}{{Zonta}, F.}, \bibinfo{author}{{Soldati}, A.},
  \bibinfo{year}{2021}.
\newblock \bibinfo{title}{{Towards the ultimate regime in Rayleigh-Darcy
  convection}}.
\newblock \bibinfo{journal}{Journal of Fluid Mechanics} \bibinfo{volume}{911},
  \bibinfo{pages}{R4}.
\newblock \DOIprefix\doi{10.1017/jfm.2020.1178}.
\bibitem[{{Sakatani} et~al.(2021){Sakatani}, {Tanaka}, {Okada}, {Fukuhara},
  {Riu}, {Sugita}, {Honda}, {Morota}, {Kameda}, {Yokota}, {Tatsumi}, {Yumoto},
  {Hirata}, {Miura}, {Kouyama}, {Senshu}, {Shimaki}, {Arai}, {Takita},
  {Demura}, {Sekiguchi}, {M{\"u}ller}, {Hagermann}, {Biele}, {Grott}, {Hamm},
  {Delbo}, {Neumann}, {Taguchi}, {Ogawa}, {Matsunaga}, {Wada}, {Hasegawa},
  {Helbert}, {Hirata}, {Noguchi}, {Yamada}, {Suzuki}, {Honda}, {Ogawa},
  {Hayakawa}, {Yoshioka}, {Matsuoka}, {Cho}, {Sawada}, {Kitazato}, {Iwata},
  {Abe}, {Ohtake}, {Matsuura}, {Matsumoto}, {Noda}, {Ishihara}, {Yamamoto},
  {Higuchi}, {Namiki}, {Ono}, {Saiki}, {Imamura}, {Takagi}, {Yano}, {Shirai},
  {Okamoto}, {Nakazawa}, {Iijima}, {Arakawa}, {Wada}, {Kadono}, {Ishibashi},
  {Terui}, {Kikuchi}, {Yamaguchi}, {Ogawa}, {Mimasu}, {Yoshikawa}, {Takahashi},
  {Takei}, {Fujii}, {Takeuchi}, {Yamamoto}, {Hirose}, {Hosoda}, {Mori},
  {Shimada}, {Soldini}, {Tsukizaki}, {Ozaki}, {Tachibana}, {Ikeda}, {Ishiguro},
  {Yabuta}, {Yoshikawa}, {Watanabe} and {Tsuda}}]{2021NatAs...5..766S}
\bibinfo{author}{{Sakatani}, N.}, \bibinfo{author}{{Tanaka}, S.},
  \bibinfo{author}{{Okada}, T.}, \bibinfo{author}{{Fukuhara}, T.},
  \bibinfo{author}{{Riu}, L.}, \bibinfo{author}{{Sugita}, S.},
  \bibinfo{author}{{Honda}, R.}, \bibinfo{author}{{Morota}, T.},
  \bibinfo{author}{{Kameda}, S.}, \bibinfo{author}{{Yokota}, Y.},
  \bibinfo{author}{{Tatsumi}, E.}, \bibinfo{author}{{Yumoto}, K.},
  \bibinfo{author}{{Hirata}, N.}, \bibinfo{author}{{Miura}, A.},
  \bibinfo{author}{{Kouyama}, T.}, \bibinfo{author}{{Senshu}, H.},
  \bibinfo{author}{{Shimaki}, Y.}, \bibinfo{author}{{Arai}, T.},
  \bibinfo{author}{{Takita}, J.}, \bibinfo{author}{{Demura}, H.},
  \bibinfo{author}{{Sekiguchi}, T.}, \bibinfo{author}{{M{\"u}ller}, T.G.},
  \bibinfo{author}{{Hagermann}, A.}, \bibinfo{author}{{Biele}, J.},
  \bibinfo{author}{{Grott}, M.}, \bibinfo{author}{{Hamm}, M.},
  \bibinfo{author}{{Delbo}, M.}, \bibinfo{author}{{Neumann}, W.},
  \bibinfo{author}{{Taguchi}, M.}, \bibinfo{author}{{Ogawa}, Y.},
  \bibinfo{author}{{Matsunaga}, T.}, \bibinfo{author}{{Wada}, T.},
  \bibinfo{author}{{Hasegawa}, S.}, \bibinfo{author}{{Helbert}, J.},
  \bibinfo{author}{{Hirata}, N.}, \bibinfo{author}{{Noguchi}, R.},
  \bibinfo{author}{{Yamada}, M.}, \bibinfo{author}{{Suzuki}, H.},
  \bibinfo{author}{{Honda}, C.}, \bibinfo{author}{{Ogawa}, K.},
  \bibinfo{author}{{Hayakawa}, M.}, \bibinfo{author}{{Yoshioka}, K.},
  \bibinfo{author}{{Matsuoka}, M.}, \bibinfo{author}{{Cho}, Y.},
  \bibinfo{author}{{Sawada}, H.}, \bibinfo{author}{{Kitazato}, K.},
  \bibinfo{author}{{Iwata}, T.}, \bibinfo{author}{{Abe}, M.},
  \bibinfo{author}{{Ohtake}, M.}, \bibinfo{author}{{Matsuura}, S.},
  \bibinfo{author}{{Matsumoto}, K.}, \bibinfo{author}{{Noda}, H.},
  \bibinfo{author}{{Ishihara}, Y.}, \bibinfo{author}{{Yamamoto}, K.},
  \bibinfo{author}{{Higuchi}, A.}, \bibinfo{author}{{Namiki}, N.},
  \bibinfo{author}{{Ono}, G.}, \bibinfo{author}{{Saiki}, T.},
  \bibinfo{author}{{Imamura}, H.}, \bibinfo{author}{{Takagi}, Y.},
  \bibinfo{author}{{Yano}, H.}, \bibinfo{author}{{Shirai}, K.},
  \bibinfo{author}{{Okamoto}, C.}, \bibinfo{author}{{Nakazawa}, S.},
  \bibinfo{author}{{Iijima}, Y.}, \bibinfo{author}{{Arakawa}, M.},
  \bibinfo{author}{{Wada}, K.}, \bibinfo{author}{{Kadono}, T.},
  \bibinfo{author}{{Ishibashi}, K.}, \bibinfo{author}{{Terui}, F.},
  \bibinfo{author}{{Kikuchi}, S.}, \bibinfo{author}{{Yamaguchi}, T.},
  \bibinfo{author}{{Ogawa}, N.}, \bibinfo{author}{{Mimasu}, Y.},
  \bibinfo{author}{{Yoshikawa}, K.}, \bibinfo{author}{{Takahashi}, T.},
  \bibinfo{author}{{Takei}, Y.}, \bibinfo{author}{{Fujii}, A.},
  \bibinfo{author}{{Takeuchi}, H.}, \bibinfo{author}{{Yamamoto}, Y.},
  \bibinfo{author}{{Hirose}, C.}, \bibinfo{author}{{Hosoda}, S.},
  \bibinfo{author}{{Mori}, O.}, \bibinfo{author}{{Shimada}, T.},
  \bibinfo{author}{{Soldini}, S.}, \bibinfo{author}{{Tsukizaki}, R.},
  \bibinfo{author}{{Ozaki}, M.}, \bibinfo{author}{{Tachibana}, S.},
  \bibinfo{author}{{Ikeda}, H.}, \bibinfo{author}{{Ishiguro}, M.},
  \bibinfo{author}{{Yabuta}, H.}, \bibinfo{author}{{Yoshikawa}, M.},
  \bibinfo{author}{{Watanabe}, S.}, \bibinfo{author}{{Tsuda}, Y.},
  \bibinfo{year}{2021}.
\newblock \bibinfo{title}{{Anomalously porous boulders on (162173) Ryugu as
  primordial materials from its parent body}}.
\newblock \bibinfo{journal}{Nature Astronomy} \bibinfo{volume}{5},
  \bibinfo{pages}{766--774}.
\newblock \DOIprefix\doi{10.1038/s41550-021-01371-7}.
\bibitem[{{Schrader} et~al.(2017){Schrader}, {Nagashima}, {Krot}, {Ogliore},
  {Yin}, {Amelin}, {Stirling} and {Kaltenbach}}]{2017GeCoA.201..275S}
\bibinfo{author}{{Schrader}, D.L.}, \bibinfo{author}{{Nagashima}, K.},
  \bibinfo{author}{{Krot}, A.N.}, \bibinfo{author}{{Ogliore}, R.C.},
  \bibinfo{author}{{Yin}, Q.Z.}, \bibinfo{author}{{Amelin}, Y.},
  \bibinfo{author}{{Stirling}, C.H.}, \bibinfo{author}{{Kaltenbach}, A.},
  \bibinfo{year}{2017}.
\newblock \bibinfo{title}{{Distribution of $^{26}$Al in the CR chondrite
  chondrule-forming region of the protoplanetary disk}}.
\newblock \bibinfo{journal}{Geochimica et Cosmochimica Acta}
  \bibinfo{volume}{201}, \bibinfo{pages}{275--302}.
\newblock \DOIprefix\doi{10.1016/j.gca.2016.06.023}.
\bibitem[{{Shibuya} et~al.(2024){Shibuya}, {Sekine}, {Kikuchi}, {Kurokawa},
  {Fukushi}, {Nakamura} and {Watanabe}}]{2024GeCoA.374..264S}
\bibinfo{author}{{Shibuya}, T.}, \bibinfo{author}{{Sekine}, Y.},
  \bibinfo{author}{{Kikuchi}, S.}, \bibinfo{author}{{Kurokawa}, H.},
  \bibinfo{author}{{Fukushi}, K.}, \bibinfo{author}{{Nakamura}, T.},
  \bibinfo{author}{{Watanabe}, S.i.}, \bibinfo{year}{2024}.
\newblock \bibinfo{title}{{Aqueous alteration in icy planetesimals: The effect
  of outward transport of gaseous hydrogen}}.
\newblock \bibinfo{journal}{Geochimica et Cosmochimica Acta}
  \bibinfo{volume}{374}, \bibinfo{pages}{264--283}.
\newblock \DOIprefix\doi{10.1016/j.gca.2024.03.022}.
\bibitem[{{Sugawara} et~al.(2024){Sugawara}, {Fujiya}, {Kawasaki}, {Sakamoto},
  {Yamaguchi} and {Yurimoto}}]{2024GeCoA.382...40S}
\bibinfo{author}{{Sugawara}, S.}, \bibinfo{author}{{Fujiya}, W.},
  \bibinfo{author}{{Kawasaki}, N.}, \bibinfo{author}{{Sakamoto}, N.},
  \bibinfo{author}{{Yamaguchi}, A.}, \bibinfo{author}{{Yurimoto}, H.},
  \bibinfo{year}{2024}.
\newblock \bibinfo{title}{{Update on the $^{53}$Mn-$^{53}$Cr ages of dolomite
  in the Ivuna CI chondrite and asteroid Ryugu sample}}.
\newblock \bibinfo{journal}{Geochimica et Cosmochimica Acta}
  \bibinfo{volume}{382}, \bibinfo{pages}{40--50}.
\newblock \DOIprefix\doi{10.1016/j.gca.2024.08.013}.
\bibitem[{{Sugita} et~al.(2019){Sugita}, {Honda}, {Morota}, {Kameda}, {Sawada},
  {Tatsumi}, {Yamada}, {Honda}, {Yokota}, {Kouyama}, {Sakatani}, {Ogawa},
  {Suzuki}, {Okada}, {Namiki}, {Tanaka}, {Iijima}, {Yoshioka}, {Hayakawa},
  {Cho}, {Matsuoka}, {Hirata}, {Hirata}, {Miyamoto}, {Domingue}, {Hirabayashi},
  {Nakamura}, {Hiroi}, {Michikami}, {Michel}, {Ballouz}, {Barnouin}, {Ernst},
  {Schr{\"o}der}, {Kikuchi}, {Hemmi}, {Komatsu}, {Fukuhara}, {Taguchi}, {Arai},
  {Senshu}, {Demura}, {Ogawa}, {Shimaki}, {Sekiguchi}, {M{\"u}ller},
  {Hagermann}, {Mizuno}, {Noda}, {Matsumoto}, {Yamada}, {Ishihara}, {Ikeda},
  {Araki}, {Yamamoto}, {Abe}, {Yoshida}, {Higuchi}, {Sasaki}, {Oshigami},
  {Tsuruta}, {Asari}, {Tazawa}, {Shizugami}, {Kimura}, {Otsubo}, {Yabuta},
  {Hasegawa}, {Ishiguro}, {Tachibana}, {Palmer}, {Gaskell}, {Le Corre},
  {Jaumann}, {Otto}, {Schmitz}, {Abell}, {Barucci}, {Zolensky}, {Vilas},
  {Thuillet}, {Sugimoto}, {Takaki}, {Suzuki}, {Kamiyoshihara}, {Okada},
  {Nagata}, {Fujimoto}, {Yoshikawa}, {Yamamoto}, {Shirai}, {Noguchi}, {Ogawa},
  {Terui}, {Kikuchi}, {Yamaguchi}, {Oki}, {Takao}, {Takeuchi}, {Ono}, {Mimasu},
  {Yoshikawa}, {Takahashi}, {Takei}, {Fujii}, {Hirose}, {Nakazawa}, {Hosoda},
  {Mori}, {Shimada}, {Soldini}, {Iwata}, {Abe}, {Yano}, {Tsukizaki}, {Ozaki},
  {Nishiyama}, {Saiki}, {Watanabe} and {Tsuda}}]{2019Sci...364..252S}
\bibinfo{author}{{Sugita}, S.}, \bibinfo{author}{{Honda}, R.},
  \bibinfo{author}{{Morota}, T.}, \bibinfo{author}{{Kameda}, S.},
  \bibinfo{author}{{Sawada}, H.}, \bibinfo{author}{{Tatsumi}, E.},
  \bibinfo{author}{{Yamada}, M.}, \bibinfo{author}{{Honda}, C.},
  \bibinfo{author}{{Yokota}, Y.}, \bibinfo{author}{{Kouyama}, T.},
  \bibinfo{author}{{Sakatani}, N.}, \bibinfo{author}{{Ogawa}, K.},
  \bibinfo{author}{{Suzuki}, H.}, \bibinfo{author}{{Okada}, T.},
  \bibinfo{author}{{Namiki}, N.}, \bibinfo{author}{{Tanaka}, S.},
  \bibinfo{author}{{Iijima}, Y.}, \bibinfo{author}{{Yoshioka}, K.},
  \bibinfo{author}{{Hayakawa}, M.}, \bibinfo{author}{{Cho}, Y.},
  \bibinfo{author}{{Matsuoka}, M.}, \bibinfo{author}{{Hirata}, N.},
  \bibinfo{author}{{Hirata}, N.}, \bibinfo{author}{{Miyamoto}, H.},
  \bibinfo{author}{{Domingue}, D.}, \bibinfo{author}{{Hirabayashi}, M.},
  \bibinfo{author}{{Nakamura}, T.}, \bibinfo{author}{{Hiroi}, T.},
  \bibinfo{author}{{Michikami}, T.}, \bibinfo{author}{{Michel}, P.},
  \bibinfo{author}{{Ballouz}, R.L.}, \bibinfo{author}{{Barnouin}, O.S.},
  \bibinfo{author}{{Ernst}, C.M.}, \bibinfo{author}{{Schr{\"o}der}, S.E.},
  \bibinfo{author}{{Kikuchi}, H.}, \bibinfo{author}{{Hemmi}, R.},
  \bibinfo{author}{{Komatsu}, G.}, \bibinfo{author}{{Fukuhara}, T.},
  \bibinfo{author}{{Taguchi}, M.}, \bibinfo{author}{{Arai}, T.},
  \bibinfo{author}{{Senshu}, H.}, \bibinfo{author}{{Demura}, H.},
  \bibinfo{author}{{Ogawa}, Y.}, \bibinfo{author}{{Shimaki}, Y.},
  \bibinfo{author}{{Sekiguchi}, T.}, \bibinfo{author}{{M{\"u}ller}, T.G.},
  \bibinfo{author}{{Hagermann}, A.}, \bibinfo{author}{{Mizuno}, T.},
  \bibinfo{author}{{Noda}, H.}, \bibinfo{author}{{Matsumoto}, K.},
  \bibinfo{author}{{Yamada}, R.}, \bibinfo{author}{{Ishihara}, Y.},
  \bibinfo{author}{{Ikeda}, H.}, \bibinfo{author}{{Araki}, H.},
  \bibinfo{author}{{Yamamoto}, K.}, \bibinfo{author}{{Abe}, S.},
  \bibinfo{author}{{Yoshida}, F.}, \bibinfo{author}{{Higuchi}, A.},
  \bibinfo{author}{{Sasaki}, S.}, \bibinfo{author}{{Oshigami}, S.},
  \bibinfo{author}{{Tsuruta}, S.}, \bibinfo{author}{{Asari}, K.},
  \bibinfo{author}{{Tazawa}, S.}, \bibinfo{author}{{Shizugami}, M.},
  \bibinfo{author}{{Kimura}, J.}, \bibinfo{author}{{Otsubo}, T.},
  \bibinfo{author}{{Yabuta}, H.}, \bibinfo{author}{{Hasegawa}, S.},
  \bibinfo{author}{{Ishiguro}, M.}, \bibinfo{author}{{Tachibana}, S.},
  \bibinfo{author}{{Palmer}, E.}, \bibinfo{author}{{Gaskell}, R.},
  \bibinfo{author}{{Le Corre}, L.}, \bibinfo{author}{{Jaumann}, R.},
  \bibinfo{author}{{Otto}, K.}, \bibinfo{author}{{Schmitz}, N.},
  \bibinfo{author}{{Abell}, P.A.}, \bibinfo{author}{{Barucci}, M.A.},
  \bibinfo{author}{{Zolensky}, M.E.}, \bibinfo{author}{{Vilas}, F.},
  \bibinfo{author}{{Thuillet}, F.}, \bibinfo{author}{{Sugimoto}, C.},
  \bibinfo{author}{{Takaki}, N.}, \bibinfo{author}{{Suzuki}, Y.},
  \bibinfo{author}{{Kamiyoshihara}, H.}, \bibinfo{author}{{Okada}, M.},
  \bibinfo{author}{{Nagata}, K.}, \bibinfo{author}{{Fujimoto}, M.},
  \bibinfo{author}{{Yoshikawa}, M.}, \bibinfo{author}{{Yamamoto}, Y.},
  \bibinfo{author}{{Shirai}, K.}, \bibinfo{author}{{Noguchi}, R.},
  \bibinfo{author}{{Ogawa}, N.}, \bibinfo{author}{{Terui}, F.},
  \bibinfo{author}{{Kikuchi}, S.}, \bibinfo{author}{{Yamaguchi}, T.},
  \bibinfo{author}{{Oki}, Y.}, \bibinfo{author}{{Takao}, Y.},
  \bibinfo{author}{{Takeuchi}, H.}, \bibinfo{author}{{Ono}, G.},
  \bibinfo{author}{{Mimasu}, Y.}, \bibinfo{author}{{Yoshikawa}, K.},
  \bibinfo{author}{{Takahashi}, T.}, \bibinfo{author}{{Takei}, Y.},
  \bibinfo{author}{{Fujii}, A.}, \bibinfo{author}{{Hirose}, C.},
  \bibinfo{author}{{Nakazawa}, S.}, \bibinfo{author}{{Hosoda}, S.},
  \bibinfo{author}{{Mori}, O.}, \bibinfo{author}{{Shimada}, T.},
  \bibinfo{author}{{Soldini}, S.}, \bibinfo{author}{{Iwata}, T.},
  \bibinfo{author}{{Abe}, M.}, \bibinfo{author}{{Yano}, H.},
  \bibinfo{author}{{Tsukizaki}, R.}, \bibinfo{author}{{Ozaki}, M.},
  \bibinfo{author}{{Nishiyama}, K.}, \bibinfo{author}{{Saiki}, T.},
  \bibinfo{author}{{Watanabe}, S.}, \bibinfo{author}{{Tsuda}, Y.},
  \bibinfo{year}{2019}.
\newblock \bibinfo{title}{{The geomorphology, color, and thermal properties of
  Ryugu: Implications for parent-body processes}}.
\newblock \bibinfo{journal}{Science} \bibinfo{volume}{364},
  \bibinfo{pages}{eaaw0422}.
\newblock \DOIprefix\doi{10.1126/science.aaw0422}.
\bibitem[{{Tanaka} et~al.(2024){Tanaka}, {Ratnayake}, {Ota}, {Miklusicak},
  {Kunihiro}, {Potiszil}, {Sakaguchi}, {Kobayashi}, {Kitagawa}, {Yamanaka},
  {Abe}, {Miyazaki}, {Nakato}, {Nakazawa}, {Nishimura}, {Okada}, {Saiki},
  {Tanaka}, {Terui}, {Tsuda}, {Usui}, {Watanabe}, {Yada}, {Yogata}, {Yoshikawa}
  and {Nakamura}}]{2024ApJ...965...52T}
\bibinfo{author}{{Tanaka}, R.}, \bibinfo{author}{{Ratnayake}, D.M.},
  \bibinfo{author}{{Ota}, T.}, \bibinfo{author}{{Miklusicak}, N.},
  \bibinfo{author}{{Kunihiro}, T.}, \bibinfo{author}{{Potiszil}, C.},
  \bibinfo{author}{{Sakaguchi}, C.}, \bibinfo{author}{{Kobayashi}, K.},
  \bibinfo{author}{{Kitagawa}, H.}, \bibinfo{author}{{Yamanaka}, M.},
  \bibinfo{author}{{Abe}, M.}, \bibinfo{author}{{Miyazaki}, A.},
  \bibinfo{author}{{Nakato}, A.}, \bibinfo{author}{{Nakazawa}, S.},
  \bibinfo{author}{{Nishimura}, M.}, \bibinfo{author}{{Okada}, T.},
  \bibinfo{author}{{Saiki}, T.}, \bibinfo{author}{{Tanaka}, S.},
  \bibinfo{author}{{Terui}, F.}, \bibinfo{author}{{Tsuda}, Y.},
  \bibinfo{author}{{Usui}, T.}, \bibinfo{author}{{Watanabe}, S.i.},
  \bibinfo{author}{{Yada}, T.}, \bibinfo{author}{{Yogata}, K.},
  \bibinfo{author}{{Yoshikawa}, M.}, \bibinfo{author}{{Nakamura}, E.},
  \bibinfo{year}{2024}.
\newblock \bibinfo{title}{{Unraveling the Cr Isotopes of Ryugu: An Accurate
  Aqueous Alteration Age and the Least Thermally Processed Solar System
  Material}}.
\newblock \bibinfo{journal}{The Astrophysical Journal} \bibinfo{volume}{965},
  \bibinfo{pages}{52}.
\newblock \DOIprefix\doi{10.3847/1538-4357/ad276a}.
\bibitem[{{Tang} et~al.(2023){Tang}, {Young}, {Tafla}, {Pack}, {Di Rocco},
  {Abe}, {Al{\'e}on}, {O'D. Alexander}, {Amari}, {Amelin}, {Bajo}, {Bizzarro},
  {Bouvier}, {Carlson}, {Chaussidon}, {Choi}, {Dauphas}, {Davis}, {Fujiya},
  {Fukai}, {Gautam}, {Haba}, {Hibiya}, {Hidaka}, {Homma}, {Hoppe}, {Huss},
  {Ichida}, {Iizuka}, {Ireland}, {Ishikawa}, {Ito}, {Itoh}, {Kawasaki}, {Kita},
  {Kitajima}, {Kleine}, {Komatani}, {Krot}, {Liu}, {Masuda}, {McKeegan},
  {Morita}, {Motomura}, {Moynier}, {Nagashima}, {Nakai}, {Nguyen}, {Nittler},
  {Onose}, {Park}, {Piani}, {Qin}, {Russell}, {Sakamoto},
  {Sch{\"o}nb{\"a}chler}, {Terada}, {Terada}, {Usui}, {Wada}, {Wadhwa},
  {Walker}, {Yamashita}, {Yin}, {Yokoyama}, {Yoneda}, {Yui}, {Zhang},
  {Nakamura}, {Naraoka}, {Noguchi}, {Okazaki}, {Sakamoto}, {Yabuta}, {Abe},
  {Miyazaki}, {Nakato}, {Nishimura}, {Okada}, {Yada}, {Yogata}, {Nakazawa},
  {Saiki}, {Tanaka}, {Terui}, {Tsuda}, {Watanabe}, {Yoshikawa}, {Tachibana} and
  {Yurimoto}}]{2023PSJ.....4..144T}
\bibinfo{author}{{Tang}, H.}, \bibinfo{author}{{Young}, E.D.},
  \bibinfo{author}{{Tafla}, L.}, \bibinfo{author}{{Pack}, A.},
  \bibinfo{author}{{Di Rocco}, T.}, \bibinfo{author}{{Abe}, Y.},
  \bibinfo{author}{{Al{\'e}on}, J.}, \bibinfo{author}{{O'D. Alexander}, C.M.},
  \bibinfo{author}{{Amari}, S.}, \bibinfo{author}{{Amelin}, Y.},
  \bibinfo{author}{{Bajo}, K.i.}, \bibinfo{author}{{Bizzarro}, M.},
  \bibinfo{author}{{Bouvier}, A.}, \bibinfo{author}{{Carlson}, R.W.},
  \bibinfo{author}{{Chaussidon}, M.}, \bibinfo{author}{{Choi}, B.G.},
  \bibinfo{author}{{Dauphas}, N.}, \bibinfo{author}{{Davis}, A.M.},
  \bibinfo{author}{{Fujiya}, W.}, \bibinfo{author}{{Fukai}, R.},
  \bibinfo{author}{{Gautam}, I.}, \bibinfo{author}{{Haba}, M.K.},
  \bibinfo{author}{{Hibiya}, Y.}, \bibinfo{author}{{Hidaka}, H.},
  \bibinfo{author}{{Homma}, H.}, \bibinfo{author}{{Hoppe}, P.},
  \bibinfo{author}{{Huss}, G.R.}, \bibinfo{author}{{Ichida}, K.},
  \bibinfo{author}{{Iizuka}, T.}, \bibinfo{author}{{Ireland}, T.R.},
  \bibinfo{author}{{Ishikawa}, A.}, \bibinfo{author}{{Ito}, M.},
  \bibinfo{author}{{Itoh}, S.}, \bibinfo{author}{{Kawasaki}, N.},
  \bibinfo{author}{{Kita}, N.T.}, \bibinfo{author}{{Kitajima}, K.},
  \bibinfo{author}{{Kleine}, T.}, \bibinfo{author}{{Komatani}, S.},
  \bibinfo{author}{{Krot}, A.N.}, \bibinfo{author}{{Liu}, M.C.},
  \bibinfo{author}{{Masuda}, Y.}, \bibinfo{author}{{McKeegan}, K.D.},
  \bibinfo{author}{{Morita}, M.}, \bibinfo{author}{{Motomura}, K.},
  \bibinfo{author}{{Moynier}, F.}, \bibinfo{author}{{Nagashima}, K.},
  \bibinfo{author}{{Nakai}, I.}, \bibinfo{author}{{Nguyen}, A.},
  \bibinfo{author}{{Nittler}, L.}, \bibinfo{author}{{Onose}, M.},
  \bibinfo{author}{{Park}, C.}, \bibinfo{author}{{Piani}, L.},
  \bibinfo{author}{{Qin}, L.}, \bibinfo{author}{{Russell}, S.S.},
  \bibinfo{author}{{Sakamoto}, N.}, \bibinfo{author}{{Sch{\"o}nb{\"a}chler},
  M.}, \bibinfo{author}{{Terada}, K.}, \bibinfo{author}{{Terada}, Y.},
  \bibinfo{author}{{Usui}, T.}, \bibinfo{author}{{Wada}, S.},
  \bibinfo{author}{{Wadhwa}, M.}, \bibinfo{author}{{Walker}, R.J.},
  \bibinfo{author}{{Yamashita}, K.}, \bibinfo{author}{{Yin}, Q.Z.},
  \bibinfo{author}{{Yokoyama}, T.}, \bibinfo{author}{{Yoneda}, S.},
  \bibinfo{author}{{Yui}, H.}, \bibinfo{author}{{Zhang}, A.C.},
  \bibinfo{author}{{Nakamura}, T.}, \bibinfo{author}{{Naraoka}, H.},
  \bibinfo{author}{{Noguchi}, T.}, \bibinfo{author}{{Okazaki}, R.},
  \bibinfo{author}{{Sakamoto}, K.}, \bibinfo{author}{{Yabuta}, H.},
  \bibinfo{author}{{Abe}, M.}, \bibinfo{author}{{Miyazaki}, A.},
  \bibinfo{author}{{Nakato}, A.}, \bibinfo{author}{{Nishimura}, M.},
  \bibinfo{author}{{Okada}, T.}, \bibinfo{author}{{Yada}, T.},
  \bibinfo{author}{{Yogata}, K.}, \bibinfo{author}{{Nakazawa}, S.},
  \bibinfo{author}{{Saiki}, T.}, \bibinfo{author}{{Tanaka}, S.},
  \bibinfo{author}{{Terui}, F.}, \bibinfo{author}{{Tsuda}, Y.},
  \bibinfo{author}{{Watanabe}, S.i.}, \bibinfo{author}{{Yoshikawa}, M.},
  \bibinfo{author}{{Tachibana}, S.}, \bibinfo{author}{{Yurimoto}, H.},
  \bibinfo{year}{2023}.
\newblock \bibinfo{title}{{The Oxygen Isotopic Composition of Samples Returned
  from Asteroid Ryugu with Implications for the Nature of the Parent
  Planetesimal}}.
\newblock \bibinfo{journal}{The Planetary Science Journal} \bibinfo{volume}{4},
  \bibinfo{pages}{144}.
\newblock \DOIprefix\doi{10.3847/PSJ/acea62}.
\bibitem[{{Tatsumi} et~al.(2021){Tatsumi}, {Sugimoto}, {Riu}, {Sugita},
  {Nakamura}, {Hiroi}, {Morota}, {Popescu}, {Michikami}, {Kitazato},
  {Matsuoka}, {Kameda}, {Honda}, {Yamada}, {Sakatani}, {Kouyama}, {Yokota},
  {Honda}, {Suzuki}, {Cho}, {Ogawa}, {Hayakawa}, {Sawada}, {Yoshioka},
  {Pilorget}, {Ishida}, {Domingue}, {Hirata}, {Sasaki}, {de Le{\'o}n},
  {Barucci}, {Michel}, {Suemitsu}, {Saiki}, {Tanaka}, {Terui}, {Nakazawa},
  {Kikuchi}, {Yamaguchi}, {Ogawa}, {Ono}, {Mimasu}, {Yoshikawa}, {Takahashi},
  {Takei}, {Fujii}, {Yamamoto}, {Okada}, {Hirose}, {Hosoda}, {Mori}, {Shimada},
  {Soldini}, {Tsukizaki}, {Mizuno}, {Iwata}, {Yano}, {Ozaki}, {Abe}, {Ohtake},
  {Namiki}, {Tachibana}, {Arakawa}, {Ikeda}, {Ishiguro}, {Wada}, {Yabuta},
  {Takeuchi}, {Shimaki}, {Shirai}, {Hirata}, {Iijima}, {Tsuda}, {Watanabe} and
  {Yoshikawa}}]{2021NatAs...5...39T}
\bibinfo{author}{{Tatsumi}, E.}, \bibinfo{author}{{Sugimoto}, C.},
  \bibinfo{author}{{Riu}, L.}, \bibinfo{author}{{Sugita}, S.},
  \bibinfo{author}{{Nakamura}, T.}, \bibinfo{author}{{Hiroi}, T.},
  \bibinfo{author}{{Morota}, T.}, \bibinfo{author}{{Popescu}, M.},
  \bibinfo{author}{{Michikami}, T.}, \bibinfo{author}{{Kitazato}, K.},
  \bibinfo{author}{{Matsuoka}, M.}, \bibinfo{author}{{Kameda}, S.},
  \bibinfo{author}{{Honda}, R.}, \bibinfo{author}{{Yamada}, M.},
  \bibinfo{author}{{Sakatani}, N.}, \bibinfo{author}{{Kouyama}, T.},
  \bibinfo{author}{{Yokota}, Y.}, \bibinfo{author}{{Honda}, C.},
  \bibinfo{author}{{Suzuki}, H.}, \bibinfo{author}{{Cho}, Y.},
  \bibinfo{author}{{Ogawa}, K.}, \bibinfo{author}{{Hayakawa}, M.},
  \bibinfo{author}{{Sawada}, H.}, \bibinfo{author}{{Yoshioka}, K.},
  \bibinfo{author}{{Pilorget}, C.}, \bibinfo{author}{{Ishida}, M.},
  \bibinfo{author}{{Domingue}, D.}, \bibinfo{author}{{Hirata}, N.},
  \bibinfo{author}{{Sasaki}, S.}, \bibinfo{author}{{de Le{\'o}n}, J.},
  \bibinfo{author}{{Barucci}, M.A.}, \bibinfo{author}{{Michel}, P.},
  \bibinfo{author}{{Suemitsu}, M.}, \bibinfo{author}{{Saiki}, T.},
  \bibinfo{author}{{Tanaka}, S.}, \bibinfo{author}{{Terui}, F.},
  \bibinfo{author}{{Nakazawa}, S.}, \bibinfo{author}{{Kikuchi}, S.},
  \bibinfo{author}{{Yamaguchi}, T.}, \bibinfo{author}{{Ogawa}, N.},
  \bibinfo{author}{{Ono}, G.}, \bibinfo{author}{{Mimasu}, Y.},
  \bibinfo{author}{{Yoshikawa}, K.}, \bibinfo{author}{{Takahashi}, T.},
  \bibinfo{author}{{Takei}, Y.}, \bibinfo{author}{{Fujii}, A.},
  \bibinfo{author}{{Yamamoto}, Y.}, \bibinfo{author}{{Okada}, T.},
  \bibinfo{author}{{Hirose}, C.}, \bibinfo{author}{{Hosoda}, S.},
  \bibinfo{author}{{Mori}, O.}, \bibinfo{author}{{Shimada}, T.},
  \bibinfo{author}{{Soldini}, S.}, \bibinfo{author}{{Tsukizaki}, R.},
  \bibinfo{author}{{Mizuno}, T.}, \bibinfo{author}{{Iwata}, T.},
  \bibinfo{author}{{Yano}, H.}, \bibinfo{author}{{Ozaki}, M.},
  \bibinfo{author}{{Abe}, M.}, \bibinfo{author}{{Ohtake}, M.},
  \bibinfo{author}{{Namiki}, N.}, \bibinfo{author}{{Tachibana}, S.},
  \bibinfo{author}{{Arakawa}, M.}, \bibinfo{author}{{Ikeda}, H.},
  \bibinfo{author}{{Ishiguro}, M.}, \bibinfo{author}{{Wada}, K.},
  \bibinfo{author}{{Yabuta}, H.}, \bibinfo{author}{{Takeuchi}, H.},
  \bibinfo{author}{{Shimaki}, Y.}, \bibinfo{author}{{Shirai}, K.},
  \bibinfo{author}{{Hirata}, N.}, \bibinfo{author}{{Iijima}, Y.},
  \bibinfo{author}{{Tsuda}, Y.}, \bibinfo{author}{{Watanabe}, S.},
  \bibinfo{author}{{Yoshikawa}, M.}, \bibinfo{year}{2021}.
\newblock \bibinfo{title}{{Collisional history of Ryugu's parent body from
  bright surface boulders}}.
\newblock \bibinfo{journal}{Nature Astronomy} \bibinfo{volume}{5},
  \bibinfo{pages}{39--45}.
\newblock \DOIprefix\doi{10.1038/s41550-020-1179-z}.
\bibitem[{{Tenner} et~al.(2019){Tenner}, {Nakashima}, {Ushikubo}, {Tomioka},
  {Kimura}, {Weisberg} and {Kita}}]{2019GeCoA.260..133T}
\bibinfo{author}{{Tenner}, T.J.}, \bibinfo{author}{{Nakashima}, D.},
  \bibinfo{author}{{Ushikubo}, T.}, \bibinfo{author}{{Tomioka}, N.},
  \bibinfo{author}{{Kimura}, M.}, \bibinfo{author}{{Weisberg}, M.K.},
  \bibinfo{author}{{Kita}, N.T.}, \bibinfo{year}{2019}.
\newblock \bibinfo{title}{{Extended chondrule formation intervals in distinct
  physicochemical environments: Evidence from Al-Mg isotope systematics of CR
  chondrite chondrules with unaltered plagioclase}}.
\newblock \bibinfo{journal}{Geochimica et Cosmochimica Acta}
  \bibinfo{volume}{260}, \bibinfo{pages}{133--160}.
\newblock \DOIprefix\doi{10.1016/j.gca.2019.06.023}.
\bibitem[{{Travis} et~al.(2018){Travis}, {Bland}, {Feldman} and
  {Sykes}}]{2018M&PS...53.2008T}
\bibinfo{author}{{Travis}, B.J.}, \bibinfo{author}{{Bland}, P.A.},
  \bibinfo{author}{{Feldman}, W.C.}, \bibinfo{author}{{Sykes}, M.V.},
  \bibinfo{year}{2018}.
\newblock \bibinfo{title}{{Hydrothermal dynamics in a CM-based model of
  Ceres}}.
\newblock \bibinfo{journal}{Meteoritics \& Planetary Science}
  \bibinfo{volume}{53}, \bibinfo{pages}{2008--2032}.
\newblock \DOIprefix\doi{10.1111/maps.13138}.
\bibitem[{{Tsuchiyama} et~al.(2024){Tsuchiyama}, {Matsumoto}, {Matsuno},
  {Yasutake}, {Nakamura}, {Noguchi}, {Miyake}, {Uesugi}, {Takeuchi}, {Okumura},
  {Fujioka}, {Sun}, {Takigawa}, {Matsumoto}, {Enju}, {Mitsukawa}, {Enokido},
  {Kawamoto}, {Mikouchi}, {Michikami}, {Morita}, {Kikuiri}, {Amano}, {Kagawa},
  {Rubino}, {Dionnet}, {Al{\'e}on-Toppani}, {Brunetto}, {Zolensky}, {Nakano},
  {Nakano}, {Yurimoto}, {Okazaki}, {Yabuta}, {Naraoka}, {Sakamoto}, {Yada},
  {Nishimura}, {Nakato}, {Miyazaki}, {Yogata}, {Abe}, {Okada}, {Usui},
  {Yoshikawa}, {Saiki}, {Tanaka}, {Nakazawa}, {Terui}, {Tachibana}, {Watanabe}
  and {Tsuda}}]{2024GeCoA.375..146T}
\bibinfo{author}{{Tsuchiyama}, A.}, \bibinfo{author}{{Matsumoto}, M.},
  \bibinfo{author}{{Matsuno}, J.}, \bibinfo{author}{{Yasutake}, M.},
  \bibinfo{author}{{Nakamura}, T.}, \bibinfo{author}{{Noguchi}, T.},
  \bibinfo{author}{{Miyake}, A.}, \bibinfo{author}{{Uesugi}, K.},
  \bibinfo{author}{{Takeuchi}, A.}, \bibinfo{author}{{Okumura}, S.},
  \bibinfo{author}{{Fujioka}, Y.}, \bibinfo{author}{{Sun}, M.},
  \bibinfo{author}{{Takigawa}, A.}, \bibinfo{author}{{Matsumoto}, T.},
  \bibinfo{author}{{Enju}, S.}, \bibinfo{author}{{Mitsukawa}, I.},
  \bibinfo{author}{{Enokido}, Y.}, \bibinfo{author}{{Kawamoto}, T.},
  \bibinfo{author}{{Mikouchi}, T.}, \bibinfo{author}{{Michikami}, T.},
  \bibinfo{author}{{Morita}, T.}, \bibinfo{author}{{Kikuiri}, M.},
  \bibinfo{author}{{Amano}, K.}, \bibinfo{author}{{Kagawa}, E.},
  \bibinfo{author}{{Rubino}, S.}, \bibinfo{author}{{Dionnet}, Z.},
  \bibinfo{author}{{Al{\'e}on-Toppani}, A.}, \bibinfo{author}{{Brunetto}, R.},
  \bibinfo{author}{{Zolensky}, M.E.}, \bibinfo{author}{{Nakano}, T.},
  \bibinfo{author}{{Nakano}, N.}, \bibinfo{author}{{Yurimoto}, H.},
  \bibinfo{author}{{Okazaki}, R.}, \bibinfo{author}{{Yabuta}, H.},
  \bibinfo{author}{{Naraoka}, H.}, \bibinfo{author}{{Sakamoto}, K.},
  \bibinfo{author}{{Yada}, T.}, \bibinfo{author}{{Nishimura}, M.},
  \bibinfo{author}{{Nakato}, A.}, \bibinfo{author}{{Miyazaki}, A.},
  \bibinfo{author}{{Yogata}, K.}, \bibinfo{author}{{Abe}, M.},
  \bibinfo{author}{{Okada}, T.}, \bibinfo{author}{{Usui}, T.},
  \bibinfo{author}{{Yoshikawa}, M.}, \bibinfo{author}{{Saiki}, T.},
  \bibinfo{author}{{Tanaka}, S.}, \bibinfo{author}{{Nakazawa}, S.},
  \bibinfo{author}{{Terui}, F.}, \bibinfo{author}{{Tachibana}, S.},
  \bibinfo{author}{{Watanabe}, S.i.}, \bibinfo{author}{{Tsuda}, Y.},
  \bibinfo{year}{2024}.
\newblock \bibinfo{title}{{Three-dimensional textures of Ryugu samples and
  their implications for the evolution of aqueous alteration in the Ryugu
  parent body}}.
\newblock \bibinfo{journal}{Geochimica et Cosmochimica Acta}
  \bibinfo{volume}{375}, \bibinfo{pages}{146--172}.
\newblock \DOIprefix\doi{10.1016/j.gca.2024.03.032}.
\bibitem[{{Tsuchiyama} et~al.(2021){Tsuchiyama}, {Miyake}, {Okuzumi},
  {Kitayama}, {Kawano}, {Uesugi}, {Takeuchi}, {Nakano} and
  {Zolensky}}]{2021SciA....7.9707T}
\bibinfo{author}{{Tsuchiyama}, A.}, \bibinfo{author}{{Miyake}, A.},
  \bibinfo{author}{{Okuzumi}, S.}, \bibinfo{author}{{Kitayama}, A.},
  \bibinfo{author}{{Kawano}, J.}, \bibinfo{author}{{Uesugi}, K.},
  \bibinfo{author}{{Takeuchi}, A.}, \bibinfo{author}{{Nakano}, T.},
  \bibinfo{author}{{Zolensky}, M.}, \bibinfo{year}{2021}.
\newblock \bibinfo{title}{{Discovery of primitive CO2-bearing fluid in an
  aqueously altered carbonaceous chondrite}}.
\newblock \bibinfo{journal}{Science Advances} \bibinfo{volume}{7},
  \bibinfo{pages}{eabg9707}.
\newblock \DOIprefix\doi{10.1126/sciadv.abg9707}.
\bibitem[{{Turcotte} and {Schubert}(2002)}]{2002gedy.book.....T}
\bibinfo{author}{{Turcotte}, D.L.}, \bibinfo{author}{{Schubert}, G.},
  \bibinfo{year}{2002}.
\newblock \bibinfo{title}{{Geodynamics}}.
\bibitem[{{Wagner} and {Pru{\ss}}(2002)}]{2002JPCRD..31..387W}
\bibinfo{author}{{Wagner}, W.}, \bibinfo{author}{{Pru{\ss}}, A.},
  \bibinfo{year}{2002}.
\newblock \bibinfo{title}{{The IAPWS Formulation 1995 for the Thermodynamic
  Properties of Ordinary Water Substance for General and Scientific Use}}.
\newblock \bibinfo{journal}{Journal of Physical and Chemical Reference Data}
  \bibinfo{volume}{31}, \bibinfo{pages}{387--535}.
\newblock \DOIprefix\doi{10.1063/1.1461829}.
\bibitem[{{Wakita} and {Sekiya}(2011)}]{2011EPS...63.1193W}
\bibinfo{author}{{Wakita}, S.}, \bibinfo{author}{{Sekiya}, M.},
  \bibinfo{year}{2011}.
\newblock \bibinfo{title}{{Thermal evolution of icy planetesimals in the solar
  nebula}}.
\newblock \bibinfo{journal}{Earth, Planets and Space} \bibinfo{volume}{63},
  \bibinfo{pages}{1193--1206}.
\newblock \DOIprefix\doi{10.5047/eps.2011.08.012}.
\bibitem[{{Walsh} et~al.(2013){Walsh}, {Delb{\'o}}, {Bottke},
  {Vokrouhlick{\'y}} and {Lauretta}}]{2013Icar..225..283W}
\bibinfo{author}{{Walsh}, K.J.}, \bibinfo{author}{{Delb{\'o}}, M.},
  \bibinfo{author}{{Bottke}, W.F.}, \bibinfo{author}{{Vokrouhlick{\'y}}, D.},
  \bibinfo{author}{{Lauretta}, D.S.}, \bibinfo{year}{2013}.
\newblock \bibinfo{title}{{Introducing the Eulalia and new Polana asteroid
  families: Re-assessing primitive asteroid families in the inner Main Belt}}.
\newblock \bibinfo{journal}{Icarus} \bibinfo{volume}{225},
  \bibinfo{pages}{283--297}.
\newblock \DOIprefix\doi{10.1016/j.icarus.2013.03.005}.
\bibitem[{{Watanabe} et~al.(2019){Watanabe}, {Hirabayashi}, {Hirata}, {Hirata},
  {Noguchi}, {Shimaki}, {Ikeda}, {Tatsumi}, {Yoshikawa}, {Kikuchi}, {Yabuta},
  {Nakamura}, {Tachibana}, {Ishihara}, {Morota}, {Kitazato}, {Sakatani},
  {Matsumoto}, {Wada}, {Senshu}, {Honda}, {Michikami}, {Takeuchi}, {Kouyama},
  {Honda}, {Kameda}, {Fuse}, {Miyamoto}, {Komatsu}, {Sugita}, {Okada},
  {Namiki}, {Arakawa}, {Ishiguro}, {Abe}, {Gaskell}, {Palmer}, {Barnouin},
  {Michel}, {French}, {McMahon}, {Scheeres}, {Abell}, {Yamamoto}, {Tanaka},
  {Shirai}, {Matsuoka}, {Yamada}, {Yokota}, {Suzuki}, {Yoshioka}, {Cho},
  {Tanaka}, {Nishikawa}, {Sugiyama}, {Kikuchi}, {Hemmi}, {Yamaguchi}, {Ogawa},
  {Ono}, {Mimasu}, {Yoshikawa}, {Takahashi}, {Takei}, {Fujii}, {Hirose},
  {Iwata}, {Hayakawa}, {Hosoda}, {Mori}, {Sawada}, {Shimada}, {Soldini},
  {Yano}, {Tsukizaki}, {Ozaki}, {Iijima}, {Ogawa}, {Fujimoto}, {Ho}, {Moussi},
  {Jaumann}, {Bibring}, {Krause}, {Terui}, {Saiki}, {Nakazawa} and
  {Tsuda}}]{2019Sci...364..268W}
\bibinfo{author}{{Watanabe}, S.}, \bibinfo{author}{{Hirabayashi}, M.},
  \bibinfo{author}{{Hirata}, N.}, \bibinfo{author}{{Hirata}, N.},
  \bibinfo{author}{{Noguchi}, R.}, \bibinfo{author}{{Shimaki}, Y.},
  \bibinfo{author}{{Ikeda}, H.}, \bibinfo{author}{{Tatsumi}, E.},
  \bibinfo{author}{{Yoshikawa}, M.}, \bibinfo{author}{{Kikuchi}, S.},
  \bibinfo{author}{{Yabuta}, H.}, \bibinfo{author}{{Nakamura}, T.},
  \bibinfo{author}{{Tachibana}, S.}, \bibinfo{author}{{Ishihara}, Y.},
  \bibinfo{author}{{Morota}, T.}, \bibinfo{author}{{Kitazato}, K.},
  \bibinfo{author}{{Sakatani}, N.}, \bibinfo{author}{{Matsumoto}, K.},
  \bibinfo{author}{{Wada}, K.}, \bibinfo{author}{{Senshu}, H.},
  \bibinfo{author}{{Honda}, C.}, \bibinfo{author}{{Michikami}, T.},
  \bibinfo{author}{{Takeuchi}, H.}, \bibinfo{author}{{Kouyama}, T.},
  \bibinfo{author}{{Honda}, R.}, \bibinfo{author}{{Kameda}, S.},
  \bibinfo{author}{{Fuse}, T.}, \bibinfo{author}{{Miyamoto}, H.},
  \bibinfo{author}{{Komatsu}, G.}, \bibinfo{author}{{Sugita}, S.},
  \bibinfo{author}{{Okada}, T.}, \bibinfo{author}{{Namiki}, N.},
  \bibinfo{author}{{Arakawa}, M.}, \bibinfo{author}{{Ishiguro}, M.},
  \bibinfo{author}{{Abe}, M.}, \bibinfo{author}{{Gaskell}, R.},
  \bibinfo{author}{{Palmer}, E.}, \bibinfo{author}{{Barnouin}, O.S.},
  \bibinfo{author}{{Michel}, P.}, \bibinfo{author}{{French}, A.S.},
  \bibinfo{author}{{McMahon}, J.W.}, \bibinfo{author}{{Scheeres}, D.J.},
  \bibinfo{author}{{Abell}, P.A.}, \bibinfo{author}{{Yamamoto}, Y.},
  \bibinfo{author}{{Tanaka}, S.}, \bibinfo{author}{{Shirai}, K.},
  \bibinfo{author}{{Matsuoka}, M.}, \bibinfo{author}{{Yamada}, M.},
  \bibinfo{author}{{Yokota}, Y.}, \bibinfo{author}{{Suzuki}, H.},
  \bibinfo{author}{{Yoshioka}, K.}, \bibinfo{author}{{Cho}, Y.},
  \bibinfo{author}{{Tanaka}, S.}, \bibinfo{author}{{Nishikawa}, N.},
  \bibinfo{author}{{Sugiyama}, T.}, \bibinfo{author}{{Kikuchi}, H.},
  \bibinfo{author}{{Hemmi}, R.}, \bibinfo{author}{{Yamaguchi}, T.},
  \bibinfo{author}{{Ogawa}, N.}, \bibinfo{author}{{Ono}, G.},
  \bibinfo{author}{{Mimasu}, Y.}, \bibinfo{author}{{Yoshikawa}, K.},
  \bibinfo{author}{{Takahashi}, T.}, \bibinfo{author}{{Takei}, Y.},
  \bibinfo{author}{{Fujii}, A.}, \bibinfo{author}{{Hirose}, C.},
  \bibinfo{author}{{Iwata}, T.}, \bibinfo{author}{{Hayakawa}, M.},
  \bibinfo{author}{{Hosoda}, S.}, \bibinfo{author}{{Mori}, O.},
  \bibinfo{author}{{Sawada}, H.}, \bibinfo{author}{{Shimada}, T.},
  \bibinfo{author}{{Soldini}, S.}, \bibinfo{author}{{Yano}, H.},
  \bibinfo{author}{{Tsukizaki}, R.}, \bibinfo{author}{{Ozaki}, M.},
  \bibinfo{author}{{Iijima}, Y.}, \bibinfo{author}{{Ogawa}, K.},
  \bibinfo{author}{{Fujimoto}, M.}, \bibinfo{author}{{Ho}, T.M.},
  \bibinfo{author}{{Moussi}, A.}, \bibinfo{author}{{Jaumann}, R.},
  \bibinfo{author}{{Bibring}, J.P.}, \bibinfo{author}{{Krause}, C.},
  \bibinfo{author}{{Terui}, F.}, \bibinfo{author}{{Saiki}, T.},
  \bibinfo{author}{{Nakazawa}, S.}, \bibinfo{author}{{Tsuda}, Y.},
  \bibinfo{year}{2019}.
\newblock \bibinfo{title}{{Hayabusa2 arrives at the carbonaceous asteroid
  162173 Ryugu{\textemdash}A spinning top-shaped rubble pile}}.
\newblock \bibinfo{journal}{Science} \bibinfo{volume}{364},
  \bibinfo{pages}{268--272}.
\newblock \DOIprefix\doi{10.1126/science.aav8032}.
\bibitem[{{Yamaguchi} et~al.(2023){Yamaguchi}, {Tomioka}, {Ito}, {Shirai},
  {Kimura}, {Greenwood}, {Liu}, {McCain}, {Matsuda}, {Uesugi}, {Imae},
  {Ohigashi}, {Uesugi}, {Nakato}, {Yogata}, {Yuzawa}, {Kodama}, {Hirahara},
  {Sakurai}, {Okada}, {Karouji}, {Nakazawa}, {Okada}, {Saiki}, {Tanaka},
  {Terui}, {Yoshikawa}, {Miyazaki}, {Nishimura}, {Yada}, {Abe}, {Usui},
  {Watanabe} and {Tsuda}}]{2023NatAs...7..398Y}
\bibinfo{author}{{Yamaguchi}, A.}, \bibinfo{author}{{Tomioka}, N.},
  \bibinfo{author}{{Ito}, M.}, \bibinfo{author}{{Shirai}, N.},
  \bibinfo{author}{{Kimura}, M.}, \bibinfo{author}{{Greenwood}, R.C.},
  \bibinfo{author}{{Liu}, M.C.}, \bibinfo{author}{{McCain}, K.A.},
  \bibinfo{author}{{Matsuda}, N.}, \bibinfo{author}{{Uesugi}, M.},
  \bibinfo{author}{{Imae}, N.}, \bibinfo{author}{{Ohigashi}, T.},
  \bibinfo{author}{{Uesugi}, K.}, \bibinfo{author}{{Nakato}, A.},
  \bibinfo{author}{{Yogata}, K.}, \bibinfo{author}{{Yuzawa}, H.},
  \bibinfo{author}{{Kodama}, Y.}, \bibinfo{author}{{Hirahara}, K.},
  \bibinfo{author}{{Sakurai}, I.}, \bibinfo{author}{{Okada}, I.},
  \bibinfo{author}{{Karouji}, Y.}, \bibinfo{author}{{Nakazawa}, S.},
  \bibinfo{author}{{Okada}, T.}, \bibinfo{author}{{Saiki}, T.},
  \bibinfo{author}{{Tanaka}, S.}, \bibinfo{author}{{Terui}, F.},
  \bibinfo{author}{{Yoshikawa}, M.}, \bibinfo{author}{{Miyazaki}, A.},
  \bibinfo{author}{{Nishimura}, M.}, \bibinfo{author}{{Yada}, T.},
  \bibinfo{author}{{Abe}, M.}, \bibinfo{author}{{Usui}, T.},
  \bibinfo{author}{{Watanabe}, S.i.}, \bibinfo{author}{{Tsuda}, Y.},
  \bibinfo{year}{2023}.
\newblock \bibinfo{title}{{Insight into multi-step geological evolution of
  C-type asteroids from Ryugu particles}}.
\newblock \bibinfo{journal}{Nature Astronomy} \bibinfo{volume}{7},
  \bibinfo{pages}{398--405}.
\newblock \DOIprefix\doi{10.1038/s41550-023-01925-x}.
\bibitem[{{Yokoyama} et~al.(2023a){Yokoyama}, {Nagashima}, {Nakai}, {Young},
  {Abe}, {Al{\'e}on}, {Alexander}, {Amari}, {Amelin}, {Bajo}, {Bizzarro},
  {Bouvier}, {Carlson}, {Chaussidon}, {Choi}, {Dauphas}, {Davis}, {Di Rocco},
  {Fujiya}, {Fukai}, {Gautam}, {Haba}, {Hibiya}, {Hidaka}, {Homma}, {Hoppe},
  {Huss}, {Ichida}, {Iizuka}, {Ireland}, {Ishikawa}, {Ito}, {Itoh}, {Kawasaki},
  {Kita}, {Kitajima}, {Kleine}, {Komatani}, {Krot}, {Liu}, {Masuda},
  {McKeegan}, {Morita}, {Motomura}, {Moynier}, {Nguyen}, {Nittler}, {Onose},
  {Pack}, {Park}, {Piani}, {Qin}, {Russell}, {Sakamoto},
  {Sch{\"o}nb{\"a}chler}, {Tafla}, {Tang}, {Terada}, {Terada}, {Usui}, {Wada},
  {Wadhwa}, {Walker}, {Yamashita}, {Yin}, {Yoneda}, {Yui}, {Zhang}, {Connolly},
  {Lauretta}, {Nakamura}, {Naraoka}, {Noguchi}, {Okazaki}, {Sakamoto},
  {Yabuta}, {Abe}, {Arakawa}, {Fujii}, {Hayakawa}, {Hirata}, {Hirata}, {Honda},
  {Honda}, {Hosoda}, {Iijima}, {Ikeda}, {Ishiguro}, {Ishihara}, {Iwata},
  {Kawahara}, {Kikuchi}, {Kitazato}, {Matsumoto}, {Matsuoka}, {Michikami},
  {Mimasu}, {Miura}, {Morota}, {Nakazawa}, {Namiki}, {Noda}, {Noguchi},
  {Ogawa}, {Ogawa}, {Okada}, {Okamoto}, {Ono}, {Ozaki}, {Saiki}, {Sakatani},
  {Sawada}, {Senshu}, {Shimaki}, {Shirai}, {Sugita}, {Takei}, {Takeuchi},
  {Tanaka}, {Tatsumi}, {Terui}, {Tsuda}, {Tsukizaki}, {Wada}, {Watanabe},
  {Yamada}, {Yamada}, {Yamamoto}, {Yano}, {Yokota}, {Yoshihara}, {Yoshikawa},
  {Yoshikawa}, {Furuya}, {Hatakeda}, {Hayashi}, {Hitomi}, {Kumagai},
  {Miyazaki}, {Nakato}, {Nishimura}, {Soejima}, {Suzuki}, {Yada}, {Yamamoto},
  {Yogata}, {Yoshitake}, {Tachibana} and {Yurimoto}}]{2023Sci...379.7850Y}
\bibinfo{author}{{Yokoyama}, T.}, \bibinfo{author}{{Nagashima}, K.},
  \bibinfo{author}{{Nakai}, I.}, \bibinfo{author}{{Young}, E.D.},
  \bibinfo{author}{{Abe}, Y.}, \bibinfo{author}{{Al{\'e}on}, J.},
  \bibinfo{author}{{Alexander}, C.M.O.{\textquoteright}.},
  \bibinfo{author}{{Amari}, S.}, \bibinfo{author}{{Amelin}, Y.},
  \bibinfo{author}{{Bajo}, K.i.}, \bibinfo{author}{{Bizzarro}, M.},
  \bibinfo{author}{{Bouvier}, A.}, \bibinfo{author}{{Carlson}, R.W.},
  \bibinfo{author}{{Chaussidon}, M.}, \bibinfo{author}{{Choi}, B.G.},
  \bibinfo{author}{{Dauphas}, N.}, \bibinfo{author}{{Davis}, A.M.},
  \bibinfo{author}{{Di Rocco}, T.}, \bibinfo{author}{{Fujiya}, W.},
  \bibinfo{author}{{Fukai}, R.}, \bibinfo{author}{{Gautam}, I.},
  \bibinfo{author}{{Haba}, M.K.}, \bibinfo{author}{{Hibiya}, Y.},
  \bibinfo{author}{{Hidaka}, H.}, \bibinfo{author}{{Homma}, H.},
  \bibinfo{author}{{Hoppe}, P.}, \bibinfo{author}{{Huss}, G.R.},
  \bibinfo{author}{{Ichida}, K.}, \bibinfo{author}{{Iizuka}, T.},
  \bibinfo{author}{{Ireland}, T.R.}, \bibinfo{author}{{Ishikawa}, A.},
  \bibinfo{author}{{Ito}, M.}, \bibinfo{author}{{Itoh}, S.},
  \bibinfo{author}{{Kawasaki}, N.}, \bibinfo{author}{{Kita}, N.T.},
  \bibinfo{author}{{Kitajima}, K.}, \bibinfo{author}{{Kleine}, T.},
  \bibinfo{author}{{Komatani}, S.}, \bibinfo{author}{{Krot}, A.N.},
  \bibinfo{author}{{Liu}, M.C.}, \bibinfo{author}{{Masuda}, Y.},
  \bibinfo{author}{{McKeegan}, K.D.}, \bibinfo{author}{{Morita}, M.},
  \bibinfo{author}{{Motomura}, K.}, \bibinfo{author}{{Moynier}, F.},
  \bibinfo{author}{{Nguyen}, A.}, \bibinfo{author}{{Nittler}, L.},
  \bibinfo{author}{{Onose}, M.}, \bibinfo{author}{{Pack}, A.},
  \bibinfo{author}{{Park}, C.}, \bibinfo{author}{{Piani}, L.},
  \bibinfo{author}{{Qin}, L.}, \bibinfo{author}{{Russell}, S.S.},
  \bibinfo{author}{{Sakamoto}, N.}, \bibinfo{author}{{Sch{\"o}nb{\"a}chler},
  M.}, \bibinfo{author}{{Tafla}, L.}, \bibinfo{author}{{Tang}, H.},
  \bibinfo{author}{{Terada}, K.}, \bibinfo{author}{{Terada}, Y.},
  \bibinfo{author}{{Usui}, T.}, \bibinfo{author}{{Wada}, S.},
  \bibinfo{author}{{Wadhwa}, M.}, \bibinfo{author}{{Walker}, R.J.},
  \bibinfo{author}{{Yamashita}, K.}, \bibinfo{author}{{Yin}, Q.Z.},
  \bibinfo{author}{{Yoneda}, S.}, \bibinfo{author}{{Yui}, H.},
  \bibinfo{author}{{Zhang}, A.C.}, \bibinfo{author}{{Connolly}, H.C.},
  \bibinfo{author}{{Lauretta}, D.S.}, \bibinfo{author}{{Nakamura}, T.},
  \bibinfo{author}{{Naraoka}, H.}, \bibinfo{author}{{Noguchi}, T.},
  \bibinfo{author}{{Okazaki}, R.}, \bibinfo{author}{{Sakamoto}, K.},
  \bibinfo{author}{{Yabuta}, H.}, \bibinfo{author}{{Abe}, M.},
  \bibinfo{author}{{Arakawa}, M.}, \bibinfo{author}{{Fujii}, A.},
  \bibinfo{author}{{Hayakawa}, M.}, \bibinfo{author}{{Hirata}, N.},
  \bibinfo{author}{{Hirata}, N.}, \bibinfo{author}{{Honda}, R.},
  \bibinfo{author}{{Honda}, C.}, \bibinfo{author}{{Hosoda}, S.},
  \bibinfo{author}{{Iijima}, Y.i.}, \bibinfo{author}{{Ikeda}, H.},
  \bibinfo{author}{{Ishiguro}, M.}, \bibinfo{author}{{Ishihara}, Y.},
  \bibinfo{author}{{Iwata}, T.}, \bibinfo{author}{{Kawahara}, K.},
  \bibinfo{author}{{Kikuchi}, S.}, \bibinfo{author}{{Kitazato}, K.},
  \bibinfo{author}{{Matsumoto}, K.}, \bibinfo{author}{{Matsuoka}, M.},
  \bibinfo{author}{{Michikami}, T.}, \bibinfo{author}{{Mimasu}, Y.},
  \bibinfo{author}{{Miura}, A.}, \bibinfo{author}{{Morota}, T.},
  \bibinfo{author}{{Nakazawa}, S.}, \bibinfo{author}{{Namiki}, N.},
  \bibinfo{author}{{Noda}, H.}, \bibinfo{author}{{Noguchi}, R.},
  \bibinfo{author}{{Ogawa}, N.}, \bibinfo{author}{{Ogawa}, K.},
  \bibinfo{author}{{Okada}, T.}, \bibinfo{author}{{Okamoto}, C.},
  \bibinfo{author}{{Ono}, G.}, \bibinfo{author}{{Ozaki}, M.},
  \bibinfo{author}{{Saiki}, T.}, \bibinfo{author}{{Sakatani}, N.},
  \bibinfo{author}{{Sawada}, H.}, \bibinfo{author}{{Senshu}, H.},
  \bibinfo{author}{{Shimaki}, Y.}, \bibinfo{author}{{Shirai}, K.},
  \bibinfo{author}{{Sugita}, S.}, \bibinfo{author}{{Takei}, Y.},
  \bibinfo{author}{{Takeuchi}, H.}, \bibinfo{author}{{Tanaka}, S.},
  \bibinfo{author}{{Tatsumi}, E.}, \bibinfo{author}{{Terui}, F.},
  \bibinfo{author}{{Tsuda}, Y.}, \bibinfo{author}{{Tsukizaki}, R.},
  \bibinfo{author}{{Wada}, K.}, \bibinfo{author}{{Watanabe}, S.i.},
  \bibinfo{author}{{Yamada}, M.}, \bibinfo{author}{{Yamada}, T.},
  \bibinfo{author}{{Yamamoto}, Y.}, \bibinfo{author}{{Yano}, H.},
  \bibinfo{author}{{Yokota}, Y.}, \bibinfo{author}{{Yoshihara}, K.},
  \bibinfo{author}{{Yoshikawa}, M.}, \bibinfo{author}{{Yoshikawa}, K.},
  \bibinfo{author}{{Furuya}, S.}, \bibinfo{author}{{Hatakeda}, K.},
  \bibinfo{author}{{Hayashi}, T.}, \bibinfo{author}{{Hitomi}, Y.},
  \bibinfo{author}{{Kumagai}, K.}, \bibinfo{author}{{Miyazaki}, A.},
  \bibinfo{author}{{Nakato}, A.}, \bibinfo{author}{{Nishimura}, M.},
  \bibinfo{author}{{Soejima}, H.}, \bibinfo{author}{{Suzuki}, A.},
  \bibinfo{author}{{Yada}, T.}, \bibinfo{author}{{Yamamoto}, D.},
  \bibinfo{author}{{Yogata}, K.}, \bibinfo{author}{{Yoshitake}, M.},
  \bibinfo{author}{{Tachibana}, S.}, \bibinfo{author}{{Yurimoto}, H.},
  \bibinfo{year}{2023}a.
\newblock \bibinfo{title}{{Samples returned from the asteroid Ryugu are similar
  to Ivuna-type carbonaceous meteorites}}.
\newblock \bibinfo{journal}{Science} \bibinfo{volume}{379},
  \bibinfo{pages}{abn7850}.
\newblock \DOIprefix\doi{10.1126/science.abn7850}.
\bibitem[{{Yokoyama} et~al.(2023b){Yokoyama}, {Wadhwa}, {Iizuka}, {Rai},
  {Gautam}, {Hibiya}, {Masuda}, {Haba}, {Fukai}, {Hines}, {Phelan}, {Abe},
  {Al{\'e}on}, {Alexander}, {Amari}, {Amelin}, {Bajo}, {Bizzarro}, {Bouvier},
  {Carlson}, {Chaussidon}, {Choi}, {Dauphas}, {Davis}, {Di Rocco}, {Fujiya},
  {Hidaka}, {Homma}, {Hoppe}, {Huss}, {Ichida}, {Ireland}, {Ishikawa}, {Itoh},
  {Kawasaki}, {Kita}, {Kitajima}, {Kleine}, {Komatani}, {Krot}, {Liu},
  {McKeegan}, {Morita}, {Motomura}, {Moynier}, {Nakai}, {Nagashima}, {Nguyen},
  {Nittler}, {Onose}, {Pack}, {Park}, {Piani}, {Qin}, {Russell}, {Sakamoto},
  {Sch{\"o}nb{\"a}chler}, {Tafla}, {Tang}, {Terada}, {Terada}, {Usui}, {Wada},
  {Walker}, {Yamashita}, {Yin}, {Yoneda}, {Young}, {Yui}, {Zhang}, {Nakamura},
  {Naraoka}, {Noguchi}, {Okazaki}, {Sakamoto}, {Yabuta}, {Abe}, {Miyazaki},
  {Nakato}, {Nishimura}, {Okada}, {Yada}, {Yogata}, {Nakazawa}, {Saiki},
  {Tanaka}, {Terui}, {Tsuda}, {Watanabe}, {Yoshikawa}, {Tachibana} and
  {Yurimoto}}]{2023SciA....9I7048Y}
\bibinfo{author}{{Yokoyama}, T.}, \bibinfo{author}{{Wadhwa}, M.},
  \bibinfo{author}{{Iizuka}, T.}, \bibinfo{author}{{Rai}, V.},
  \bibinfo{author}{{Gautam}, I.}, \bibinfo{author}{{Hibiya}, Y.},
  \bibinfo{author}{{Masuda}, Y.}, \bibinfo{author}{{Haba}, M.K.},
  \bibinfo{author}{{Fukai}, R.}, \bibinfo{author}{{Hines}, R.},
  \bibinfo{author}{{Phelan}, N.}, \bibinfo{author}{{Abe}, Y.},
  \bibinfo{author}{{Al{\'e}on}, J.}, \bibinfo{author}{{Alexander}, C.M.O.D.},
  \bibinfo{author}{{Amari}, S.}, \bibinfo{author}{{Amelin}, Y.},
  \bibinfo{author}{{Bajo}, K.i.}, \bibinfo{author}{{Bizzarro}, M.},
  \bibinfo{author}{{Bouvier}, A.}, \bibinfo{author}{{Carlson}, R.W.},
  \bibinfo{author}{{Chaussidon}, M.}, \bibinfo{author}{{Choi}, B.G.},
  \bibinfo{author}{{Dauphas}, N.}, \bibinfo{author}{{Davis}, A.M.},
  \bibinfo{author}{{Di Rocco}, T.}, \bibinfo{author}{{Fujiya}, W.},
  \bibinfo{author}{{Hidaka}, H.}, \bibinfo{author}{{Homma}, H.},
  \bibinfo{author}{{Hoppe}, P.}, \bibinfo{author}{{Huss}, G.R.},
  \bibinfo{author}{{Ichida}, K.}, \bibinfo{author}{{Ireland}, T.},
  \bibinfo{author}{{Ishikawa}, A.}, \bibinfo{author}{{Itoh}, S.},
  \bibinfo{author}{{Kawasaki}, N.}, \bibinfo{author}{{Kita}, N.T.},
  \bibinfo{author}{{Kitajima}, K.}, \bibinfo{author}{{Kleine}, T.},
  \bibinfo{author}{{Komatani}, S.}, \bibinfo{author}{{Krot}, A.N.},
  \bibinfo{author}{{Liu}, M.C.}, \bibinfo{author}{{McKeegan}, K.D.},
  \bibinfo{author}{{Morita}, M.}, \bibinfo{author}{{Motomura}, K.},
  \bibinfo{author}{{Moynier}, F.}, \bibinfo{author}{{Nakai}, I.},
  \bibinfo{author}{{Nagashima}, K.}, \bibinfo{author}{{Nguyen}, A.},
  \bibinfo{author}{{Nittler}, L.}, \bibinfo{author}{{Onose}, M.},
  \bibinfo{author}{{Pack}, A.}, \bibinfo{author}{{Park}, C.},
  \bibinfo{author}{{Piani}, L.}, \bibinfo{author}{{Qin}, L.},
  \bibinfo{author}{{Russell}, S.}, \bibinfo{author}{{Sakamoto}, N.},
  \bibinfo{author}{{Sch{\"o}nb{\"a}chler}, M.}, \bibinfo{author}{{Tafla}, L.},
  \bibinfo{author}{{Tang}, H.}, \bibinfo{author}{{Terada}, K.},
  \bibinfo{author}{{Terada}, Y.}, \bibinfo{author}{{Usui}, T.},
  \bibinfo{author}{{Wada}, S.}, \bibinfo{author}{{Walker}, R.J.},
  \bibinfo{author}{{Yamashita}, K.}, \bibinfo{author}{{Yin}, Q.Z.},
  \bibinfo{author}{{Yoneda}, S.}, \bibinfo{author}{{Young}, E.D.},
  \bibinfo{author}{{Yui}, H.}, \bibinfo{author}{{Zhang}, A.C.},
  \bibinfo{author}{{Nakamura}, T.}, \bibinfo{author}{{Naraoka}, H.},
  \bibinfo{author}{{Noguchi}, T.}, \bibinfo{author}{{Okazaki}, R.},
  \bibinfo{author}{{Sakamoto}, K.}, \bibinfo{author}{{Yabuta}, H.},
  \bibinfo{author}{{Abe}, M.}, \bibinfo{author}{{Miyazaki}, A.},
  \bibinfo{author}{{Nakato}, A.}, \bibinfo{author}{{Nishimura}, M.},
  \bibinfo{author}{{Okada}, T.}, \bibinfo{author}{{Yada}, T.},
  \bibinfo{author}{{Yogata}, K.}, \bibinfo{author}{{Nakazawa}, S.},
  \bibinfo{author}{{Saiki}, T.}, \bibinfo{author}{{Tanaka}, S.},
  \bibinfo{author}{{Terui}, F.}, \bibinfo{author}{{Tsuda}, Y.},
  \bibinfo{author}{{Watanabe}, S.i.}, \bibinfo{author}{{Yoshikawa}, M.},
  \bibinfo{author}{{Tachibana}, S.}, \bibinfo{author}{{Yurimoto}, H.},
  \bibinfo{year}{2023}b.
\newblock \bibinfo{title}{{Water circulation in Ryugu asteroid affected the
  distribution of nucleosynthetic isotope anomalies in returned sample}}.
\newblock \bibinfo{journal}{Science Advances} \bibinfo{volume}{9},
  \bibinfo{pages}{eadi7048}.
\newblock \DOIprefix\doi{10.1126/sciadv.adi7048}.
\bibitem[{{Young} et~al.(2003){Young}, {Zhang} and
  {Schubert}}]{2003E&PSL.213..249Y}
\bibinfo{author}{{Young}, E.D.}, \bibinfo{author}{{Zhang}, K.K.},
  \bibinfo{author}{{Schubert}, G.}, \bibinfo{year}{2003}.
\newblock \bibinfo{title}{{Conditions for pore water convection within
  carbonaceous chondrite parent bodies - implications for planetesimal size and
  heat production}}.
\newblock \bibinfo{journal}{Earth and Planetary Science Letters}
  \bibinfo{volume}{213}, \bibinfo{pages}{249--259}.
\newblock \DOIprefix\doi{10.1016/S0012-821X(03)00345-5}.
\bibitem[{{Zhang}(2023)}]{2023ApJ...956L..25Z}
\bibinfo{author}{{Zhang}, Z.}, \bibinfo{year}{2023}.
\newblock \bibinfo{title}{{Ice Sublimation in Planetesimals Formed at the
  Outward Migrating Snowline}}.
\newblock \bibinfo{journal}{The Astrophysical Journal Letters}
  \bibinfo{volume}{956}, \bibinfo{pages}{L25}.
\newblock \DOIprefix\doi{10.3847/2041-8213/acfdaa}.

\end{thebibliography}






\end{document}